\documentclass[a4paper,11pt]{article}
\pdfoutput=1 % if your are submitting a pdflatex (i.e. if you have
\usepackage{jheppub,tikz} % for details on the use of the package, please
\usetikzlibrary{shapes.geometric}

\usepackage[T1]{fontenc} % if needed  

\newcommand{\nl}{\nonumber \\} 
 
\newcommand{\beq}{\begin{equation}}
\newcommand{\eeq}{\end{equation}}
\newcommand{\bea}{\begin{eqnarray}}
\newcommand{\eea}{\end{eqnarray}}
\newcommand{\beas}{\begin{eqnarray*}}
\newcommand{\eeas}{\end{eqnarray*}}

\newcommand{\bquo}{\begin{quote}}
\newcommand{\enqu}{\end{quote}}
\newcommand{\p}{\partial}
\def\stroke{\vrule height8pt width0.4pt depth-0.1pt}
\def\topfleck{\vrule height8pt width0.5pt depth-5.9pt}
\def\botfleck{\vrule height2pt width0.5pt depth0.1pt}
\def\Zmath{\vcenter{\hbox{\numbers\rlap{\rlap{Z}\kern 0.8pt\topfleck}\kern
2.2pt
                   \rlap Z\kern 6pt\botfleck\kern 1pt}}}
\def\Qmath{\vcenter{\hbox{\upright\rlap{\rlap{Q}\kern
                   3.8pt\stroke}\phantom{Q}}}}
\def\Nmath{\vcenter{\hbox{\upright\rlap{I}\kern 1.7pt N}}}
\def\Cmath{\vcenter{\hbox{\upright\rlap{\rlap{C}\kern
                   3.8pt\stroke}\phantom{C}}}}
\def\Rmath{\vcenter{\hbox{\upright\rlap{I}\kern 1.7pt R}}}
\def\Z{\ifmmode\Zmath\else$\Zmath$\fi}
\def\Q{\ifmmode\Qmath\else$\Qmath$\fi}
\def\N{\ifmmode\Nmath\else$\Nmath$\fi}
\def\C{\ifmmode\Cmath\else$\Cmath$\fi}
\def\R{\ifmmode\Rmath\else$\Rmath$\fi}
\def\2{{1\over 2}}

\def\4N{${\cal N}=4$}

\def\beq{\begin{equation}}
\def\eeq{\end{equation}}
\def\ba{\beq\new\begin{array}{c}}
\def\ea{\end{array}\eeq}

\def\a{\alpha}

\def\b{\beta}

\def\l{\lambda}

\def\s{\sigma}

\def\ep{\epsilon}

\def\ba{{\bar{a}}}
\def\p{\partial}

\def\N{\nabla}

\def\le{\left(}
\def\ri{\right)}

\def\Q{\tilde{Q}}

\def\hhz{{\hat{z}}}

\def\dia{
\raisebox{-0.1cm}{\!\!\resizebox{0.039\textwidth}{!}{
\begin{tikzpicture} 
\draw[ultra thick] (-1,0) -- (0,1);
\draw[ultra thick] (0,1) -- (1,0); 
\draw[ultra thick] (1,0) -- (0,-1); 
\draw[ultra thick] (0,-1) -- (-1,0); 
\end{tikzpicture} }}
}
 
\def\pentau{ \raisebox{-0.1cm}{\!\!
\resizebox{0.04\textwidth}{!}{
\begin{tikzpicture}
\draw[ultra thick] (-1,0) -- (-0.4,0.6);
\draw[ultra thick] (-0.4,0.6) -- (0.4,0.6);
\draw[ultra thick] (0.4,0.6) -- (1,0); 
\draw[ultra thick] (1,0) -- (0,-1); 
\draw[ultra thick] (0,-1) -- (-1,0); 
\end{tikzpicture} }}
}

\def\pentad{\raisebox{-0.1cm}{\!\!
\resizebox{0.04\textwidth}{!}{
\begin{tikzpicture}
\draw[ultra thick] (-1,0) -- (0,1);
\draw[ultra thick] (0,1) -- (1,0); 
\draw[ultra thick] (1,0) -- (0.4,-0.6);
\draw[ultra thick ] (0.4,-0.6) -- (-0.4,-0.6);
\draw[ultra thick] (-0.4,-0.6) -- (-1,0); 
\end{tikzpicture} }}}

\def\exa{\raisebox{-0.05cm}{\!\!
\resizebox{0.042\textwidth}{!}{
\begin{tikzpicture}
\draw[ultra thick] (-1,0) -- (-0.4,0.6);
\draw[ultra thick ] (-0.4,0.6) -- (0.4,0.6);
\draw[ultra thick] (0.4,0.6) -- (1,0); 
\draw[ultra thick] (1,0) -- (0.4,-0.6);
\draw[ultra thick ] (0.4,-0.6) -- (-0.4,-0.6);
\draw[ultra thick] (-0.4,-0.6) -- (-1,0); 
\end{tikzpicture} }}}

\title{\boldmath  % Complexity as a bridge over troubled geometries \\ or \\
On the time dependence of holographic complexity for charged AdS black holes with scalar hair 
}

\author[a]{Roberto Auzzi,}
\author[b]{Stefano Bolognesi,}
\author[c,d]{Eliezer Rabinovici,}
\author[d]{Fidel I. Schaposnik Massolo,}
\author[e]{Gianni Tallarita}

\affiliation[a]{Dipartimento di Matematica e Fisica, Universit\`a Cattolica del Sacro Cuore, Via Musei 41, 25121 Brescia, Italy,
2 and 
INFN Sezione di Perugia, Via A. Pascoli, 06123 Perugia, Italy}
\affiliation[b]{Department of Physics  E. Fermi, University of Pisa
and INFN Sezione di Pisa \\
Largo Pontecorvo, 3, Ed. C, 56127 Pisa, Italy}
\affiliation[c]{Racah Institute of Physics, The Hebrew University of Jerusalem, 91904, Israel}
\affiliation[d]{Institut des Hautes Etudes Scientifiques, 35 route de Chartres, 91440 Bures-sur-Yvette, France}
\affiliation[e]{Departamento de Ciencias, Facultad de Artes Liberales,
Universidad Adolfo Ib\'a\~nez, \\  Santiago 7941169, Chile}

\emailAdd{roberto.auzzi@unicatt.it}
\emailAdd{stefano.bolognesi@unipi.it}
\emailAdd{eliezer@mail.huji.ac.il}
\emailAdd{fidels@ihes.fr}
\emailAdd{gianni.tallarita@uai.cl}

\abstract{
In the presence of a scalar hair perturbation, the Cauchy horizon of
a Reissner-Nordstr\"om black hole disappears and is replaced
by the rapid collapse of the Einstein-Rosen bridge,
which leads to a Kasner singularity \cite{Hartnoll:2020rwq,Hartnoll:2020fhc}.
We study the time-dependence of holographic complexity, both for the volume
and for the action proposals, in a class of models with hairy black holes.
Volume complexity can only probe a portion of the black hole
interior that remains far away from the Kasner singularity.
We provide numerical evidence that the Lloyd bound is
 satisfied by the volume complexity rate in all the parameter space
 that we explored. Action complexity can instead  probe a portion of the spacetime 
closer to the singularity. In particular,  the complexity rate diverges 
at the critical time $t_c$ for which the Wheeler-DeWitt patch touches the singularity.
After the critical time the action complexity rate approaches a constant.
We find that the Kasner exponent does not directly affect the details
of the divergence of the complexity rate at $t=t_c$
 and the late-time behaviour of the complexity.
 The Lloyd bound is violated by action complexity
 at finite time, because the complexity rate diverges at $t=t_c$.
We find that the Lloyd bound is satisfied by the asymptotic action complexity
 rate in all the parameter space that we investigated.
}

\begin{document} 
\maketitle
\flushbottom

\section{Introduction and conclusions}
\label{sec:intro}

The AdS/CFT correspondence  constitutes a tool one can use to relate
the  problem of quantum gravity in asymptotically anti de Sitter (AdS) spacetimes
to the study of a Conformal Field Theory (CFT) defined on the boundary of the spacetime. 
Quantum information concepts such as entanglement entropy \cite{Ryu:2006bv,Hubeny:2007xt}
play an important role in reconstructing the geometry close to event horizons.
However, despite many advances, we do not currently have
a clear picture of the emergence of various aspects of the spacetime geometry from
the physics of the CFT living on the boundary.
For example, it turns out that the entanglement entropy is not enough \cite{Susskind:2014moa}
to probe the growth of the Einstein-Rosen Bridge  (ERB) 
inside the event horizon of a Black Hole (BH).

In  \cite{Susskind:2014rva}
it was proposed that computational complexity
may play an important role in understanding the black hole interior.
Quantum computational complexity is a quantum information 
concept.
For a quantum system, one can heuristically define complexity
as the minimal number of elementary operations that are needed to prepare
a given state starting from a reference one. A continuous and bounded version
of complexity, following Nielsen \cite{Nielsen1},
 can be defined  in terms of geodesics in the space of unitary Hamiltonians.
 There is a great deal of ambiguity in the definition of complexity, due to
the choice of the reference state and of the computational cost of the elementary 
operations. It is not yet clear which one of the many 
possible choices of these details of computational complexity might be relevant in its precise definition for holography.
For systems with a finite number of degrees of freedom, it is expected
that the maximal complexity scales exponentially with the number of qubits.
 Negative curvature \cite{Brown:2016wib,Brown:2017jil,Brown:2019whu,Auzzi:2020idm,Brown:2021euk,Basteiro:2021ene} 
is an important ingredient to realise this property.
The study of complexity for quantum systems with an infinite number of degrees of freedom
is still in its preliminary stages, with much progress being made in defining complexity for free field theories
 \cite{Jefferson:2017sdb,Chapman:2017rqy,Khan:2018rzm,Hackl:2018ptj}. However, a definition of complexity in interacting CFT is still lacking,
 see \cite{Caputa:2017urj,Caputa:2018kdj,Erdmenger:2020sup,Flory:2020eot,Chagnet:2021uvi,Koch:2021tvp}
 for some advances in this direction.
 
 Two main quantities have been proposed as bulk holographic duals to computational complexity.
The Complexity=Volume (CV) proposal \cite{Stanford:2014jda} relates the complexity to the maximum volume $V$ of the 
codimension-one surface anchored to the given boundary time,
\beq
C_V = \frac{V}{G L} \, ,
\label{CV-formula}
\eeq
where $G$ is the Newton constant and $L$ is the AdS radius. 
The Complexity=Action (CA) conjecture \cite{Brown:2015bva,Brown:2015lvg} proposes instead that complexity is proportional
to the classical action $I_{WDW}$ evaluated on the Wheeler-DeWitt patch, which is defined as
the bulk domain of dependence of the maximal slice 
attached to the boundary time, \emph{i.e.}
\beq
C_A = \frac{I_{WDW}}{\pi} \, .
\label{CA-formula}
\eeq
Due to the large amount of arbitrariness in the definition of computational complexity,
it could be that both of these holographic duals, and also possible generalisations thereof \cite{Couch:2016exn,Belin:2021bga}, may
correspond to different prescriptions to define the
computational complexity on the dual CFT, perhaps sharing some similarities.
  
In recent years both the CV and CA proposals
 have been tested in a variety of holographic models and in different backgrounds. 
There are a few universal properties that we would require from any notion of complexity to be acceptable,
and which are  reproduced by both prescriptions, such as  the switchback effect \cite{Stanford:2014jda,Susskind:2014jwa},
the structure of the  UV divergencies \cite{Carmi:2016wjl,Reynolds:2016rvl,Akhavan:2019zax,Omidi:2020oit} and
 the linear growth for a parametrically large time after thermalization.
In particular, it is expected that quantum complexity increases
linearly for a time which is exponential in the entropy of the system \cite{Susskind:2015toa}.
This regime is reproduced by the classical gravity dual 
\cite{Stanford:2014jda,Brown:2015bva,Brown:2015lvg,Cai:2016xho,Carmi:2017jqz,Yang:2017czx,Auzzi:2018zdu,Auzzi:2018pbc,Bernamonti:2021jyu}, which gives
a late-time growth which is  linear in the boundary  time $t_b$
\beq
\lim_{t_b \to \infty } \frac{d \, C_{V}}{ d t_b} = W_{V} \, , \qquad
\lim_{t_b \to \infty }  \frac{d \, I_{WDW}}{ d t_b} = W_{A} \, ,
\label{coefficienti-crescita-complessita}
\eeq
where the coefficients $W_{V}$ and $W_A$ are proportional, up to an order-one coefficient,
 to $T S$, where $T$ is the temperature and $S$ the entropy of the system. 
After this linear growth epoch, complexity  should experience a plateau \cite{Susskind:2015toa}
and then, after the recurrence time (which is double exponential in the entropy of the system)
complexity should become small again.  This very late-time behavior
is probably related to gravitational quantum corrections,
as studied in \cite{Iliesiu:2021ari} for two-dimensional gravity.

 In this work we will test both the volume and the action
 conjectures in certain backgrounds of charged black holes with scalar fields, where
  the inner Cauchy horizon  disappears due to a small perturbation of the BH exterior
   and the singularity becomes space-like and of Kasner type. 
We will focus on the case of an eternal black hole with two disconnected boundaries,
which on the field theory side is dual to the thermofield double state \cite{Maldacena:2001kr},
which has the following schematic form
 \beq
|\Psi_{TFD} \rangle   \propto   \sum_n  e^{-E_n \beta/2- i E_n \, t_b} | E_{n} \rangle_R  | E_{n} \rangle_L \, ,
\eeq
where  $\beta$ is the inverse temperature and
 $| E_{n} \rangle_{L,R}$ are the energy eigenstates of left and right boundary theories, $t_b$ being again the boundary time.

In section \ref{section-model} we will introduce the model
 and describe various solutions of charged black holes.     This section mostly consists of a review of previous works. We rederive the numerical solutions in order to use them in the rest of the paper. 
We will work with Einstein-Maxwell theory in four dimensions with negative
 cosmological constant, and thus an asymptotic AdS$_4$ background.
 This is dual to a strongly-coupled $3$-dimensional CFT with a $U(1)$ symmetry. 
 We will consider black holes in the Poincar\'e patch with the metric
 \beq
ds^2= \frac{1}{z^2} \le -f e^{-\chi} dt^2 +\frac{dz^2}{f} +dx^2 +dy^2 \ri \, ,
\eeq
where $f$ and $\chi$ are functions of the radial AdS coordinate $z$.
The Reissner-Nordstr\"om (RN) black hole solution provides a gravity dual 
to  the CFT states  with finite temperature and chemical potential. 
As in the case of asymptotically flat spacetime \cite{Simpson:1973ua,Chandrasekhar-Hartle}, 
the Cauchy horizon inside the event horizon of the black hole 
is unstable w.r.t. small perturbations. In order to detect this instability 
one may consider a scalar perturbation  \cite{Hartnoll:2020rwq,Hartnoll:2020fhc},
which on the CFT side is dual to an operator whose dimension is related to the mass.

The deformation by a neutral scalar field was studied in detail in \cite{Hartnoll:2020rwq}.
 Turning  on a relevant deformation in the CFT  by adding a source for the scalar field has a dramatic
  effect in the inner structure on the black hole. The Cauchy horizon is no longer present and 
   the causal structure of the BH becomes closer to that of an eternal Schwarzschild BH.
    The singularity is of Kasner type, with a Kasner
     coefficient that depends on the various external parameters: the temperature, the chemical potential 
     and the magnitude of the source for the scalar operator. For small values of the source, the black
      hole remains close to the RN solution only up to the ``would be'' Cauchy horizon. In this context, another interesting 
      new phenomenon that has been observed in \cite{Hartnoll:2020rwq} is that the scalar field condensation becomes
       important and backreacts on the geometry, the most prominent consequence of this being an exponential decay
        of $g_{tt}$. This phenomenon was called the ``collapse of the Einstein-Rosen (ER) bridge''. 
         After this period the metric flows to a Kasner singularity, with
         \beq
f=-f_0 z^{3+\a^2} \, , 
\qquad
\chi=2 \a^2 \log z + \chi_0 \, ,
\label{metrica-kasner-intro}
\eeq
where $\chi_0$ and $f_0$ are integration constants and $\a$ 
can be used to parametrize the Kasner exponents, \emph{i.e.}
\beq
p_t=\frac{\a^2-1}{3+\a^2} \, , \qquad p_{x}=\frac{2}{3+\a^2} \, .
\eeq
whose definition will be recalled later in (\ref{defkasmetric}).

For a charged scalar field,  even without an external source, the phenomenon of spontaneous condensation below a certain temperature is observed, corresponding to the holographic superconductor \cite{Hartnoll:2008vx,Hartnoll:2008kx}.   
In this case there is also a resolution of the Cauchy horizon into a Kasner singularity, and the collapse of the ER bridge \cite{Hartnoll:2020fhc}. 
Other interesting phenomena have been observed inside the horizon, such as Josephson-type oscillations of the scalar order parameter or
the inversion of the Kasner exponent close to the singularity. As in the neutral case, the metric nearby the singularity is again of the form \eqref{metrica-kasner-intro}.

The regime where the scalar hair is small outside the Cauchy horizon
 is most challenging for numerical approaches, since it is in these conditions where the collapse
 of the bridge is more rapid and abrupt.
 The case of charged scalar is particularly interesting due to the possible presence of Kasner inversions,
 in which the Einsten-Rosen bridge can experience a period of expansion
 followed by a final contraction towards the singularity. In the limit
 where the scalar is small outside the Cauchy horizon, the number of Kasner
 inversions depends in a very sensitive way on the scalar hair.
Close to the critical temperature of the holographic superconductor this 
also occurs in the absence of an external source.
 For fine-tuned values of the scalar, an infinite number of transitions 
 can take place, and the solution has a fractal-like behaviour in the external source
 and temperature parameters \cite{Hartnoll:2020fhc}.   Here we also present some new results regarding the complete phase diagram of the charged scalar field case also in presence of a source, see Figure \ref{para-spazio-charged}.

A natural question to ask is how the echos of this  bulk gravitational chaotic behaviour  inside the event horizon
may be reflected in the physics of the conformal field theory on the boundary.
Complexity is then a natural probe to consider.
  For an eternal black hole, the CV proposal implies the computation of
   the volume of a maximal spatial slice which is anchored at the two sides of the black hole. 
 Going from one side to the other inevitably requires crossing the ER bridge,
 and thus allows us to glance inside the horizon of the black hole.
   The linear growth of volume complexity at late times is indeed related to the growth of the ER bridge,
   and it is therefore natural to wonder if it can detect the collapse of the bridge.

 In section \ref{section-volume} we investigate the volume conjecture and find that
 this proposal does not probe the collapse of the bridge and the Kasner behavior,
 because the extremal codimension one surface gets stuck far away from the would be Cauchy 
   horizon, well before the new phenomena discovered in \cite{Hartnoll:2020rwq,Hartnoll:2020fhc}
   start to take place. Volume compexity for holographic superconductors have been studyed before in   \cite{Yang:2019gce}.  Respect to  \cite{Yang:2019gce} we investigate a more general class of solutions which includes the neutral scalar case  
and the charged scalar case with external  source.
We find that the Lloyd bound \cite{Lloyd,Brown:2015lvg}  is always satisfied 
in the parameter space that we investigate, both with Dirichlet and Neumann 
definition of the mass.

Section \ref{section-action-complexity} contains the study of action complexity in this class of backgrounds. This is the main new result of the paper.  
  The action prescription for complexity  requires the computation of the action of the Wheeler-DeWitt (WDW) 
  wedge which is anchored at the two sides of the black hole.
   It can be thought of as the union of all possible spatial slices with the given boundary conditions. 
       In contrast to the maximal spatial slice, the
    WDW wedge can touch the singularity and thus, 
    in principle, can also probe the region nearby the black hole singularity. 
   In the discussion of action complexity we must distinguish two cases, 
   according to the shape of the Penrose diagram which describes the causal structure of the black hole solution  \cite{Fidkowski:2003nf}.
Indeed, the Penrose diagram can be schematically drawn as a square, where the vertical sides correspond
    to the left and right boundaries where each of the entangled CFTs  of the thermofield double state live.
    The spacetime singularity can then bend the top and bottom sides inwards or outwards, compared to the horizontal side of the square
    inside the black hole horizon.   This will be reviewed in Appendix \ref{penrose-appendix}.  This distinction is a conformally invariant
    property of the black hole solution which does not depend on the arbitrary choice
    of functions used in the conformal mapping that was used to construct the diagram. 
        In this paper we will denote a solution with a lower bending of the singularity as being of ``type $D$'',
    while one with an upper bending will be referred to as of ``type $U$''.
For example, the Schwarzschild solution in AdS$_d$ with  $d>3$ is of type $D$.
  In the case of the black holes with scalar hair discussed in this paper, both type $D$  
and type $U$ solutions can be realised. 
 \begin{figure}[h]
\begin{center}
\includegraphics[scale=1.]{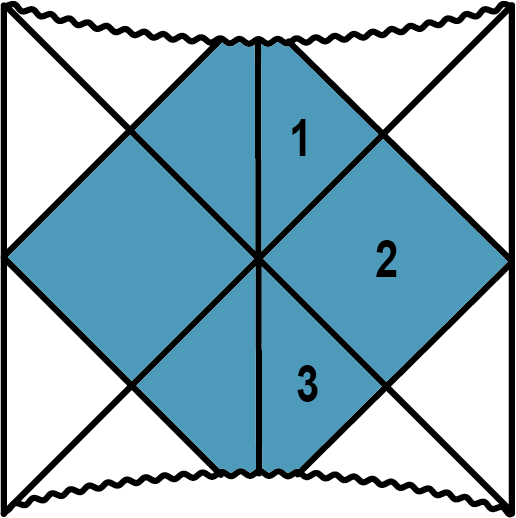} \qquad
\includegraphics[scale=1.]{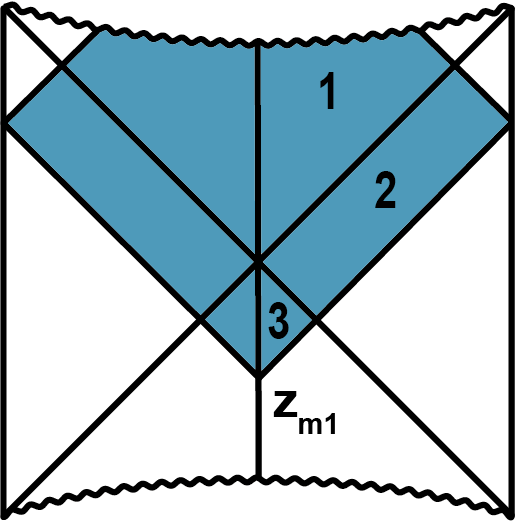}
\caption{
WDW patch for a type $D$ solution. Left: before $t_c$. Right: after $t_c$.
}
\label{fig-caso1}
\end{center}
\end{figure}

The eternal black hole solution is symmetric under time reflection $t_b \to -t_b$, and so it is enough
   to study the behaviour of complexity for $t_b>0$. The time dependence of the WDW patch for 
   type $D$ solutions is shown in figure \ref{fig-caso1}.
   At $t_b=0$, the WDW patch has a hexagonal shape  \exa.
   As the boundary time increases, at some critical time $t_b=t_c$
   the WDW wedge experiences a discontinuous transition \exa $\to$\pentau.
   For $t<t_c$, the complexity rate is identically zero.
   When the tip of the WDW (a null-like joint) forms at the singularity inside the white hole horizon,
    there is a peculiar logarithmic divergence in the complexity rate
    \beq
\frac{d I_{\rm WDW}}{d t_b} \approx  \frac{V_0}{16 \pi G} A(\infty) \, \log ( t_b - t_c) \, , \qquad t_b>t_c \, ,
\eeq
where $V_0$ is the volume of the boundary theory and  $A(\infty)>0$ is given by eq. (\ref{A-asintotico}).
When the limit  $t_b \to t_c$ is approached from $t_b>t_c$, the complexity 
rate tends to $- \infty$. This is the same kind of behaviour
as in the Schwarzschild case \cite{Carmi:2017jqz,Yang:2017czx}.

\begin{figure}[h]
\begin{center}
\includegraphics[scale=1.]{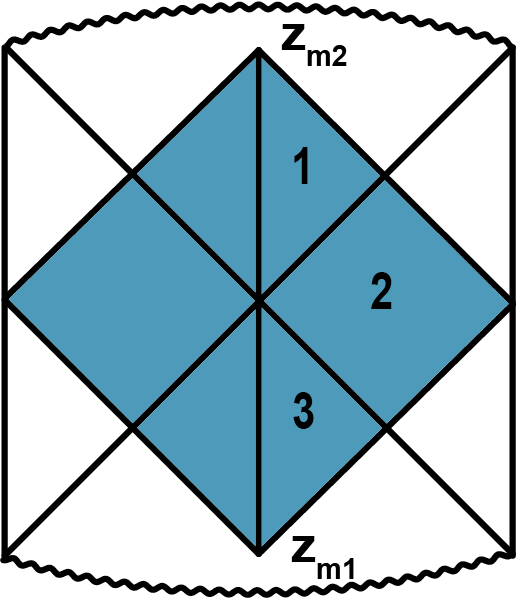} \qquad
\includegraphics[scale=1.]{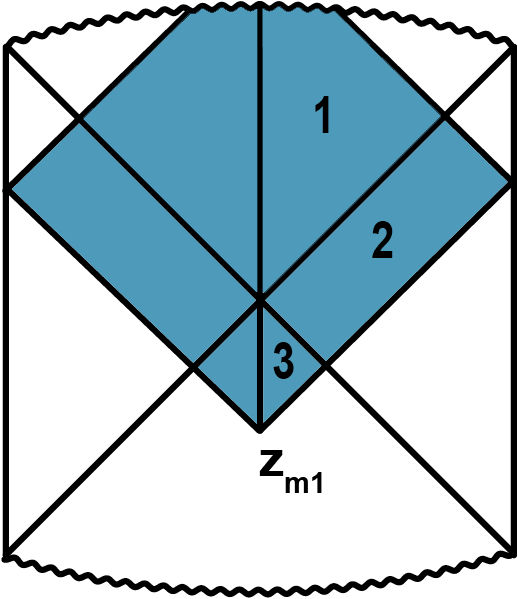}
\caption{
WDW patch for type $U$.
Left: before $t_c$. Right: after $t_c$.
}
\label{fig-caso2}
\end{center}
\end{figure}

A sketch of the time dependence of the WDW patch for the 
   type $U$ solution is shown in figure \ref{fig-caso2}.
   In this case, at $t_b=0$ the WDW has the shape \dia.
   At a critical time $t_b=t_c$ 
    the shape of the WDW changes abruptly as \dia $\to$\pentau.
   The limit in which the scalar perturbation vanishes outside the Cauchy horizon
corresponds to an extreme type $U$ case, in which the singularity 
approaches the location of the would be Cauchy horizon in the Penrose diagram.
In this limit we have that the critical time tends to infinity.
Denoting by $\phi_h$ the value of the scalar on the event horizon,
we find (see appendix \ref{appe-stima-tempo-critico}) that the critical time scales as
\beq
t_c =O\left( \frac{1}{\phi_h^2} \right)
 \, .
\eeq
See figure \ref{critical-time} for a plot of the critical time as a function of $\phi_h$. 
    
 When the tip of the  WDW touches  the singularity there is a logarithmic divergence in the complexity rate,
  \beq
\frac{d I_{\rm WDW }}{d t_b} \approx - \frac{V_0}{16 \pi G} A(\infty) \, \log (t_c -t_b ) \, , \qquad t_b <t_c \, .
\label{singolarita-rate-tipo-U}
\eeq 
If we approach the limit $t_b \to t_c$ from $t_b < t_c$, the complexity rate diverges to $+ \infty$.
For $t_b>t_c$ the complexity rate is instead finite.
In the $\phi_h \to 0$ limit, the value of $A(\infty)$  in eq. (\ref{singolarita-rate-tipo-U})
can be determined analytically:
\begin{itemize}
\item For a neutral scalar $q=0$, 
 the quantity $A(\infty)$  tends to a finite value for $\phi_h \to 0$,
\beq
A( \infty) \to - \frac{16 \pi G}{V_0} T_c S_c \, ,
\eeq
where $S_c$ and $T_c$  are the formal  entropy and negative temperature
computed on the Cauchy horizon, see eq.  \eqref{TS-cauchy}.
In this limit the Kasner parameter $\a$ diverges.
\item
For $q \neq 0$, we find that  $A(\infty)$  vanishes for $\phi_h \to 0$,
so that the divergence of the complexity rate at the critical time tends to disappear.
In this case the final Kasner parameter $\a$ oscillates an infinite number of times
as the limit  $\phi_h \to 0$ is approached.
\end{itemize}
Thus, there is no direct relation between the divergence 
in the complexity rate in eq. (\ref{singolarita-rate-tipo-U}) 
and the Kasner parameter $\a$.
Also, the divergence of the complexity rate at critical time
in eq. (\ref{singolarita-rate-tipo-U}) shows that the Lloyd bound
can not hold at finite time in this class of models.

We determined analytically the asymptotic action complexity rate in the $\phi_h \to 0$ limit.
We find that the behavior of the asymptotic action complexity 
rate is discontinuous  for $q \to 0$:
\begin{itemize}
\item
For the neutral scalar case $q=0$,  the  asymptotic complexity
tends to that of the RN black hole, i.e.
\beq
\lim_{t_b \to \infty} \frac{d I_{\rm WDW}}{d t_b } 
= T \, S \, - T_c \, S_c  \, ,
\eeq
where $T$ and $S$ are the temperature and entropy.
\item
In the charged scalar case $q \neq 0$, we find
\beq
\lim_{t_b \to \infty} \frac{d I_{\rm WDW}}{d t_b } = T \, S  \, .
\eeq
\end{itemize}
We numerically  checked that the Lloyd bound holds 
for the asymptotic action complexity rate $W_A$
in all the parameter space that we investigated.

In geometries with a naked Kasner singularity, 
the structure of the time dependence of complexity  is such that 
 complexity decreases as the singularity  is approached \cite{Barbon:2015ria,Bolognesi:2018ion}.
In that case, however, the singularity is not hidden behind a protective horizon
as in the present setup. Eventually, one would expect that there is an intermediate 
setup for which the complexity ceases to increase 
for a long time without decreasing, allowing for a parametrically long
period of large and constant complexity.
This was not achieved in this work.
However, we do sustain the believe that such a setup will be eventually found,
either semiclassically or perhaps including higher genus effects  \cite{Iliesiu:2021ari}.

We now resume the main new results contained in the paper:
\begin{itemize}
\item We numerically compute the Kasner exponent $p_t$
as a function of $\phi_0$ and $T$ 
for the charged scalar and we determine if solution
is type  $U$ or type $D$. See figure \ref{para-spazio-charged}.
\item We investigate the time dependence of volume complexity
 in a large class of hairy black hole solutions 
and we provide 
numerical evidence that
the  Lloyd bound  \cite{Lloyd,Brown:2015lvg}  is always satisfied
in this model.
\item We study the time dependence of action complexity in the same class of hairy black hole solutions.
The time dependence of action complexity can distinguish between 
type $D$ and type $U$ solutions.
For type $D$ solutions, we find a similar behavior to the Schwarzschild case studied in \cite{Carmi:2017jqz}.
 For type $U$ solutions, we find that the complexity rate diverges to $+ \infty$ at the critical time $t_c$.
 This is how complexity feels the Kasner singularity.
\item 
The divergence of complexity rate 
at critical time shows that the Lloyd bound does not hold for action complexity 
at finite time  in this class of solutions.
We provide numerical evidence that the Lloyd bounds holds 
for the asymptotic  action complexity rate.
\item We find that the Kasner exponents nearby the singularity
are not directly correlated with the coefficient of the linear growth at late time $W_A$ 
and with the coefficient of the leading singular behavior of action complexity at the critical time.
\end{itemize}

{\bf Note added:} While we were finishing to write this work,
Ref. \cite{An:2022lvo}  appeared on the arXiv.
There is  some overlap with the results presented in this work.
    
\section{The model and the black hole solutions}
 \label{section-model}

\subsection{Theoretical setting}

We consider the Einstein-Maxwell model with a scalar field, with action:
\beq
S= \frac{1}{16 \pi G} 
\int d^4 x \sqrt{-g} \left\{  R -2 \Lambda  
 -\frac{1}{4} F_{\mu \nu} F^{\mu \nu} 
-g^{\mu \nu} D_\mu \phi (D_\nu \phi)^* -m^2 \phi \phi^*
 \right\} \, ,
 \label{lagrangiana-modello}
\eeq
where the cosmological constant is
\beq
\Lambda=-\frac{3}{L^2} \, ,
\eeq
and $L$ is the AdS radius.
The covariant derivative is
\beq
D_\mu \phi =\p_\mu \phi + i q A_\mu \phi \, .
\eeq
We will consider both the case of neutral scalar field $q=0$ and charged scalar field $q \neq 0$.
 We will take the scalar mass to be above the Breitenlohner-Freedman (BF) bound \cite{Breitenlohner:1982jf}
\beq
 m^2=-\frac{2}{L^2} \,,
\eeq
and from now on we will set $L=1$ for simplicity.

A generic metric for a planar black hole is 
\beq
\label{metric1}
ds^2= \frac{1}{z^2} \le -f e^{-\chi} dt^2 +\frac{dz^2}{f} +dx^2 +dy^2 \ri \, ,
\eeq
where $f$ and $\chi$ are functions of $z$ and the ansatz for the scalar and gauge field is
\beq
\label{ansatz1}
\phi=\phi(z) \, ,  \qquad A=a(z) dt \, .
\eeq

The equations of motion are
\bea
&&f' - \left(  \frac{   \chi' } {2 }  + \frac{ 3  } { z} \right) f+ \frac{ 3 } { z} -
    \frac{  e^{\chi } z^3 \left(a'\right)^2 }{4  }
           -\frac{  m^2 \phi ^2}{2 z}  =0 \, , \nl
&&  \chi '-\frac{  z a^2 q^2 \phi ^2}{f^2} e^{\chi }  - z  (\phi' )^2  = 0 \, ,  \nl
 &&  a''+ \frac{ \chi '}{2  } a'  -\frac{2    q^2 \, \phi^2}{ f z^2} a  = 0  \, , \nl
  &&   \phi ''
- \left( \frac{ \chi ' }{2 }
+\frac{2   }{  z}
 -\frac{    f' }{ f }\right) \phi '
 +\left( \frac{ a^2 q^2 e^{\chi }   }{ f^2 } - \frac{  m^2  }{ f z^2}\right) \phi= 0\,,
     \label{full-system-eom}
\eea
with asymptotic conditions at the boundary
\beq\label{bcs00}
f|_{z \rightarrow 0}=1 \, , \qquad \chi|_{z \rightarrow 0}=0 \, , \qquad
a|_{z \rightarrow 0}= \mu \, ,
\eeq
where $\mu$ is chemical potential.
With our choice of scalar mass, the expansion of the
scalar field $\phi$ near the boundary reads
\beq
\phi= \phi_0 z + \phi_1 z^2 + \dots \, .
\label{expansion-phi-boundary}
\eeq
With Dirichlet boundary conditions,  $\phi_0$ can be identified with the source
and $\phi_1$ with the expectation value of the operator with dimension $\Delta_+=2$.
On the other hand, with Neumann boundary conditions $\phi_1$ is proportional to the source and $\phi_0$ to the expectation value of an operator with dimension $\Delta_-=1$.
We will mainly work with the choice $ \Delta_-$ that corresponds to Neumann boundary conditions for the scalar field. 
 The empty AdS solution  of (\ref{full-system-eom}) is
\beq
f=1\, , \qquad  \chi= a=\phi=0 \ .
\eeq

\subsection{The Reissner-Nordstr\"om black hole}

For the Reissner-Nordstr\"om (RN) solution we have $\phi,\chi=0$ and 
\bea
&& f_{\rm RN}(z)= 1 - \le \frac{z}{z_h} \ri^3 \le 1+\frac{\rho^2 z_h^4}{4} \ri
+\frac{\rho^2}{4} z^4 \, , \qquad \rho=\frac{\mu}{z_h} \, , \nl  
&& a(z)=\mu \le 1-\frac{z}{z_h} \ri \, ,
\label{fRN1}
\eea
where $z_h$ is the external horizon, $\mu$ is the chemical potential as before, and  $\rho$
is the charge density.
The RN black hole has a Cauchy horizon at $z = z_c$. 
In terms of $(z_h,z_c)$ the solution is:
\beq
f_{\rm RN}(z)=1 - z^3 \frac{(z_h+z_c)(z_c^2+z_h^2)}{z_h^3 z_c^3}+ z^4 \frac{z_h^2+z_c^2+z_h z_c }{z_h^3 z_c^3} \, , 
\qquad \rho=\frac{2 \sqrt{z_h^2+z_c^2+z_h z_c } }{z_h^{\frac{3}{2}} z_c^{\frac{3}{2}}} \,,
\label{fRN2}
\eeq
and has energy density
\beq
\mathcal{E}= \frac{1}{16 \pi G}  \frac{2 (z_h+z_c)(z_c^2+z_h^2)}{z_h^3 z_c^3} \, .
\eeq
The total black hole charge, mass and entropy are
\beq
Q= \rho \, V_0 \, , \qquad M= \mathcal{E} \,  V_0 \, , \qquad
S=\frac{V_0}{4 G} \frac{1}{z_h^2} \,,
\label{carica-totale}
\eeq
the black hole temperature being in turn
\beq
T=- \frac{1}{4 \pi} f'(z_h) 
=\frac{(z_c-z_h)(3 z_c^2+2 z_c z_h +z_h^2)}{4 \pi z_c^3 z_h} \, .
\label{RN-temp}
\eeq

\begin{figure}[h]
\begin{center}
\includegraphics[width=0.6\textwidth]{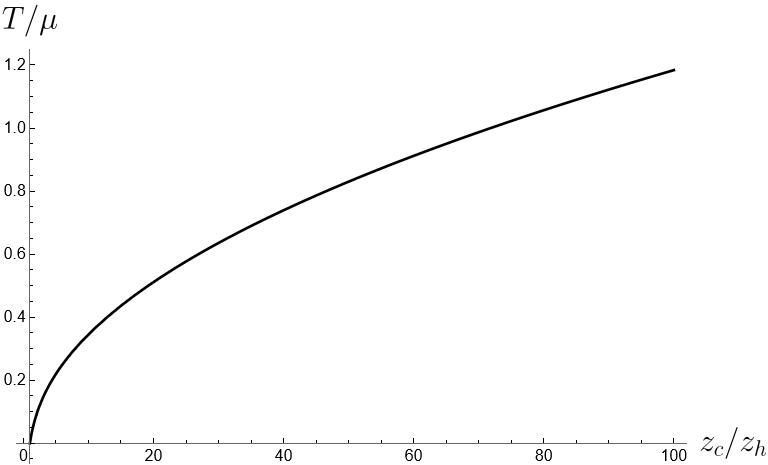}
\caption{  $T/\mu$ as a function of $z_c/z_h$ for the Reissner-Nordstr\"om AdS spacetime.}
\label{T-mu-zc-zh-RN}
\end{center}
\end{figure}

If we send $\rho \rightarrow 0$, we have $z_c \to \infty$ and in this limit we recover the Schwarzschild black hole.
The extremal limit corresponds to 
\beq
\mu_e=\frac{2 \sqrt{3}}{z_h} \,,
\eeq
for which $z_c=z_h$ and the temperature vanishes.
It is useful to express $T/\mu$ as a function of $y=z_c/z_h$, giving
\beq
\frac{T}{\mu} = \frac{1}{8 \pi} \frac{(y-1) (3 y^2 + 2 y +1 )} {y^{3/2} \sqrt{1+y+y^2} } \, ,
\eeq
see figure \ref{T-mu-zc-zh-RN} for a plot.

When we include the scalar backreaction,  the inner horizon 
of the RN black hole, which is  unstable, disappears and  there is 
singularity at $z \rightarrow \infty$, just as for the Schwarzchild black hole.
This was proved in various theoretical setting in
 \cite{Hartnoll:2020rwq,Hartnoll:2020fhc,Cai:2020wrp,An:2021plu}.

\subsection{Approximate solution at large $z$: the Kasner limit}

The presence of the scalar $\phi$ gives rise to interesting dynamical phenomena, 
which have been investigated in detail in  several theoretical setting in 
\cite{Hartnoll:2020rwq,Hartnoll:2020fhc,Frenkel:2020ysx,Cai:2020wrp,Henneaux:2022ijt,
Mansoori:2021wxf,Sword:2021pfm,Caceres:2022smh}.
The full system in eq.  (\ref{full-system-eom}) can in general only be solved numerically.
To make the analysis more robust, it is therefore useful to introduce analytical approximations.

In particular, it is convenient to consider
a regime in which the scalar $\phi$ scales logarithmically. Numerical analysis confirms that this is a good approximation when  $z$ is   large enough.
Let us assume that we can  neglect  the electric field $a'$, the cosmological constant term, the scalar mass term $m^2$ and its charge $q$.
 The system  in eq  (\ref{full-system-eom}) is then approximately
\bea
  f'-\left(  \frac{ \chi '  }{2 } + \frac{3 }{ z}\right)f = 0 \, , \qquad    \chi ' - z  (\phi' )^2  &=& 0 \, ,
\nl
 \phi ''- \frac{ \chi ' \phi '}{2 }+ \left( \frac{f' }{f} -\frac{2  }{z}\right) \phi '&=& 0   \, .
 \label{proxy-system}
\eea
The equations above admit the following exact solution:
\beq
\label{kasner-approx}
f=-f_0 z^{3+\a^2} \, , \qquad \phi=\a \, \sqrt{2} \log z \, , \qquad
\chi=2 \a^2 \log z + \chi_0 \, ,
\eeq
where $\a$, $f_0$, $\chi_0$ are integration constants.
The behaviour for the metric component $g_{tt}$ is
\beq
g_{tt}=-f_0 e^{-\chi_0} z^{1-\a^2} \, .
\eeq
With the change of variables 
\beq
\tau=\frac{2}{ (3+ \a^2) \sqrt{f_0} } z^{-\frac{3+\a^2}{2}} \, ,
\eeq
where $\tau \to 0$ corresponds to $z \to \infty$,
the metric with $f$ and $\chi$ given by eq.~(\ref{kasner-approx}) can be put in the Kasner form
\beq
ds^2=- d \tau^2 + c_t \tau^{2 p_t} dt^2 + c_x \tau^{2 p_{x}} (dx^2+dy^2) \, ,
\label{defkasmetric}
\eeq
where $c_t, c_x$ are constants and the Kasner exponents are
\beq
p_t=\frac{\a^2-1}{3+\a^2} \, , \qquad p_{x}=\frac{2}{3+\a^2} \, .
\eeq
where   $p_t$ and $p_x$   obey the modified Kasner constraints 
\beq
p_t + 2p_x =1 \, , \qquad
p_t^2 + 2 p_x^2 =1 -\frac{4 \a^2}{(3+\a^2)^2 } \, .
\eeq 
The Schwarzchild solution close to the singularity is of Kasner-type, with $\a=0$ and $p_t=-\frac{1}{3}$. 
The exponent $p_x$ is always positive, and so the transverse direction $x,y$
always experiences a crunch inside the BH in the regime described by eq. \eqref{kasner-approx}.
The exponent $p_t$ is in turn only positive for $\a>1$, which  gives a contracting
ER bridge. It is negative for $\a <1$, which corresponds instead to an expanding bridge.

The Kasner approximation in eq. (\ref{kasner-approx}) is useful in the regime of large $z$.
In particular, for $\a >1$  the approximation is stable: if it is satisfied for some $z$,
the fields go in a direction such that the approximation becomes even better at larger $z$.
In contrast, for $\a <1$ the approximation is not stable for a charged black hole.
In this case, at large enough $z$ we are forced to consider in eq. \eqref{full-system-eom} the backreaction of the 
electric field $F_{zt}=a'(z)$ on the metric.
From the numerical study in \cite{Hartnoll:2020fhc}
we know that, also for $\a <1$, there often exists a parametrically large regime in $z$
for which eq. (\ref{kasner-approx}) provides a good approximation. However, for $\a <1$
we cannot extrapolate the approximation all the way to $z \to \infty$, because at some point 
the backreaction due to the electric field will be important.

\subsection{The Kasner inversion}
\label{kasner-inversion-section}

For $\a =\a_0<1$, if we take into account the backreaction of the electric field
the solution in eq. (\ref{kasner-approx})  experiences  a Kasner inversion, \emph{i.e.} at some point 
the exponent $\a$  jumps from $\a_0$ to its inverse $\a_f = 1/\a_0$,
\beq
\a_0 \qquad \to \qquad \a_f=\frac{1}{\a_0} \, . 
\label{kasner-jump}
\eeq
For this reason, in the limit $z \to \infty$ we always expect $\a > 1$.
In terms of the Kasner exponent $p_t$, the transition is
\beq
p_t \to - \frac{p_t}{2 p_t +1} \, ,
\eeq
and transforms a growing ER bridge ($p_t <0$) to a contracting one $(p_t>0)$.

This transition could in principle happen both in the
case of neutral or charged scalars, because it does not require a
direct coupling between the scalar $\phi$ and the gauge field $A_\mu$.
However, for $q=0$ it seems that this transition is not realised in the parameter
space of the model, which was studied in \cite{Hartnoll:2020rwq}. 
On the other hand, for a charged scalar this transition is indeed realised  in many numerical examples \cite{Hartnoll:2020fhc}.
 Due to non-linear effects in the full system of equations of motion (\ref{full-system-eom}),
   multiple Kasner transitions are also possible for $q \neq 0$ when the values of the parametersare fine-tuned .
 
The Kasner transition in eq. (\ref{kasner-jump})
can be derived  introducing a more refined approximation to the system  in eq.~(\ref{full-system-eom}),
where we keep also the term which is responsible for the backreaction of the Maxwell field strength on the metric, \emph{i.e.}
\bea
f' - \left(  \frac{   \chi' } {2 }  + \frac{ 3  } { z} \right) f -
    \frac{  e^{\chi } z^3 \left(a'\right)^2 }{4 \, L^2 }
          &=& 0 \, ,  \qquad
  \chi '  - z  (\phi' )^2  = 0 \, ,  
  \nl
  \phi ''
- \left( \frac{ \chi ' }{2 }
+\frac{2   }{  z}
 -\frac{    f' }{ f }\right) \phi' &=&  0\, .
 \qquad
  a''+ \frac{ \chi '}{2  } a'  = 0  \, .
     \label{sistema-Kasner-transition}
\eea
Here we still neglect the cosmological constant term and the effects due to the mass $m$ and the charge $q$
of the scalar field. We can integrate the  equation for $a$ as follows
\beq
a'=E_0 e^{-\frac{\chi}{2}} \, , \qquad a=A_0 + E_0 \int e^{-\frac{\chi}{2}} \, dz \, ,
\label{a-gauge-approx}
\eeq
where $E_0$ and $A_0 $ are integration constants.
With this relation, the  equation for $f$ can be put in the form
\beq
   f'    - f \le \frac{   \chi' } {2 }
     + \frac{3  } { z}  \ri
     - \frac{ E_0^2 z^3  }{4  }  = 0 \, ,
\label{f-Kasner-transition}
\eeq
and assuming the Kasner behaviour for $f$ and $\chi$ given by eq. \eqref{kasner-approx},
then the new term  $    { E_0^2 z^3  }/{4  }$ dominates the others at large $z$ only for $\a <1$.

Equation (\ref{f-Kasner-transition}) can be solved as follows
\beq
\le \frac{f e^{-\chi/2}}{z^3} \ri'  =\frac{E_0^2}{4 } e^{-\chi/2} \, , \qquad
\frac{f e^{-\frac{\chi}{2}}}{z^3} = C_0 + \frac{E_0^2}{4 } \int e^{-\frac{\chi}{2}} \, dz \,  ,
\label{tecnica1}
\eeq
where $C_0$ is an integration constant.
We can rewrite the equation involving the second  derivative of $\phi$  as
\beq
\le \frac{ e^{-\frac{\chi}{2}} f \phi' }{z^2} \ri' =0 \, .
\label{tecnica2}
\eeq
Combining eqs. (\ref{tecnica1}) and (\ref{tecnica2}) we find
\beq
\le \frac{E_0^2}{4  } e^{-\frac{\chi}{2}} \ri   z \phi'  +  \le C_0 +\frac{E_0^2}{4  } \int e^{-\frac{\chi}{2}} \, dz \,   \ri
(\phi'+ z \phi'' )= 0 \, .
\label{passaggio}
\eeq
Following  \cite{Hartnoll:2020fhc}, it is convenient to write $\phi$ in terms of an auxiliary function $\psi$ defined by
\beq
\phi=\sqrt{2} \int \frac{\psi(z)}{z} dz \, .
\eeq
Note that a constant $\psi(z)=\a$ mimics the solution without inversion in eq. (\ref{kasner-approx}).
The Kasner inversion correponds to a solution which interpolates from an initial $\psi=\a_0$
to a final  $\psi=\a_f$. Here the physically interesting regime is $0 < \a_0 <1$, since otherwise there is no inversion because the backreaction due to the electric field remains small.

In terms of $\psi$, it is possible to show that eq. (\ref{passaggio}) implies the following differential equation \cite{Hartnoll:2020fhc}
\beq
\psi'' -\frac{2 (\psi')^2}{\psi} + \frac{\psi^2 \psi'}{z}=0 \, ,
\eeq
whose solution is
\beq
\psi \le \psi -\a_0 \ri^{-\frac{1}{1-\a_0^2}} \le \frac{1}{\a_0} -\psi \ri^{-\frac{1}{1-\frac{1}{a_0^2}}} =\frac{z_0}{z} \, ,
\eeq
where $z_0$ is an integration constant.
In this case $\psi \to \a_0$ for $z \ll z_0$ and $\psi \to \a_f= 1/ \a_0$  for $z \gg z_0$.
This is indeed the behaviour described in eq. (\ref{kasner-jump}).

\subsection{Neutral scalar field }

Let us now discuss the $q=0$ model, which was studied in detail in  \cite{Hartnoll:2020rwq}.
In this case, we can solve explicitly (\ref{full-system-eom}) for the gauge field
\beq
a'=- \rho \, e^{-\frac{\chi}{2}} \, ,
\label{a-eq}
\eeq
where $\rho$ is the charge density.
Indeed, eq. (\ref{a-eq}) can be physically derived from the conservation of
the electric flux, see appendix \ref{stokes-appendix}.
We end up with a system composed of  the remaining differential equations
\bea
&&- f'+\left(\frac{z \left(\phi '\right)^2 }{2} + \frac{3 }{ z}\right)f+\frac{\rho ^2 z^3}{4 } -\frac{ \phi ^2}{ z} -\frac{3}{ z}=0 \, ,
\nl
%&& f'-\left(\frac{z \left(\phi '\right)^2 }{2} + \frac{3 }{ z}\right)f+\frac{\rho ^2 z^3}{4 } -\frac{ \phi ^2}{ z} -\frac{3}{ z}=0 \, ,
%\nl
&& \phi ''-\frac{z}{2}  \left(\phi
   '\right)^3 + \left( \frac{f' }{f} -\frac{2  }{z}\right) \phi '+\frac{2  }{f z^2} \phi= 0 \, , \qquad \chi'=z (\phi')^2 \, .
   \label{eq-diff-carica-zero}
\eea
Evaluating the equations above at the horizon $z=z_h$, which is defined by the condition
 $f(z_h)=0$, we obtain using the expansion  $f(z) \approx f'(z_h) (z-z_h)$,
\bea
&& -f'(z_h) +\frac{\rho ^2 z_h^3}{4 } -\frac{ \phi^2(z_h)}{ z_h} -\frac{3}{ z_h}=0 \, ,
\nl
&& \frac{1 }{z-z_h} \le \phi'(z_h)+\frac{2  }{ f'(z_h)  z_h^2} \phi(z_h)  \ri +{\rm finite}= 0 \, .
\eea
In order to get  a smooth solution, we must then  impose
the following condition at the horizon
\beq
\phi(z_h)= \phi_h \, , \qquad
\phi'(z_h)= -\frac{8 \phi _h}{z_h \left(\rho ^2 z_h^4-4 \phi _h^2-12\right)} \, .
\eeq
These boundary conditions, together with $f(z_h)=0$ and the value of $\rho$, are enough to solve the problem. 
A solution is then determined by the values of $z_h$, $\rho$ and $\phi_h$.

Alternatively, a given solution can be determined in terms 
of the temperature $T$ and the sources $\mu$ and $\phi_0$ in the dual CFT.
The temperature is:
\beq
T=- \frac{1}{4 \pi} f'(z_h) e^{-\frac{\chi(z_h)}{2}} \, ,
\eeq
and the chemical potential is
\beq
\mu=a(0)=\rho \int_0^{z_h} e^{-\frac{\chi}{2}} \, dz \, .
\label{mu-q=zero}
\eeq

The limit $\phi_h \to 0$ can be studied analytically.
In this case, the backreaction of the scalar is negligible up to 
 the location of the Cauchy horizon $z=z_c$ in the unperturbed RN solution,
for which the scalar field $\phi$ identically vanishes.
More precisely, the solution is with good approximation the same as the $RN$
for $z< z_i$, where
\beq
 z_i=z_c-\epsilon\, , \qquad \epsilon \to 0 \qquad  {\rm for}  \qquad  \phi_h \to 0 \, .
 \label{definizione-zeta-i}
 \eeq
For $z \approx z_i$, the dynamics  enters in a highly non-linear regime in which $g_{tt}$ drops exponentially towards zero.
This regime was called in \cite{Hartnoll:2020rwq} 
the collapse of the Einstein-Rosen (ER) bridge.\footnote{In some sense the collapse is also there in the
 RN solution, because $g_{tt} \to 0$
at the location of the Cauchy horizon. The main difference here is that the inner horizon disappears.}
As can be checked \emph{a posteriori}, in this limit we can neglect the mass of the scalar field $\phi$.
With this approximation, the equations of motion  (\ref{eq-diff-carica-zero})  take the following form
\beq
 \le \frac{e^{-\chi/2} f}{z^3} \ri'
=e^{-\chi/2}  \le
    \frac{     \rho^2 }{4  }
    -  \frac{ 3 } { z^4} 
           \ri  \, , \qquad
  \chi ' =  z  (\phi' )^2   \, ,  \qquad
 \le \frac{e^{-\chi/2} f \phi' }{z^2} \ri'  
 = 0\, .
\eeq
The collapse of the ER bridge happens in a small range of the coordinate $z$, and so it is consistent
to set $z \approx z_i$ in the equations of motion above, keeping  $f$, $\chi$ and $\phi$ as functions of $z-z_i$, \emph{i.e.}
\beq
 \le {e^{-\chi/2} f} \ri'
=e^{-\chi/2}  \le
    \frac{     \rho^2 }{4  } z_i^3
    -  \frac{ 3 } { z_i} 
           \ri  \, , \qquad
  \chi ' =  z_i  (\phi' )^2   \, ,  \qquad
 \le {e^{-\chi/2} f \phi' } \ri'  
 = 0\, .
\eeq
The equations for the metric functions can be written as
\beq
\chi' f^2 e^{-\chi} =\tilde{A} \, , \qquad f'-\frac{\chi'}{2} f = \tilde{B} \, ,
\label{sistema-stoppaccioso}
\eeq
where $\tilde{A}$ is an integration constant and
\beq
\tilde{B}=  \frac{     \rho^2 }{4  } z_i^3
    -  \frac{ 3 } { z_i}  \approx
    \frac{z_h^2 z_c +z_c^3 +z_h z_c^2 -3 z_h^3}{z_h^3 z_c} >0 \, .
    \label{B-tilde}
\eeq
Introducing 
\beq
H= z_i^2 g_{tt}= - f e^{-\chi} \, , \qquad G=e^{-\chi} \, , 
\eeq
the solution to \eqref{sistema-stoppaccioso} takes the form
\beq
\frac{\tilde{A}}{2 \tilde{B}} \log H + H = - \tilde{C} (z-z_0) \, , \qquad G=\frac{(H')^2}{ \tilde{B} \tilde{C}} \, ,
\label{soluzione-stoppacciosa}
\eeq
where $z_0 = z_c$ and $\tilde{C}$ are integration constants.
In the limit $\phi_h \to 0$, we expect that $\tilde{A} \to 0$ while
$\tilde{C} \approx f'_{RN} (z_c)=O(\phi_h^0)$.
The scalar $\phi$ is given by
\beq
%\phi'=\frac{\tilde{A}}{z_i^{1/2}} \frac{e^{\chi/2}}{f} \, \qquad
\phi=-\frac{\tilde{A}^{1/2}}{(\tilde{B} \tilde{C} )^{1/2} }  \frac{1}{z_i^{1/2}} \log (H \tilde{D}) \, ,
\label{scalar-profile-q-zero}
\eeq
where $\tilde{D}$ is another integration constant.
In the limit $\phi_h \to 0$, the equation for the scalar is approximately linear for $z<z_i$,
because the scalar is small. Then  the amplitude of the scalar in eq. (\ref{scalar-profile-q-zero})
must scale linearly in $\phi_h$, \emph{i.e.}
\beq
\tilde{A}^{1/2}=O(\phi_h) \, , \qquad {\rm for} \qquad \phi_h \to 0 \, .
\label{relazione-utile-limite-piccolo-phi-h}
\eeq
For $z \gg z_0$ we find that $g_{tt}$ approaches to zero exponentially,
\beq
g_{tt}=\frac{H}{z_i^2}  \approx \frac{1}{z_i^2} \exp \le -\frac{2 \tilde{B} \tilde{C}}{\tilde{A}} (z-z_0) \ri \, ,
\eeq

This approximation can then be matched with the Kasner approximation, which is valid at large $z$.
In particular, matching eqs. (\ref{kasner-approx})  and (\ref{soluzione-stoppacciosa}) we find
\beq
e^{-\chi_0/2} f_0 = - z^3 \, e^{-\chi/2} f =z^3 \frac{H}{G^{1/2}} \approx
z_c^3 \frac{\tilde{A}}{2 ( \tilde{B} \tilde{C})^{1/2}} \, .
\label{robe-piccolette}
\eeq
From eq. (\ref{relazione-utile-limite-piccolo-phi-h}) we find that
$e^{-\chi_0/2} f_0 =O(\phi_h^2) $ for $\phi_h \to 0$.

\begin{figure}[h]
\begin{center}
\includegraphics[width=0.49\textwidth]{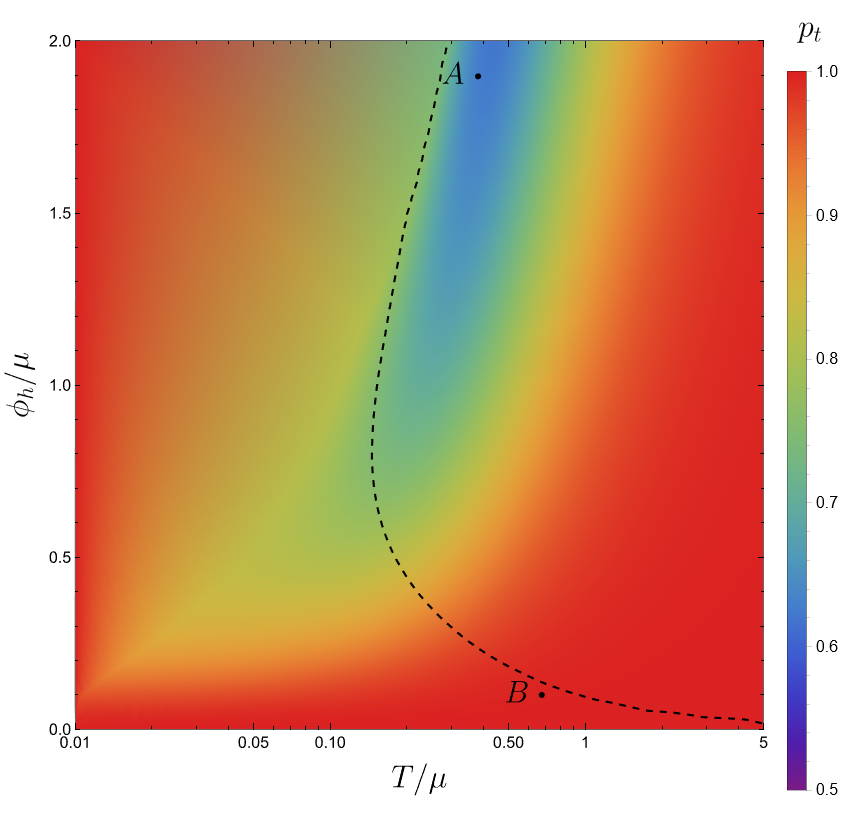}
\includegraphics[width=0.49\textwidth]{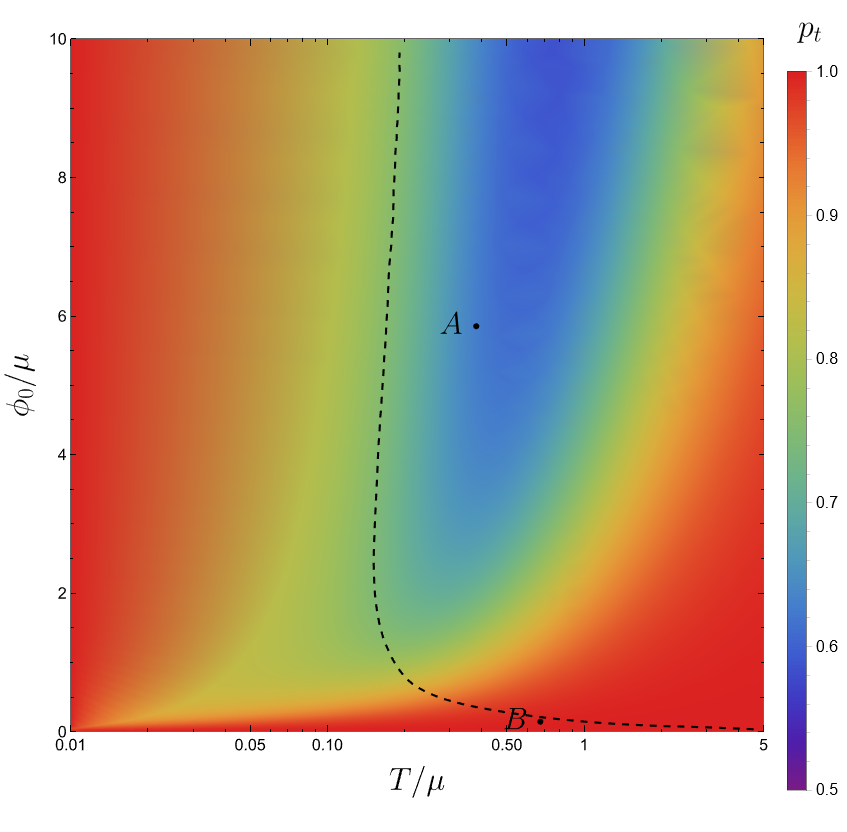} 
\caption{ 
Kasner exponent $p_t$  for the neutral scalar case ($q=0$).
 Left: $p_t$ as a function of  ${T}/{\mu}$ and ${\phi_h}/{\mu}$. Right: $p_t$ 
 as a function of  ${T}/{\mu}$ and ${\phi_0}/{\mu}$.
Here we have $p_t>0$ and $\a>1$ in all the parameter space, and so there is no Kasner inversion.
The points marked with letters A and B correspond to the specific examples shown 
in fig.~\ref{uncharged-sol-uno}.
}
\label{uncharged-para-sol}
\end{center}
\end{figure}

A numerical approach (see appendix \ref{numerical-appendix} for our numerical techniques)
can be used to solve \eqref{eq-diff-carica-zero} also
away from the limit $\phi_h \to 0$.  In general, there is a crossover at $z\approx z_c$
from a solution qualitatively similar to the RN for $z \ll z_c$ and a Kasner regime for $z \gg z_c$.
In the intermediate region, the  metric $g_{tt}$ drops exponentially to a value very close to zero.
This drop is very sudden in the $\phi_h \to 0$ limit.
In all the parameter space investigated in  \cite{Hartnoll:2020rwq},
the solution stabilizes to a Kasner  regime with  $\a>1$ just after the collapse
of the ER bridge.
For this reason, the Kasner inversion in eq. (\ref{kasner-jump}) never takes place.
In figure \ref{uncharged-para-sol} we reproduce the numerical analysis
of  \cite{Hartnoll:2020rwq}, showing how the Kasner coefficient
 $p_t$ depends on the parameter space of solutions. 

For the $q=0$ case it is necessary to introduce a source to get solutions with non-zero scalar profile, \emph{i.e.} the $\phi_h =0$ line corresponds to the
 $\phi_0 =0$ line in the two diagrams in figure \ref{uncharged-para-sol}. There is a region at very small temperatures
$T/\mu   \approx 10^{-4}$ which is not explored in fig.~\ref{uncharged-para-sol}. 
In the extremal RN limit,
the near horizon geometry is described by AdS$_2$ $\times \mathbb{R}^2$ 
with a different AdS radius, and  the scalar with mass $m^2=-{2}/{L^2}$ is below its 
BF bound, such that condensation can occur 
\cite{Hartnoll:2008kx}.

\begin{figure}[h]
\begin{center}
\begin{tabular}{cc}
\includegraphics[width=0.48\textwidth]{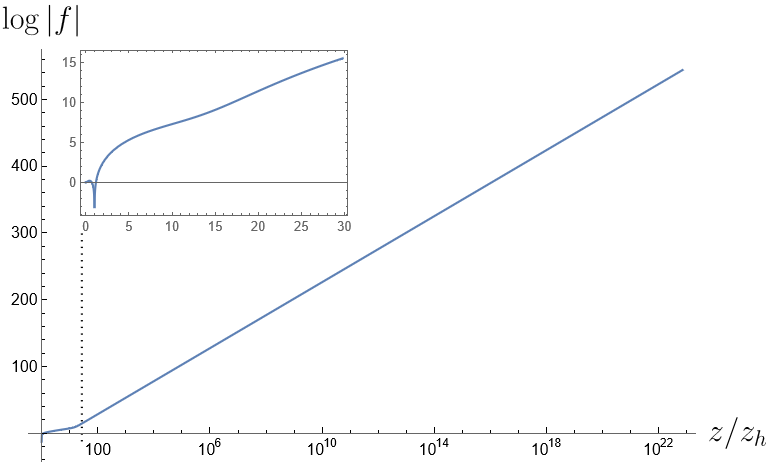} & \includegraphics[width=0.48\textwidth]{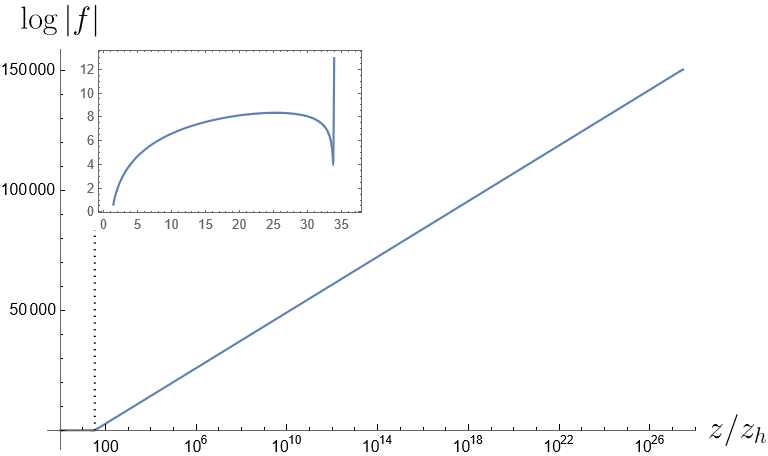}\\
\includegraphics[width=0.48\textwidth]{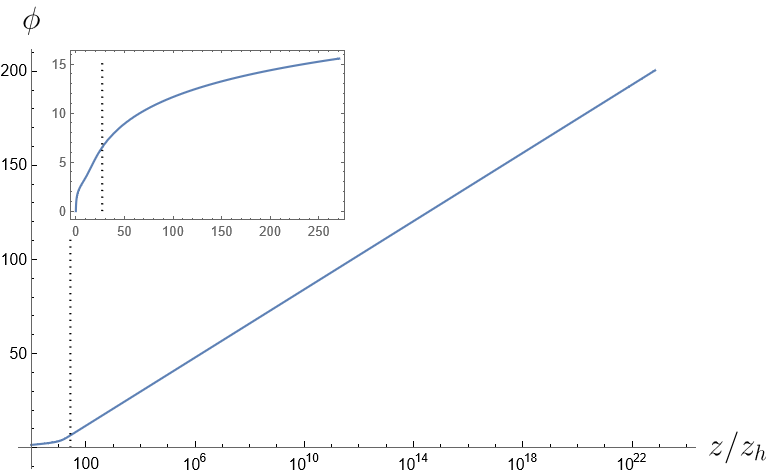} & \includegraphics[width=0.48\textwidth]{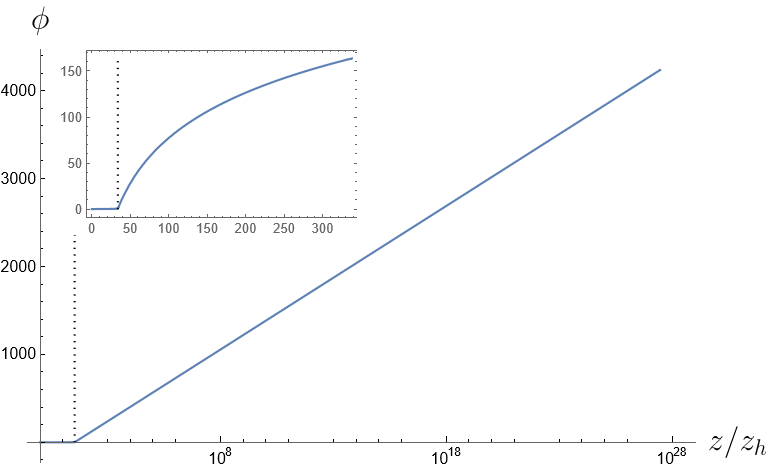}\\
\includegraphics[width=0.48\textwidth]{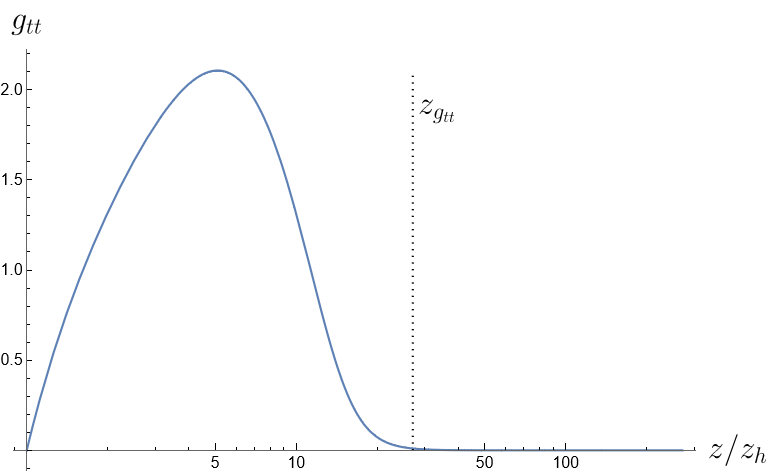} & \includegraphics[width=0.48\textwidth]{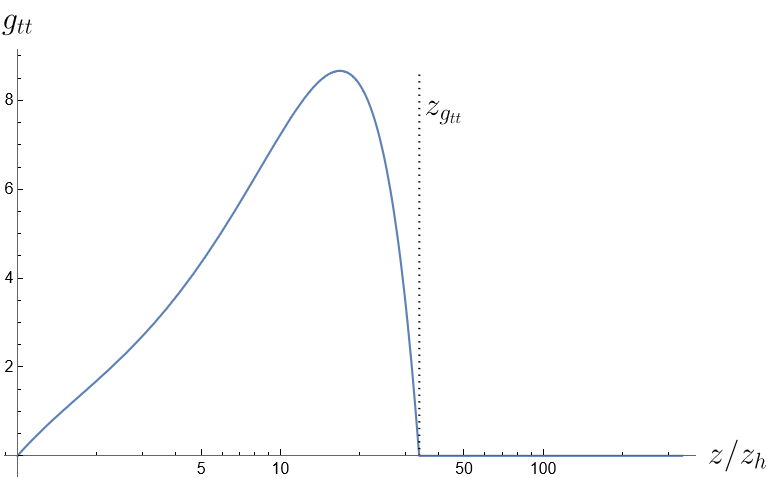}\\
(A) & (B)
\end{tabular}
\caption{Sample solutions for the neutral $q=0$ case.
For illustration purposes, we have chosen the values $(z_h,\rho,\phi_h)=(1, 0.9, 1.45)$ for (A) and $(1, 0.35, 0.035)$ for (B), where the letters match the ones figure \ref{uncharged-para-sol}. The vertical dotted line is the ``would be'' Cauchy horizon $z_{g_{tt}}$.}
\label{uncharged-sol-uno}
\end{center}
\end{figure}

Two examples of numerical solutions are presented in figure \ref{uncharged-sol-uno}.
In the example (A) we take $\phi_h$ rather large and far away 
from the limit $\phi_h \to 0$.
Example (B) is closer to the $\phi_h \to 0$ limit, and indeed we see that the collapse of the
Einstein-Rosen bridge is very fast in the coordinate $z$.
Various interesting features can be seen in these plots. It is clear that for all of the solutions the late Kasner
 behaviour predicted in (\ref{kasner-approx}) is present, with $\alpha$ controlling the linear behaviour of $\log f$ and $\phi$ vs. $\log z$. 
The example (B) is a typical case of small $\phi_h$ which is almost identical to RN up to $z=z_c$.
In the $\phi_h \to 0$ limit, the backreaction of the scalar is negligible only
up to the ``would be'' Cauchy horizon  $z_c$ of the RN solution. 
 At $z_c$ there is a transition to the Kasner regime, which is very sharp in the $z$ coordinate.
Approaching  $\phi_h = 0$ is not a continuous limit for $z>z_c$, since in this limit we  have  $\a \to \infty$ and $p_t \to 1$.
Then there is a 
crossover between
the RN solution  and the 
Kasner regime discussed in eq.~(\ref{kasner-approx}).
 This transition becomes very sharp around $z_c$ as $\phi_h \to 0$.

In order to describe the causal structure of the solution, it is convenient
to use Penrose diagrams. In these diagrams there is always an ambiguity
in the choice of the function used in the conformal mapping, so for completeness we explain our conventions in appendix \ref{penrose-appendix}.
There is an important conformally invariant property of the Penrose diagrams which is universal
and does not depend on this arbitrary choice \cite{Fidkowski:2003nf}.
Schematically, the Penrose diagram of an eternal asymptotically AdS 
black hole with a spacelike singularity can be drawn as a square,
where the vertical sides are the boundaries of AdS.
The spacelike singularity can then bend inwards or outwards the horizontal
sides of the square inside the black hole horizon. The direction of this bending is a conformally invariant property, which can be established as follows: let us prepare a couple of ingoing
radial lightlike geodesics  from the left and the right boundaries 
at time $t_b=0$, and ask whether 
they meet before reaching the singularity or not.
If the singularity coincides with the horizontal line of the square, the two light rays 
meet exactly at the singularity. If instead the singularity is bending upwards the top side of the square, the two rays meet
before intersecting the singularity; finally, if the singularity is bending the top side of the square downwards,
the two light rays never meet. This shows that the bending of the singularity
is a physical property and does not depend on the choice of conventions used to construct the Penrose diagram.

 For the non-rotating BTZ black hole in AdS$_3$ \cite{Banados:1992wn},
the singularity lies exactly on the horizontal line.
For Schwarzschild AdS$_d$ black holes with $d>3$
the singularity bends the top side downwards. In the case of the black holes discussed
in this section, we can in general achieve in the parameter space both upwards and downwards bending of the top side of the Penrose diagram.
The kind of bending of the singularity in the Penrose diagram, as we will find in section \ref{section-action-complexity}, will influence 
the time dependence of the action complexity.   For convenience, we will denote  a solution
with the singularity with upper bending as type $U$ and one with the singularity with lower bending as 
type $D$.

\begin{figure}[h]
\begin{center}
\begin{tabular}{cc}
\includegraphics[width=0.48\textwidth]{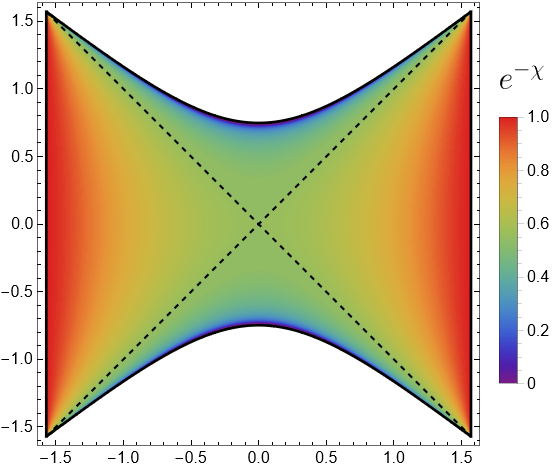} & \includegraphics[width=0.47\textwidth]{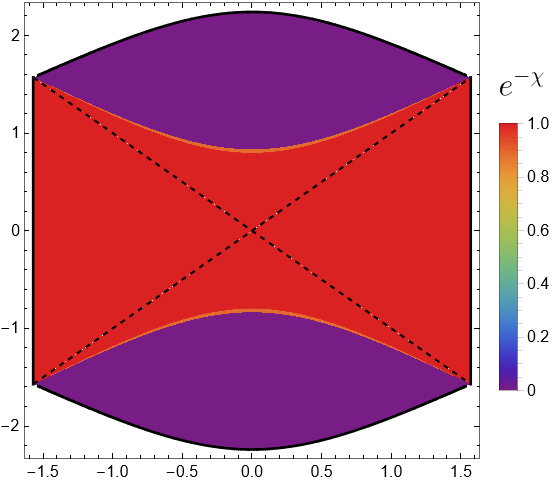} \\
(A) & (B)
\end{tabular}
\caption{Penrose diagrams for the sample solutions A and B, marked in fig.~\ref{uncharged-para-sol} and whose field
profiles are shown in fig.~\ref{uncharged-sol-uno}. On the right, we see that for case B, which is close to the limit $\phi_h\to0$, the collapse of the ER bridge is very fast, as mentioned in the main text. We refer to appendix \ref{penrose-appendix} for the definition of the coordinates of the Penrose diagram. }
\label{penrose-examples}
\end{center}
\end{figure}

The black dashed line in figure \ref{uncharged-para-sol} 
separates two shapes of Penrose diagram: the solutions on the left 
side of the line are of type $U$, while the ones on the right side are of type $D$.
In particular, in the $\phi_h \to 0$ limit, the singularity approaches the Cauchy horizon
of the RN black hole, and so the bending is up.  
In figure \ref{penrose-examples} we show two examples of these 
Penrose diagram; we plot on the diagram the value of $e^{-\chi}$,
which gives a measure of how much the solution is different from the RN
limit, which has $\chi=0$.

\subsection{Charged scalar field }

The explicit examples studied in  \cite{Hartnoll:2020fhc} correspond to the zero source 
limit, which is known as for holographic superconductors \cite{Hartnoll:2008vx,Hartnoll:2008kx}.
In the charged case, the main term which  can induce condensation of the scalar close to the horizon
is the term in the Lagrangian coming from scalar covariant derivative,
\beq
-q^2 g^{tt} (A_t)^2 \phi \phi^* \, .
\eeq
This term induces a tachyonic mass term for the scalar 
in the region laying just outside the horizon, where $g^{tt} \to - \infty$ and $A_t \to 0$.
This is the reason for the condensation of the scalar in the bulk to be rather
common in the charged case, even with a vanishing boundary source. Two well studied boundary conditions for holographic superconductors are
\begin{itemize}
\item Dirichlet, which
is realised for  $\phi_0=0$ in eq. (\ref{expansion-phi-boundary}), with expectation value $\phi_1$.
\item Neumann, which is realised by  $\phi_1=0$ in eq. (\ref{expansion-phi-boundary}), with expectation value $\phi_0$.
\end{itemize}
For both boundary conditions above, at fixed $\rho$ there is a critical temperature $T_c$ such that the expectation values are non-zero for $T<T_c$.
By dimensional analysis, the critical temperature must be proportional to  $\sqrt{\rho}$, \emph{i.e.}
\beq
T_c= C(q) \, {\rho}^{1/2} \, ,
\eeq
where $C(q)$ is a function of the charge.
For a plot of $C(q)$ for Neumann and Dirichlet boundary conditions, see  fig. 2 of \cite{Hartnoll:2008kx}.
In the regime of large $q$, we have
\bea
C(q) &\approx& 0.12 \, \sqrt{q}  \qquad {\rm for \, Dirichlet}\,, \nl
C(q) &\approx& 0.23 \,  \sqrt{q}   \qquad  {\rm for \, Neumann} \, .
\eea
Just below the critical temperature $T_c$ the condensate is negligible
on the solution, at least outside the Cauchy horizon. Then, very close to $T_c$
the solution is given in good approximation by the RN solution.
Using expressions \eqref{RN-temp} and \eqref{fRN2} to express $(T,\rho)$ in terms 
of $(z_c,z_h)$, we find that the critical temperature $T_c$ corresponds to
the  value of $y=z_c/z_h$ obtained by solving the following equation
\beq
\frac{3 y^3-y^2-y-1}{4 \sqrt{2} \pi  \sqrt[4]{y^9 \left(y^2+y+1\right)}} = C(q) \, .
\eeq
This expression, in combination with the plot of $C(q)$ in \cite{Hartnoll:2008kx},
 can be use to find the holographic superconductor regime in the parameter
space.

As for the uncharged case, the $\phi_h \to 0$ limit can also be studied analytically when $q \neq 0$.
Once more, in this limit the backreaction of the scalar is negligible up to $z = z_i$,
see eq. (\ref{definizione-zeta-i}). For $z \approx z_i$, we have again the collapse of 
the Einstein-Rosen bridge.
As can be checked \emph{a posteriori}, to study the collapse of the bridge we can neglect 
 the mass of the scalar field $\phi$
and the electric source term in Maxwell's equation.
In particular, the gauge field $a$ still satisfies with good approximation
eq. (\ref{a-eq}).
The equations of motion (\ref{full-system-eom}) then take the form
\bea
&&  \le \frac{e^{-\chi/2} f}{z^3} \ri' =  {e^{-\chi/2} }
 \le    \frac{    \rho^2 }{4  } - \frac{ 3 } { z^4}  \ri  \, , \qquad
  \chi' = \frac{  z a^2 q^2 \phi ^2}{f^2} e^{\chi }  + z  (\phi' )^2   \, ,  \nl
&&  \le \frac{e^{-\chi/2} f \phi' }{z^2} \ri'  = - 
   \frac{1  }{z^2 }   \frac{ a^2 q^2 e^{\chi/2 }   }{ f } \phi \, .
\eea
The collapse takes place in a small range of the coordinate $z$, so that as in the uncharged case it is consistent
to set $z \approx z_i$ in the equations of motion above.
In correspondence with the collapse, the field $\chi$ rapidly becomes large
so that from eq. \eqref{a-eq} the gauge field $a$ can be taken to be practically constant, \emph{i.e.} $a(z)=a_0$.
With these simplifications, we must solve the system
\bea
&&  \le {e^{-\chi/2} f} \ri' =  {e^{-\chi/2} }
 \le    \frac{    \rho^2 }{4  } z_i^3  - \frac{ 3 } { z_i}  \ri  \, , \qquad
  \chi' = z_i \le \frac{   a_0^2 q^2 \phi ^2}{f^2} e^{\chi }  +  (\phi' )^2 \ri  \, ,  \nl
&&  \le {e^{-\chi/2} f \phi' } \ri'  = - 
    \frac{ a_0^2 q^2 e^{\chi/2 }   }{ f } \phi \, .
\eea
It turns out that the solution for $f$ and $\chi$ has the same functional form
as for $q=0$. The functional form of the scalar field $\phi$ is now different, however.
We can solve for $\phi$ in terms of $\chi$ and $f$,
\beq
\phi(z )= \Phi_0 \, \cos \le q \, a_0 \, \int_{z_i}^z  \frac{e^{\chi/2}}{f} dz + \varphi_0 \ri \, ,
\eeq
where $ \Phi_0$ and $ \varphi_0$ are integration constants.
Using the relation
\beq
q^2 \phi^2 a_0^2 + e^{-\chi} f^2 (\phi')^2 = q^2 \Phi_0^2 a_0^2 \, ,
\eeq
we get the same system as for $q=0$, see eq. (\ref{sistema-stoppaccioso}), 
with $\tilde{B}$ given by eq. (\ref{B-tilde}) and 
\beq
\tilde{A}=z_i  q^2 \Phi_0^2 a_0^2 \, .
\eeq
The solution is given by eq. (\ref{soluzione-stoppacciosa}), with eq. (\ref{robe-piccolette}) still holding. Note that in the $\phi_h \to 0 $ limit we still have
$e^{-\chi_0/2} f_0 =O(\phi_h^2) $,
because $\Phi_0=O(\phi_h)$.

The system can be solved numerically also away from the
limit $\phi_h \to 0$, as done in \cite{Hartnoll:2020fhc}.
See appendix \ref{numerical-appendix} for a description of our numerical methods.
 \newline

In particular, if we consider the limit $\phi_h \to 0$ keeping $T$ and $\rho$
fixed, we find that after the collapse of the ER bridge the solution
flows to a Kasner regime, with parameter $\alpha(\phi_h)$. As a function of
$\phi_h$,  the Kasner parameter $\alpha(\phi_h)$ is not continuous
for $\phi_h \to 0$, and it oscillates between values of $\a$
which can be bigger or lower than one.
In particular, if the value of $\a$ after the collapse is less than
one we see that there is a Kasner inversion,
which in the approximation discussed in \ref{kasner-inversion-section} brings $\a \to 1/\a$.
Something special happens in the limit $\a \to 0$: in this case
the approximations used in section \ref{kasner-inversion-section}
fail, because we should also take into account terms involving 
the charge of the scalar field. In this case multiple Kasner inversions
are possible, see \cite{Henneaux:2022ijt} for a recent study with the billiard approach.

\begin{figure}[h]
\begin{center}
\begin{tabular}{cc}
\includegraphics[width=0.48\textwidth]{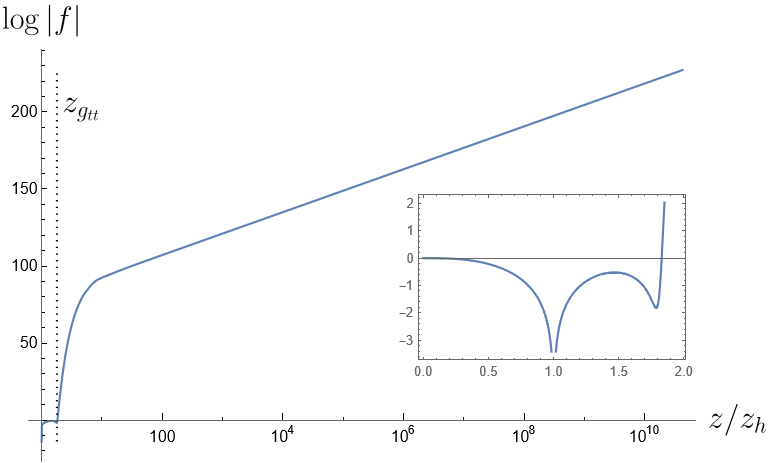} & \includegraphics[width=0.48\textwidth]{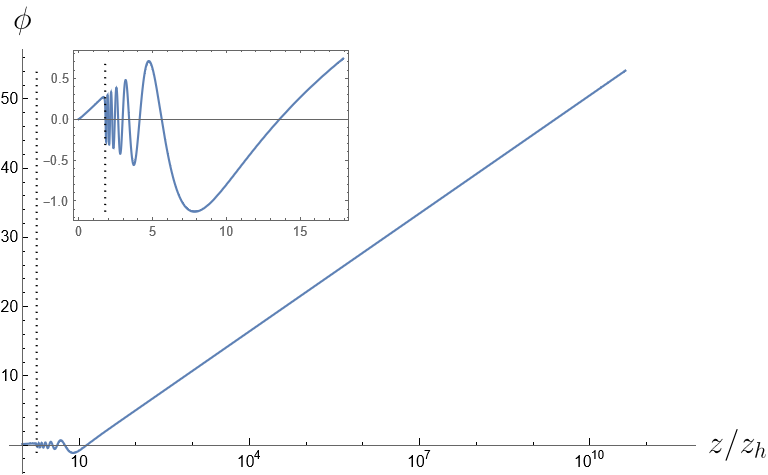}\\
\includegraphics[width=0.48\textwidth]{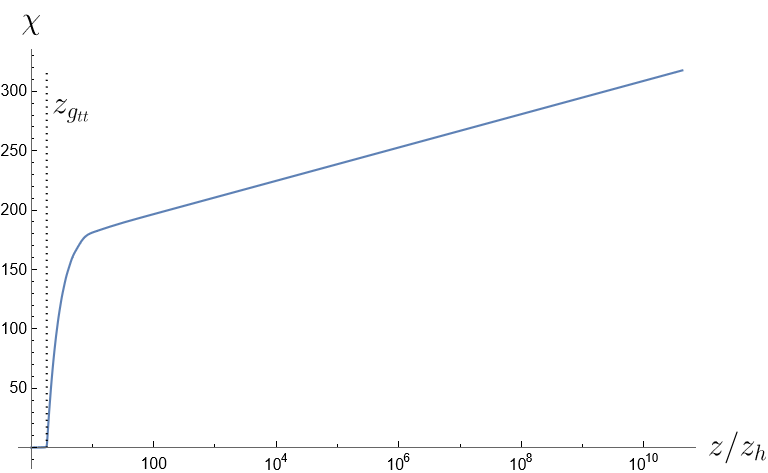} & \includegraphics[width=0.48\textwidth]{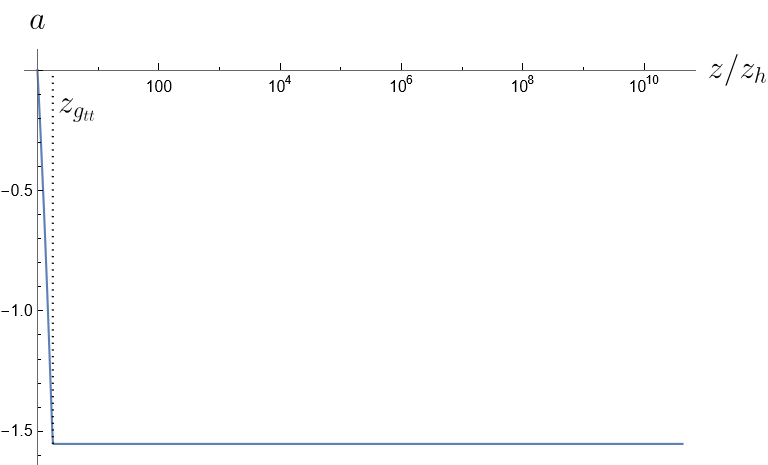}
\end{tabular}
\caption{Sample solution for the charged case, without Kasner inversion. For illustration purposes, we have chosen the values $\rho = 2, \phi_h = 0.16, z_h = 1, q = 1$ and $\chi_h = 0.1$. The vertical dotted line is the ``would be'' Cauchy horizon $z_{g_{tt}}$.  In the little boxes we display a magnification of the plots near $z_h =1$. }
\label{charged-sol}
\end{center}
\end{figure}

\begin{figure}[h]
\begin{center}
\begin{tabular}{cc}
\includegraphics[width=0.48\textwidth]{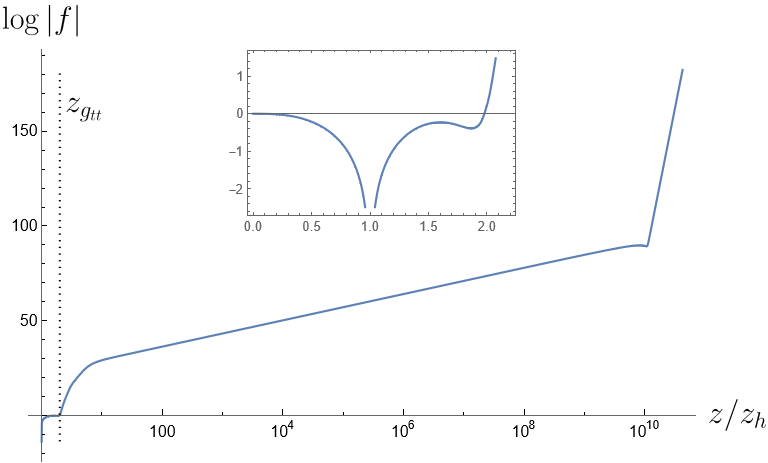} & \includegraphics[width=0.48\textwidth]{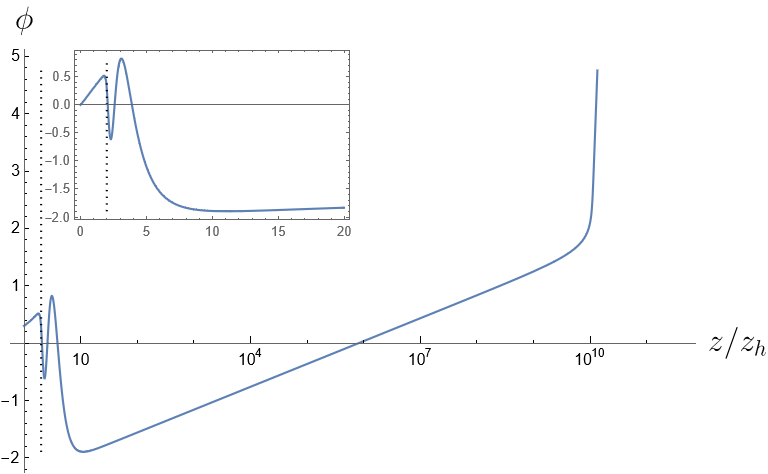}\\
\includegraphics[width=0.48\textwidth]{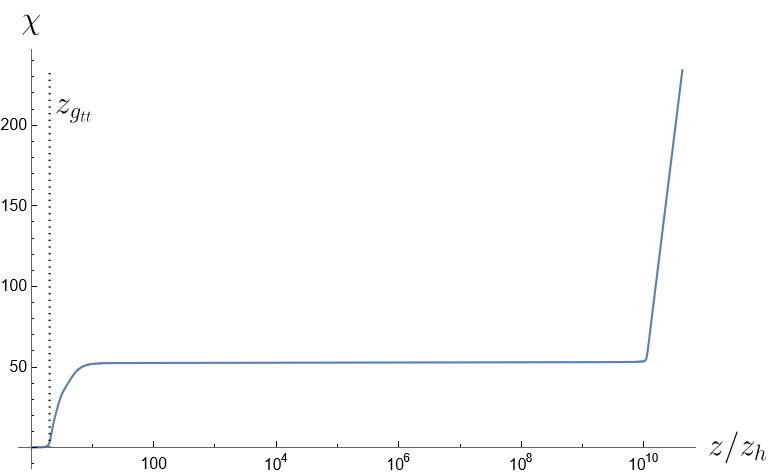} & \includegraphics[width=0.48\textwidth]{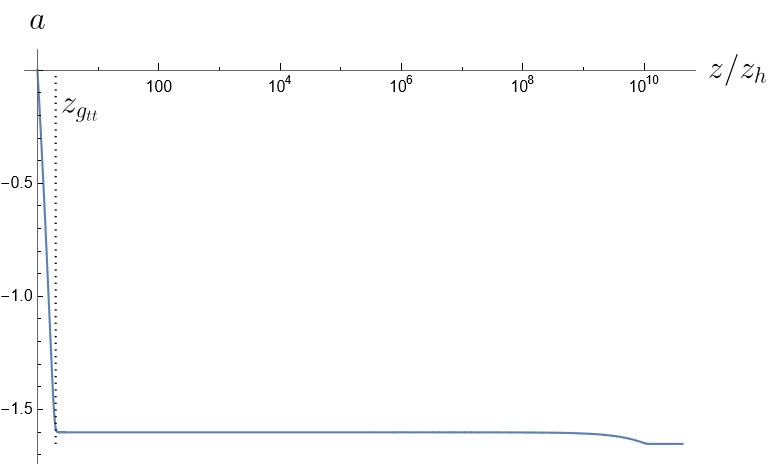}
\end{tabular}
\caption{Sample solution for the charged case, with a single Kasner inversion. For illustration purposes, we have chosen the values $\rho = 2, \phi_h = 0.3, z_h = 1, q = 1$ and $\chi_h = 0.1$. The vertical dotted line is the ``would be'' Cauchy horizon $z_{g_{tt}}$.}
\label{charged-sol-kinv}
\end{center}
\end{figure}

\begin{figure}[h]
\begin{center}
\begin{tabular}{cc}
\includegraphics[width=0.48\textwidth]{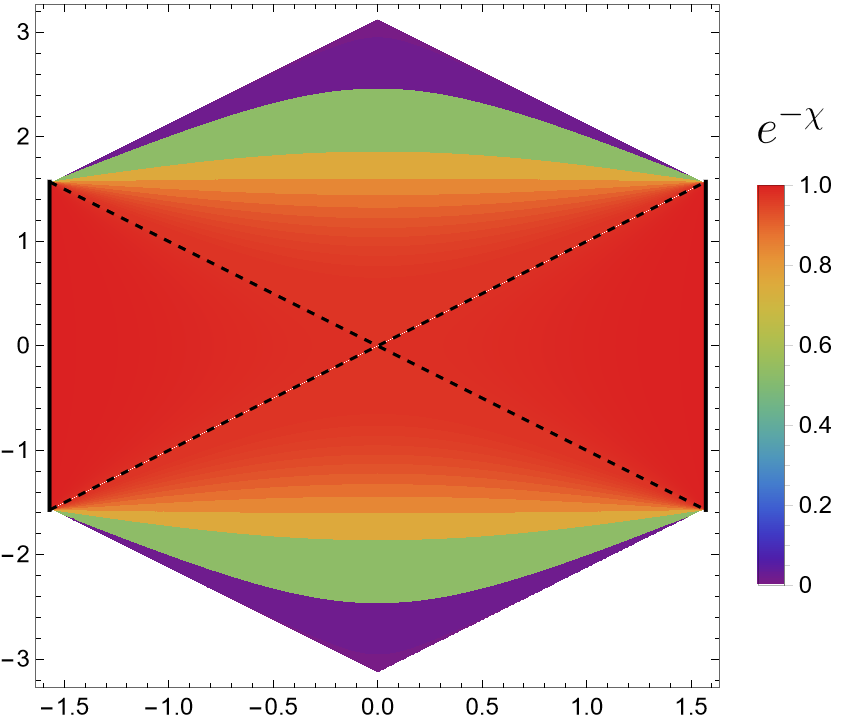} & \includegraphics[width=0.47\textwidth]{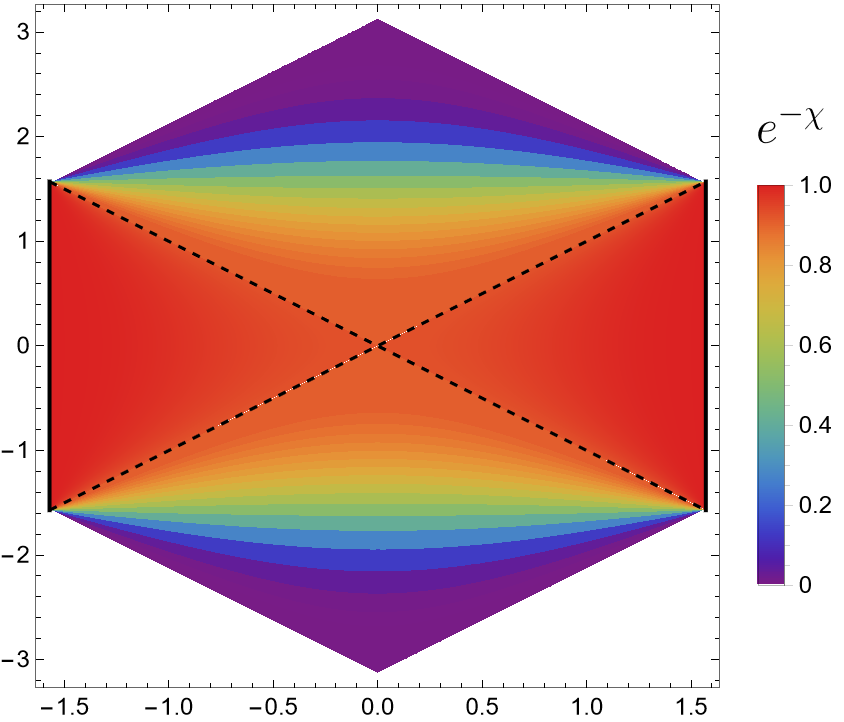} 
\end{tabular}
\caption{Penrose diagrams for the charged solutions shown in Figures \ref{charged-sol} left, and \ref{charged-sol-kinv} right. }
\label{penrose-examplescharged}
\end{center}
\end{figure}

In figure \ref{charged-sol} we show an example of a solution without inversions,
while in figure \ref{charged-sol-kinv} we show an example with one inversion.
Penrose diagrams for these solutions are shown in figure \ref{penrose-examplescharged}.

\begin{figure}[h]
\begin{center}
\includegraphics[scale=0.4]{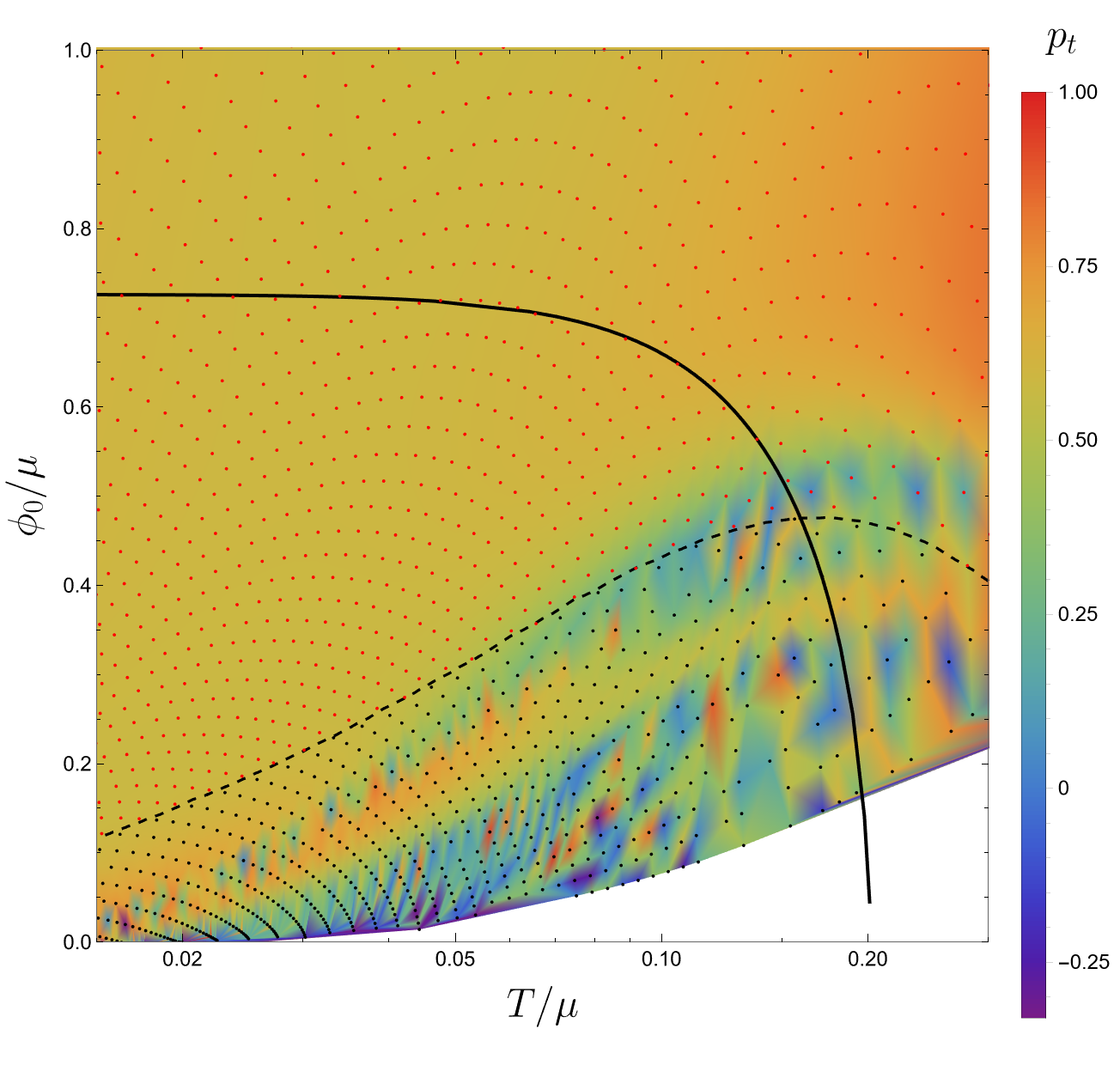} 
\caption{
The Kasner exponent $p_t$ as a function of the parameter space for $q=1$. The black solid line is the line of zero source $\phi_1=0$ with spontaneous condensation. The white region in this plot is where our numeric procedure didn't converge. The plot shows the last value of $p_t$ displayed by our solutions at the maximum possible value of $z$. The dashed line divides Type U (below) and type D (above) solutions.
}
\label{para-spazio-charged}
\end{center}
\end{figure}

Figure \ref{para-spazio-charged} displays a sketch of the Kasner 
exponent as a function of $\phi_0/\mu$ and $T/\mu$. Something similar could be obtained using $\sqrt{q\phi_1}/\mu$ instead, however the value of this condensate is very large for $q=1$ as previously found in \cite{Hartnoll:2008kx}, meaning the plot becomes less clear to visualize. For this reason we choose not to display it here.

\subsection{A conserved quantity}

Inserting the ansatz (\ref{metric1},\ref{ansatz1}) into the action \eqref{lagrangiana-modello} we obtain
\bea
\label{azione-ansatz}
S &=&  \frac{V_0}{16 \pi G}  \int d z   \left\{ - \frac{  e^{-\frac{\chi}{2}} f \chi'  }{z^3}   - \frac{ 12 f e^{-\frac{\chi}{2}} }{z^4}+  \frac{4 f' e^{-\frac{\chi}{2}} }{z^3}   +  \left(  \frac{   e^{-\frac{\chi}{2}}   f \chi'}{z^2}    -   \frac{   e^{-\frac{\chi}{2}}   f'  }{z^2}  \right)'  \right.\\
&& \qquad \qquad \quad  \left. + \frac{6e^{-\frac{\chi}{2}} }{z^4}   + \frac{e^{\frac{\chi}{2}} (a')^2 }{2}  - \frac{e^{-\frac{\chi}{2}}   f |\phi'|^2  }{z^2}  + \frac{e^{-\frac{\chi}{2}} }{z^4}   \left( \frac{z^2e^{\chi} q^2 a^2  }{ f }  -  m^2 \right) |\phi|^2\right\} .\nonumber
\eea
This action  is invariant under the symmetry \cite{Gubser:2009cg}
\beq
z \to \lambda z \, , \qquad \chi \to \chi -6 \log \lambda \, , \qquad a \to \lambda^2 a \, , 
\eeq
with fixed $f$ and $\chi$.  From this invariance, it follows that the quantity
\beq
\Q=
e^{\frac{\chi}{2}} \le   \frac{(f e^{-\chi})'}{z^2} -a a' \ri
\, ,
\label{Qconserved}
\eeq
is conserved, i.e. $\tilde{Q}'=0$,
 as can be checked from the equations of motion  (\ref{full-system-eom}).
For the RN solution, the conserved quantity  in eq.~(\ref{Qconserved}) is
\beq
\Q=\frac{z_h \rho^2}{4} -\frac{3}{z_h^3} \, .
\label{Q-conserved-RN}
\eeq
In this case $\Q$ is generically negative and vanishes in the extremal limit. \newline

In order to compute $\Q$  in the Kasner regime eq. (\ref{kasner-approx}),
let us first discuss the screening of electric charge.
We can write in general $a'$ as
\beq
a'=- \tilde{\rho}(z) \, e^{-\frac{\chi}{2}} \, ,
\label{rho-tilda}
\eeq
in such a way that  $\tilde{\rho} (0) =\rho$.
We can introduce  $\rho_f= \tilde{\rho} (\infty)$.
Using the equations of motion, we find:
\beq
\tilde{\rho}'(z) = - 2 L^2 q^2 \, a \, \frac{\phi^2 e^{\frac{\chi}{2}}}{f z^2} \, , \qquad
\tilde{\rho}(z)=\rho - 2 L^2 q^2 \, \int_{0}^z a \, \frac{\phi^2 e^{\frac{\chi}{2}}}{f \tilde{z}^2} \, d \tilde{z} \, .
\eeq
If we insert the approximation in eq. (\ref{kasner-approx})
in the conserved quantity in eq. (\ref{Qconserved}), we get
\beq
\tilde{Q}=f_0 e^{-\frac{\chi_0}{2}} (\a^2 -3 )+ \tilde{\rho}(z) a(z)  \, .
\label{Q-largez}
\eeq
We will see that this relation plays an important role in determining the asymptotic
complexity rate for the action conjecture.

In particular, it is interesting to consider the $\phi_h \to 0$ limit.
In this case, for $z < z_i$ defined in eq. (\ref{definizione-zeta-i}),
the profiles $f$, $\chi$  and $a$ are given by the RN solution.
We can then compare the conserved charge $\Q$ of the RN solution in eq.~(\ref{Q-conserved-RN})
with the one of the final Kasner in eq.~(\ref{Q-largez}), i.e. 
\beq
\Q=\frac{z_h \rho^2}{4} -\frac{3}{z_h^3}
=f_0 e^{-\frac{\chi_0}{2}} (\a^2 -3 )+ \rho_f \, a(\infty)  
 \, .
 \label{Qtilde-constarint-chargeless}
\eeq
In the limit $\phi_h \to 0$,
from the analytic solutions in eq. (\ref{soluzione-stoppacciosa}), we have that
 $\chi$ becomes suddenly very large 
at the would be Cauchy horizon $z_c$.
Moreover, from the equations of motion eq. (\ref{full-system-eom})
 we know that $\chi$ is a monotonic function of $z$.
Then, from eq. (\ref{rho-tilda}), we
have that $a$ approaches a constant for $z>z_c$. 
Then it follows that
\beq
a(\infty) \approx a_{\rm RN} (z_c) = -(z_c-z_h) \rho \, .
\eeq

Let us now distinguish the two cases:
\begin{itemize}
\item For $q=0$, there is no screening of electric charge, i.e.  $\tilde{\rho}(z)=\rho=\rho_f$. 
Using this result in eq. (\ref{Qtilde-constarint-chargeless}), we find 
a useful relation between $\a$, $f_0$ and $\chi_0$ valid in the $\phi_h \to 0$ limit:
\beq
f_0 e^{-\frac{\chi_0}{2}} (\a^2 -3 )  =
\frac{ (z_c -z_h )(z_c^2 + 2 z_h z_c+3 z_h^2)}{z_h^{3} z_c^{3}}   \, .
 \label{relazione-di-conservazione-q0}
\eeq
 From eq. (\ref{robe-piccolette}) we find 
 \beq
 f_0 e^{-\frac{\chi_0}{2}} =O(\phi_h^2)\, .
 \label{f0-chi0-A}
 \eeq
 As a consequence, $ \a  = O(1/\phi_h)$, as can be checked from numerical calculations.
We can  approximate the previous relation for $\phi_h \to 0$ as follows
\beq
\label{relazione-alpha-f0-chi}
 \a^2    = \frac{1}{f_0 e^{-\frac{\chi_0}{2}}} 
\frac{ (z_c -z_h )(z_c^2 + 2 z_h z_c+3 z_h^2)}{z_h^{3} z_c^{3}}  
 \, .
\eeq

\item 
For the first Kasner region we have that  eq. (\ref{f0-chi0-A}) still holds
because the solution in eq. (\ref{soluzione-stoppacciosa})
and the estimate  eq. (\ref{robe-piccolette}) still hold also for $q \neq 0$.
During a Kasner inversion, we expect that the value of $f_0 e^{-\frac{\chi_0}{2}}$
stays almost constant, because from equations of motion we have 
\beq
 \le \frac{e^{-\chi/2} f}{z^3} \ri' =  {e^{-\chi/2} }
 \le    \frac{    \tilde{\rho}^2 }{4  } - \frac{ 3 } { z^4}  \ri \, ,
 \label{f0-chi0-B}
 \eeq
 and, also, we have that $\chi$ is large in the $\phi_h \to 0$ limit.
The exponent $\a$ does not tend to infinity; instead,  the value of
$\a$ after the final inversion oscillates wildly for $\phi_h \to 0$ among all possible values from $1$ to $\infty$.
As a consequence of eq. (\ref{Qtilde-constarint-chargeless}) we have
\beq
\frac{\rho_f}{\rho}= \frac{1+2 y +3 y^2}{4 (1+y+y^2)} \, , \qquad y=\frac{z_c}{z_h} \, .
\eeq

\end{itemize}

\section{Volume complexity}
 \label{section-volume}
 
 In this section we will first treat with a unified approach different 
 extremal surfaces with different dimensions $k$, and then we will focus
 on the $k=3$ case, which is the one relevant for the volume
 conjecture.   Our results are consistent with \cite{Yang:2019gce}, 
 which focuses on the charged scalar case with vanishing external sources.

\subsection{Extremal bulk surfaces}

It is interesting to probe the asymptotically AdS$_4$
black hole geometry with extremal bulk surfaces
with different dimensions.
 In particular,  the length of dimension one spacelike 
 curves  is related to correlators of heavy scalar
 operators in the WKB approximation
 \cite{Balasubramanian:1999zv,Fidkowski:2003nf,Kraus:2002iv,Festuccia:2005pi,Festuccia:2008zx}.
 The area of dimension two extremal surfaces is the holographic dual 
of  the entanglement entropy \cite{Ryu:2006bv,Hubeny:2007xt,Hartman:2013qma}.
 In the CV conjecture, the complexity is dual
  to the volume of dimension three extremal bulk surfaces \cite{Susskind:2014rva,Stanford:2014jda,Carmi:2017jqz}.
In this section we will briefly discuss how to evaluate these three functionals 
in the black hole solution that we discussed in section \ref{section-model}.

It is useful to change variables, introducing  the lightcone coordinate $v$ defined by 
\beq
dv=dt-\frac{e^{\frac{\chi}{2}} }{f} dz \, ,
\label{dtime}
\eeq
where $v = {\rm  constant}$ correspond to the ingoing radial null geodesics. 
The metric (\ref{metric1}) in the coordinates $(v,z,x,y)$ reads 
\beq
ds^2= \frac{1}{z^2} \le -f e^{-\chi} dv^2 -2 e^{-\frac{\chi}{2}} dv dz +dx^2 + dy^2\ri \, .
\eeq
Let us then consider the following particular cases of extremal surfaces with different dimensions:
\begin{itemize}
\item
A spacelike radial geodesic (with constant $x$ and $y$), parametrized by  $z= z(\l)$ and $v= v(\l)$.
The length is
\beq
\ell=  \int  d \l  \, \frac{1}{z} \sqrt{- 2 \dot{v} \dot{z} e^{-\frac{\chi}{2}}
 -f e^{-\chi} \dot{v}^2 }  \, .
\eeq
\item 
A dimension two surface with constant $y$, parametrized by $(\l,x)$.
The area functional is
\beq
A= L_x \int   d \l \, \frac{1}{z^2} \sqrt{- 2 \dot{v} \dot{z} e^{-\frac{\chi}{2}}
 -f e^{-\chi} \dot{v}^2 }   \, .
\eeq
where $L_x = \int dx$ is a cutoff length in the $x$ direction.
%%%
\item
A dimension three surface, parametrized by  $(\l,x,y)$.
The volume functional is
\beq
V= V_{0} \int  d \l  \, \frac{1}{z^3} \sqrt{- 2 \dot{v} \dot{z} e^{-\frac{\chi}{2}}
 -f e^{-\chi} \dot{v}^2 }  \, .
 \label{volume-functional-originale}
\eeq
 where $V_{0} = \int dx dy$ is a cutoff area in the $x,y$ plane.
\end{itemize}

\subsection{Equations of motion}

We can treat the three functionals as particular cases of the same functional
 \beq
 S_k=\int  d \l  \, \frac{1}{z^k} \sqrt{- 2 \dot{v} \dot{z} e^{-\frac{\chi}{2}}
 -f e^{-\chi} \dot{v}^2 }   =
 \int d \l  \, \mathcal{L}_k \, .
 \label{funzionale-S-kappa}
 \eeq
Since $ S_k$ is  translationally invariant in $v$, we can obtain the  conserved quantity 
\beq
E_k=-\frac{\p \mathcal{L}_k}{\p \dot{v}}=\frac{1}{z^k} \frac{  \dot{z} \, e^{-\frac{\chi}{2}}+ \dot{v}  f e^{-\chi} }
{ \sqrt{- 2\dot{v} \dot{z} e^{-\frac{\chi}{2}}  -f e^{-\chi} \dot{v}^2 } } \, .
\label{energy-efficace}
\eeq
Following  \cite{Belin:2021bga},
it is convenient to choose the parameter $\l$ in this way
\beq
{z^k} \sqrt{- 2 \dot{v} \dot{z} e^{-\frac{\chi}{2}}
 -f e^{-\chi} \dot{v}^2 }=e^{-\frac{\chi}{2}}  \, .
 \label{normalizzando}
\eeq
With this choice, from eq. (\ref{energy-efficace}) we obtain
\beq
\dot{v}=(E_k-\dot{z}) \frac{e^{\frac{\chi}{2}}}{f} \, ,
\eeq
which, inserted in eq. (\ref{normalizzando}) gives
\beq
\dot{z}^2 +V_k(z) =  E_k^2 \, , \qquad
V_k(z)= - \frac{e^{-\chi} \,  f}{z^{2 k }}  \, .
\label{conserva}
\eeq
The problem is recasted as the motion of a classical
non-relativistic particle in a  potential $V_k(z)$.

\subsection{The time dependence}

Using eqs. (\ref{normalizzando}, \ref{conserva}), 
the functional $S_k$ in eq. (\ref{funzionale-S-kappa}) can be written as follows
\beq
S_k =  \int_{z_{\ep}}^{z_{t}}  \frac{dz}{z^k \sqrt{ f+E_k^2 z^{2 k} e^{\chi}}} \, .
\label{S-kappa-2}
\eeq
$z_{\ep}$  is the UV cutoff  and
  $z_{t}$  is the turning point.
 The value of $z_{t}$ can be obtained setting $\dot{z}=0$
 in eq.~(\ref{conserva}):
 \beq
 V_k(z_{ t}) =  E_k^2 \, . 
\label{zeta-massimo}
\eeq
The turning point  $z_{t}$ requires $f(z_{t}) \leq 0$,
and so it is inside the horizon $z_h$.
The quantity $E_k$ is function of the boundary time $t_b$ at which the 
probe is anchored.
Note also that for $z_{t}=z_h$,
we have that $E_k=0$. As a convention,  we set $t_b=0$ for $E_k=0$.

The difference in $v$ coordinates 
\beq
v\left(z_{t }\right)-v\left(z_{\ep }\right)
 = \int_{z_{\ep }}^{z_{t }}  \frac{  d z }{f} \le 
 \frac{e^{\chi} z^k E_k }{ \sqrt{ f + e^\chi E_k^2 z^{2k}}} - e^{\frac{\chi}{2}}  \ri \, .
\eeq
The boundary time can be obtained by integrating eq.~(\ref{dtime})
\beq
\int_{z_{\ep}}^{z_{t}} dt =\int_{z_{\ep}}^{z_{t}} dv + \int_{z_{\ep}}^{z_{t}} \frac{dz}{f} e^{\frac{\chi}{2}} \, .
\eeq
which gives
\beq
t(z_{t})-t(z_{\ep})=v(z_{t})-v(z_{\ep}) + \int_{z_{\ep}}^{z_{t}} \frac{dz}{f} e^{\frac{\chi}{2}} \, ,
\eeq
By symmetry argument at the turning point 
$t(z_{t})=0$.  We can then solve for the boundary time:
\beq
t_b =t(z_{\ep})=
- \int_{z_{\ep }}^{z_{t }}  \frac{  d z }{f}  
 \frac{e^{\chi} z^k E_k }{ \sqrt{ f + e^\chi E_k^2 z^{2 k}}} 
  \, .
 \label{tempo-sul-bordo}
\eeq
Using eq. (\ref{tempo-sul-bordo}), we can  rewrite $S_k$  in eq. (\ref{S-kappa-2}) as follows
\beq
S_k 
=\int_{z_{\ep}}^{z_{t}}  \frac{\sqrt{ f+E_k^2 z^{2 k} e^{\chi}}}{z^k f  } dz + E_k t_b
\label{S-kappa-3}
\eeq
Taking time derivative of (\ref{S-kappa-3}) and using eq. (\ref{zeta-massimo}), we find 
\beq
\frac{d  S_k }{d t_b}  = E_k \, .
\eeq
This proves that the derivative of $S_k$  with respect $t_b$ is  exactly  $E_k$.
Using the potential $V_k$ is defined in eq. (\ref{conserva}),
we can express the time derivative of $S_k$ as a function of $z_{t}$ as follows:
\beq
\frac{d  S_k }{d t_b}=
\sqrt{V_k(z_{t})}\, .
\eeq
The relation between boundary time $t_b$ and $z_{t}$
must be determined numerically, by integrating eq.~(\ref{tempo-sul-bordo}).

\begin{figure}[h]
\begin{center}
\includegraphics[scale=0.4]{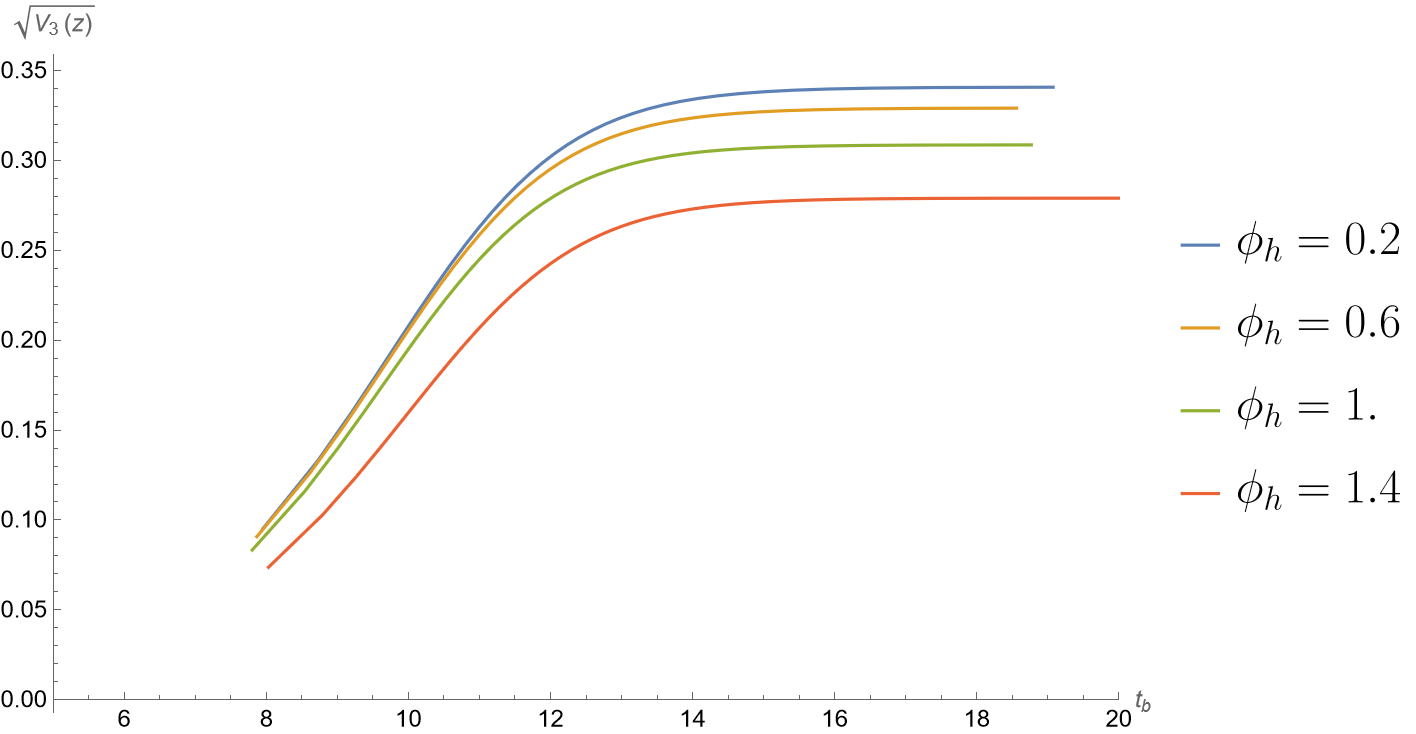} 
\caption{Volume complexity rate 
as a function of time  for a  black hole with $\rho=2$, $z_h=1$ and $q=1$.
}
\label{volcomp}
\end{center}
\end{figure}

We can write eq.~(\ref{tempo-sul-bordo}) in the following form
\beq
t_b =
 \int_{z_{\ep }}^{z_{t}}   \, \frac{  1 }{(-f e^{-\chi/2})}  
\frac{1}{\sqrt{E_k^2-V_k(z)}} \, dz 
  \, .
 \label{tempo-sul-bordo-2}
\eeq
 Figure \ref{volcomp} shows a plot of $\sqrt{V_3}$ as a function of $t_b$.

Note that the integrand in eq. (\ref{tempo-sul-bordo-2})
is diverging for $z=z_{t}$, since
$E_k^2=V_k( z_{t} )$.  In order for the result of the integral to diverge
logarithmically, we have to impose that $E_k^2-V_k(z)$ is quadratic in $z$
for $z=z_{t}$. For this reason, the late time  limit $t_b \to \infty$
corresponds to a maximum of $V_k(z_{t})$. 

For the Schwarzschild case, 
 $V_{2,3}$ have always a unique extremal point, which is a maximum.
On the other hand $V_1$ is a monotonic function
which  has no extremal points.
For this reason, spacelike geodesic can pass arbitrarily near
the singularity \cite{Fidkowski:2003nf}. Instead dimension two and three
extremal surfaces are respectively stuck at the value of $z$ corresponding to the maximum
of  $V_{2}$ and $V_{3}$.

In the RN case, we have that $V_k$ has always a unique
extremal point, which is a maximum,
at a finite  value of $z<z_c$.
So, no extremal surface can approach the singularity.

If we consider the model in eq. (\ref{lagrangiana-modello}), 
we have that, in the $\phi_h \to 0$ limit, the field profiles 
tend to the RN solution for $z < z_c$.
So, in this limit, the potential is the same as in the RN case
and it has a maximum for $z_h < z < z_c$.
The potential may have additional extremal points for
$z >z_c$.
Using the Kasner approximation in eq. (\ref{kasner-approx}), we get that at large $z$
\beq
V_k(z)=f_0 e^{-\chi_0} z^{3- 2 k - \a^2} \, .
\eeq
For $k>1$,  we get that $V_k(z) $ has no maximum in the Kasner region. 
As pointed out in  \cite{Hartnoll:2020fhc}, for geodesics ($k=1$)
we can get another maximum of $V_k(z)$
in correspondence of the Kasner inversion.
  In this case there exist values of the boundary time
  for which multiple extremal surfaces are possible.
It would be interesting to understand the physical meaning of these multiple solutions.

\subsection{Asymptotic  complexity rate}

The asymptotic volume complexity rate is given by
\beq
W_V= \lim_{t_b \to \infty}
\frac{d \, C_V}{d \, t_b }=\frac{V_0}{G } \sqrt{V_3(z_{\rm max})} \, ,
\label{wwwc}
\eeq
where $z_{\rm max}$ is the position of the maximum of $V_3(z)$,
see eq. (\ref{conserva}).
For the Schwarzchild solution we have
\beq
W_V
=\frac{V_0}{2 G } \frac{1}{z_h^3}  
=\frac{8 \pi}{3}  \, T S \, .
\label{volume-rate-Schwarzchild}
\eeq
In the extremal limit, for the RN solution
the rate vanishes and at the first order in $z_c-z_h$ is 
\beq
W_V
\approx  \sqrt{\frac32}  \,  \frac{V_0}{G }\frac{z_c-z_h}{z_h^4}  
=\sqrt{\frac32}  \, \frac{8 \pi}{3}  \, T S \, .
\label{volume-rate-extremal}
\eeq

\begin{figure}[h]
\begin{center}
\includegraphics[scale=0.3]{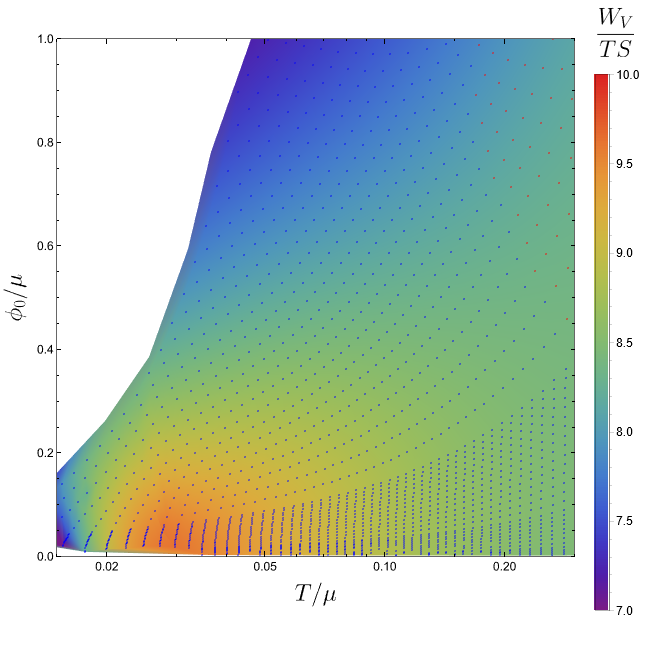}
\qquad
\includegraphics[scale=0.3]{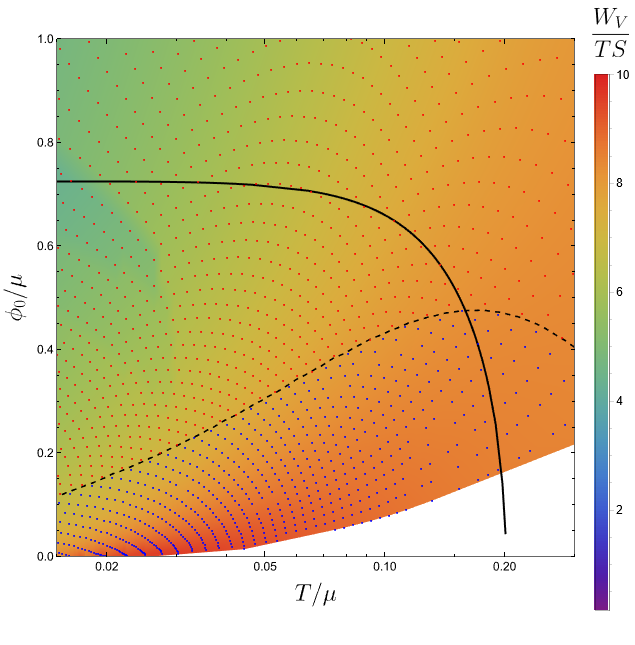}
\caption{ 
Asymptotic volume complexity rate  $ W_V$, as defined in eq. (\ref{wwwc}),
in units of $T S$.  Left: Neutral scalar case ($q=0$).
 Blue ploints denote type U solutions, red points denote type D solutions. 
Right: Charged scalar case with $q=1$. 
Blue points denote type U solutions, yellow points denote type D solutions. 
In both cases, there is no direct correlation between the complexity 
 rate and the coefficient $p_t$ plotted in figures
  \ref{uncharged-para-sol}  and  \ref{para-spazio-charged}.
   The white region of the plots is not covered by our numerics.
  }
\label{2d-asintotica-volumecomplexity}
\end{center}
\end{figure}

\begin{figure}
\begin{center}
\begin{tabular}{cc}
\includegraphics[width=0.45\textwidth]{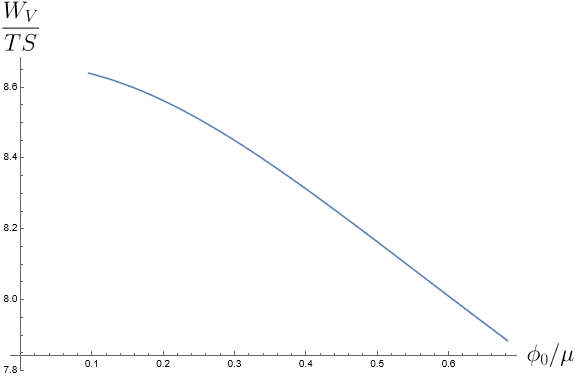}
&
\includegraphics[width=0.45\textwidth]{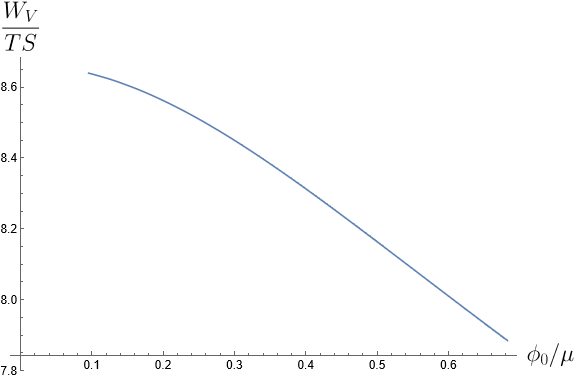}
\end{tabular}
\caption{
Left: Asymptotic volume complexity for charged solutions with $q=1$ at constant $T/\mu=0.15$. 
Right: Asymptotic volume complexity for uncharged solution at $T/\mu = 0.25$.
}
\label{asymptoticvolatconstT}
\end{center}
\end{figure}

We expect that the asymptotic complexity rate, in units of $TS$,
is a slowly varying function of the parameters of the model.
This is consistent with the numerical results shown in figure
\ref{2d-asintotica-volumecomplexity}.
In figure \ref{asymptoticvolatconstT} we show the behavior at constant temperature 
of the asymptotic volume complexity, as a function of $\phi_0/\mu$.

%%%%%%%%%%%%%%%%
%%%%%%%%%%%%
%%%%%%%%%%%%%%%%%%%

The Lloyd bound \cite{Lloyd} conjectures that the rate of computation of a quantum
computer is bounded by a quantity which is proportional to the total energy of the system.
In the context of holography, in \cite{Brown:2015lvg} it was conjectured that
the Lloyd bound for the complexity rate
is saturated by the uncharged planar BH in AdS$_{d+1}$, i.e.
 \beq
\frac{d \, C_V}{d t_b } \leq \frac{8 \pi}{d-1} M \, ,
\eeq
where $M$ is the mass of the black hole.
In the complexity=volume conjecture, the complexity rate
is a monotonically increasing function of the time
and so it is enough to check it at late time.

\begin{figure}[h]
\begin{center}
\includegraphics[scale=0.25]{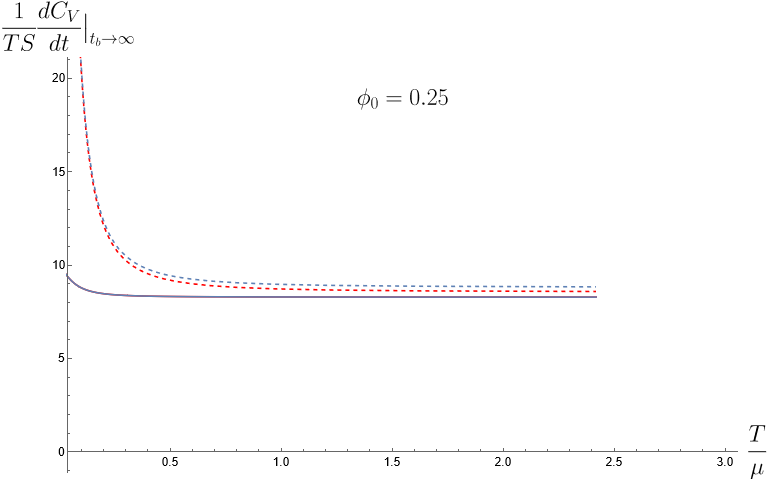}
\qquad
\includegraphics[scale=0.25]{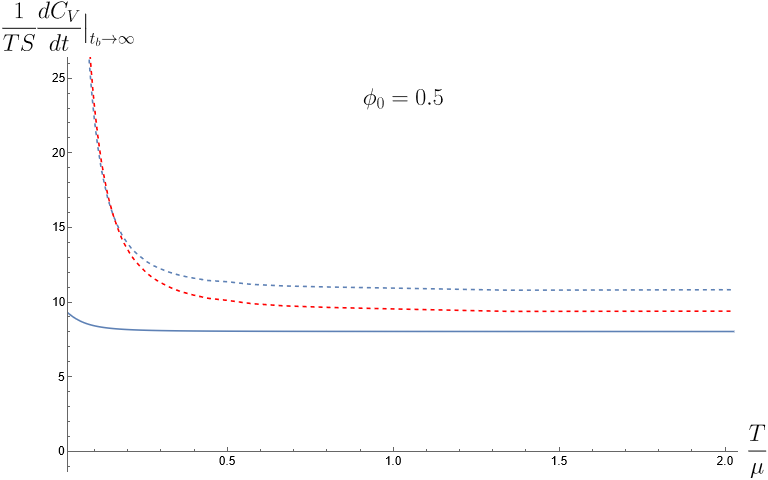}
\caption{  
In these pictures,   we show $W_V/(T S)$ in solid line, $(4 \pi M_{D})/(TS)$ in blue dashed line
and $(4 \pi M_{N})/(TS)$ in red dashed line 
(see eqs. (\ref{wwwc}),(\ref{lloyd-bound-CV}))
as a function of $T/\mu$ for a fixed $\phi_0$.
Left: neutral scalar. Right: charged $q=1$ scalar.
The Lloyd bound is always satisfied for both the Dirichlet and the Neumann definition of mass.
  }
\label{lloyd-volume}
\end{center}
\end{figure}

In our setting there is an extra subtlety, because the value of the black hole 
mass depends on the choice 
of the boundary conditions, see appendix \ref{app-masses} for details.
We checked the Lloyd bound both for the
 Dirichlet  and Neumann   choices of boundary conditions, that for $d=3$ is
 \beq
W_V  \leq 4 \pi M_{D,N} \, ,
\label{lloyd-bound-CV}
\eeq
where $M_{D,N}$ are in eqs. (\ref{mass-D},\ref{mass-N}).
We find that the bound is satisfied in all the parameter space that 
we explored, see fig. \ref{lloyd-volume} for some sample plots
for fixed value of $\phi_0$.
This is consistent with \cite{Yang:2019gce} , who numerically checked
the Lloyd bound in the case of zero sources
(for which the Neumann and Dirichlet masses coincide).

%%%%%%%%%%%%%%%%
%%%%%%%%%%%%

 \subsection{Generalised volume functionals}
 A broader class of complexity duals were proposed in \cite{Belin:2021bga}.
In particular, a larger class of functionals was considered, i.e.
\beq
\hat{S}=\int d^3 \sigma \, \sqrt{h} \,  F (g_{\mu \nu}, X^\mu) \, ,
\label{generalised-functional}
\eeq
 where $F$ is a scalar function of the background metric and of the embedding
 $X^\mu (\s^a)$ of the codimension-one surface.
 These  functionals may provide an infinite  class of alternative 
 holographic definitions of complexity; the CV conjecture is recovered
 by setting  $F$ equal to a constant.
 
 If we consider a surface parameterized by $(\l,x,y)$ in our metric ansatz, we have that
 a  functional of the form eq. (\ref{generalised-functional}) can be written as follows
  \beq
\hat{S}=\int  d \l  \, \frac{a(z)}{z^3} \sqrt{- 2 \dot{v} \dot{z} e^{-\frac{\chi}{2}}
 -f e^{-\chi} \dot{v}^2 }   = \int d \l  \, \hat{\mathcal{L}} 
 \, ,
 \label{funzionale-S-kappa-generalizzato}
 \eeq 
 where  the function $a(z)$ can be found by evaluating $F (g_{\mu \nu}, X^\mu)$
 on the surface.
 
 In a similar way to eq. (\ref{energy-efficace}),
due to translation invariance in $v$, we can again define the conserved quantity
\beq
E=-\frac{\p \hat{\mathcal{L}}}{\p \dot{v}}=\frac{a(z)}{z^k} \frac{  \dot{z} \, e^{-\frac{\chi}{2}}+ \dot{v}  f e^{-\chi} }
{ \sqrt{- 2\dot{v} \dot{z} e^{-\frac{\chi}{2}}  -f e^{-\chi} \dot{v}^2 } } \, .
\eeq
As in eq (\ref{normalizzando}), it is convenient 
to fix the $\l$ parameterization as follows
\beq
{z^3} \sqrt{- 2 \dot{v} \dot{z} e^{-\frac{\chi}{2}}
 -f e^{-\chi} \dot{v}^2 }=a(z) \, e^{-\frac{\chi}{2}}  \, .
 \label{normalizzando2}
\eeq
In this gauge, the conserved quantity $E$ takes the form
\beq
E=  \dot{z}  + \dot{v}  f e^{-\chi/2} 
\, , \qquad \dot{v}=(E-\dot{z}) \frac{e^{\frac{\chi}{2}}}{f} \, ,
\eeq
which, inserted back in (\ref{normalizzando2}) gives
an effective potential as in eq. (\ref{conserva})
\beq
\dot{z}^2 +V (z) =  E^2 \, , \qquad
V(z)= - \frac{e^{-\chi} \,  f \, a^2}{z^{6 }}  \, .
\label{potenziale-funzionale-generalizzato}
\eeq
 
 The details of complexity evolution depend crucially on the choice of 
 $ F (g_{\mu \nu}, X^\mu) $.
 Let us briefly discuss  an example.  Let us choose, as in the example explicitly discussed in \cite{Belin:2021bga},
 the following function
 \beq
 F(g_{\mu \nu}, X^\mu)=1+ \kappa \, W_{\mu \nu \rho \s} W^{\mu \nu \rho \s} \, ,
 \label{effe-weyl}
 \eeq
 where $\kappa$ is a constant and $W_{\mu \nu \rho \s}$ is the spacetime Weyl tensor.
 In this case, we find
 \beq
a=1 + \kappa \, \frac{z^4}{12} \le 2 f'' -3 f' \chi' + f (\chi')^2 - 2 f \chi'' \ri \, .
\eeq
If we specialize to the RN back hole case as in eq. (\ref{fRN1}), we find:
\beq
a=1+ \kappa \,
\frac{12 z^6 \left(\left(z_c+z_h\right) \left(z_c^2+z_h^2\right)-2 z \left(z_c
   z_h+z_c^2+z_h^2\right)\right){}^2}{z_c^6 z_h^6} \, .
   \label{a-RN}
\eeq

\begin{figure}
\begin{center}
\includegraphics[scale=0.5]{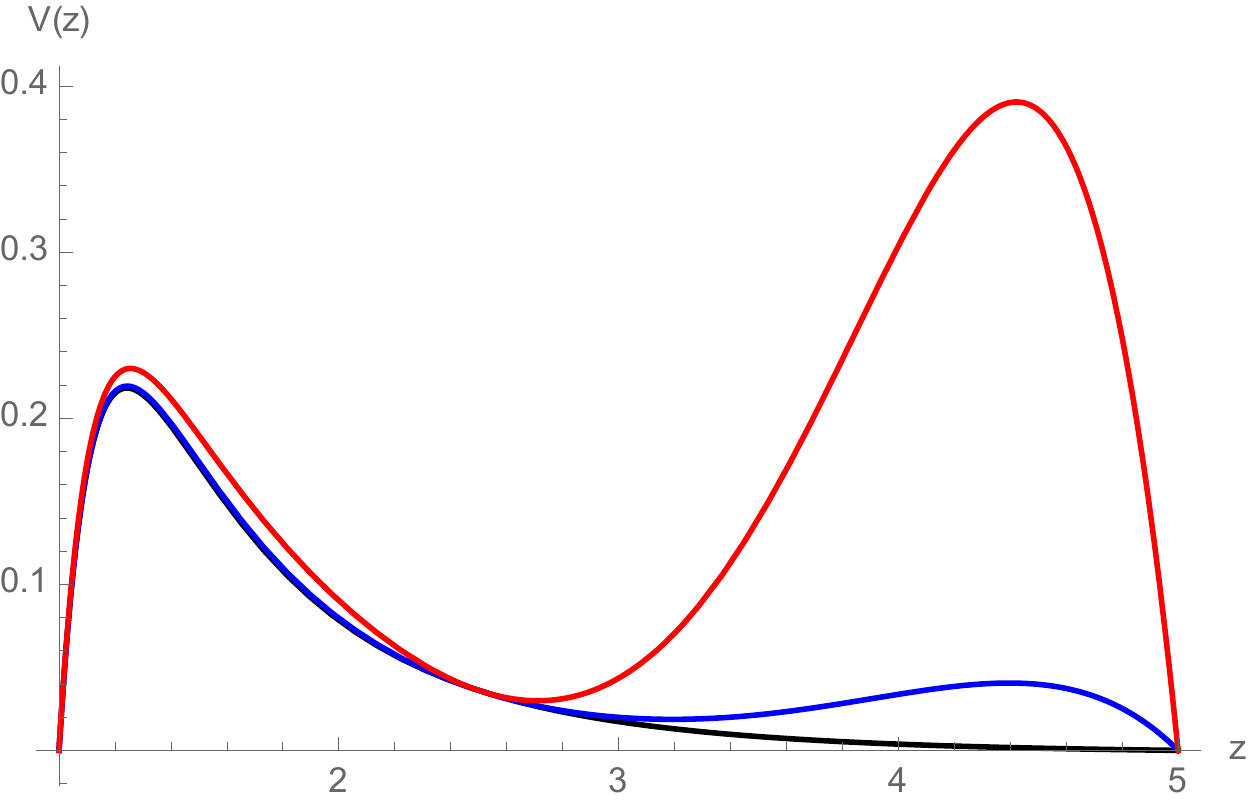} 
\caption{ Examples of potentials $V(z)$ the RN black solution for with $z_h=1$, $z_c=5$.
 In the black curve, we choose the volume functional, i.e. $F=1$.
 In the blue and red curves, we choose the functional in eq. (\ref{effe-weyl}) 
 with  $\kappa=0.0003$ and  $\kappa=0.003$, respectively.
}
\label{potenziale-generalizzato}
\end{center}
\end{figure}
We plot some examples of potentials in figure \ref{potenziale-generalizzato}.
In these cases the potential in eq. (\ref{potenziale-funzionale-generalizzato})
has two maxima.
Each of these maxima correspond to an extremal surface of the given functional
which is attached to the boundary at a time $t_b \to \infty$.

This shows that, changing  $ F (g_{\mu \nu}, X^\mu) $,   for a given value of the time 
we can have multiple extremal surfaces. 
The asymptotic complexity rate for each of the 
extremal surfaces is proportional to the value of $\sqrt{V(z)}$
evaluated on each of the maxima.
With a generic choice of  the parameter $\kappa$,
%$ F (g_{\mu \nu}, X^\mu) $, 
we expect that,
in the hairy black hole case, the extremal surface remains generically
far from the would be Cauchy horizon.
It would be interesting to study other choices of $ F (g_{\mu \nu}, X^\mu) $
in a more systematic way. We leave this as a topic for further investigation.

%
%If we perturb this solution with a small scalar hair, we have that the 
%RN solution changes in a negligible way for $z < z_c - \epsilon$,
%and then develops a spacelike singularity at $z \to \infty$

 %By fine tuning   $ F (g_{\mu \nu}, X^\mu) $ it is 
%possible that we can approach the Kasner singularity,
%but this would require a somewhat fine-tuned  choice 
%of the complexity functional.
%We leave this as a topic for further investigation.

%%%%%%%%%%%%%%%%%%

\section{Action complexity}

 \label{section-action-complexity}

To compute the action complexity we must first identify the Wheeler-DeWitt 
(WDW) patch \cite{Brown:2015bva}, which is given by the union of all the spatial 
slices that can be attached to the left and right boundaries
at some given pair of times $t_L$ and $t_R$.
The Killing vector corresponding to a translation in $t$
in the metric (\ref{metric1}) shifts
\beq
t_L \to t_L + \delta t \, , \qquad t_R \to t_R + \delta t \, .
\eeq
For this reason, if we take the left and the right time translation in opposite direction, 
the thermofield double state is time-independent. 
In order to study the time dependence of the complexity of the thermofield double, 
we take as boundary conditions for the WDW patch
\beq
t_L=t_R=\frac{t_b}{2} \, .
\eeq
The boundary of the WDW patch can be obtained
 by sending null rays from  the left and right boundary.
  
 Various shapes of the WDW region are possible.  If the future light rays meet at a point we have a future joint, 
 otherwise the future light rays end in the singularity and thus we have a space-like side of the WDW. 
 The same is true for the past light rays so  in total we have four possible shapes: 
 the diamond \dia, a five-sides polygon with one side on the singularity of the  future 
  \pentau or the past \pentad, or a six- sides polygon with sides on both singularities \exa. 

Once the WDW patch is obtained for a given boundary time $t_b$, we need to compute the action
\cite{Lehner:2016vdi}
 which  is given by the sum of the following contributions
 \beq
  I_{\rm WDW} =   I_{\rm V}  +  I_{\rm GHY}  +  I_{\rm J} + I_{\rm b} +  I_{\rm ct} \, , \label{actionfour}
\eeq
where $I_{\rm V}$ is the bulk actions in eq. (\ref{lagrangiana-modello}),
 $I_{\rm GHY}$ is the Gibbons-Hawking-York boundary term eq. (\ref{ghyformula}), 
$I_{\rm J}$ is the joint term (\ref{jjoints}),
$ I_{\rm b}$ is the  null boundary term (\ref{boundary-action})
 and $I_{\rm ct}$ its counter term (\ref{counter-term-action}).
Details on how to compute these four terms are given in appendix \ref{appterms}.
 We will always compute the derivative of the action with respect  the boundary time $\frac{d I_{\rm WDW}}{d t_b }$. 
 This removes various UV divergences\footnote{In our setting divergencies are time-independent because 
they come from the region of spacetime outside the horizon,
and so they are invariant under the Killing vector $\p_t$
(which is timelike for $z<z_h$ and spacelike for $z>z_h$).
This is different from the case where the metric  have no timelike Killing vector
 \cite{Bolognesi:2018ion}.}
 present in $I_{\rm WDW}$.

In addition to the gravitational constant $G$,  the action (\ref{actionfour}) contains
a scale $L_{\rm ct}$ which is needed in a  counterterm
that restores the reparameterization invariance of the action.
This term was introduced in \cite{Lehner:2016vdi} with the motivation to remove 
an ambiguity in the parameterization of the null hypersurfaces which delimit the WDW patch.
In many static situations, such as in  \cite{Carmi:2017jqz},
 the asymptotic complexity rate at $t_b \to \infty$ is independent
on $L_{ct}$, which affects only the action rate during a finite transient period.
In out-of-equilibrium situations, such as  Vaidya spacetime, the counterterm is
important in order to reproduce many expected properties of complexity,
such as the late-time growth and the switchback effect \cite{Chapman:2018dem,Chapman:2018lsv}.
The counterterm scale introduces an ambiguity in defining holographic complexity
that could be related to the details of how complexity is defined in the dual field theory,
such as the reference state or the choice of gates. 

In order to evaluate the action of the WDW patch,
it is convenient to introduce the tortoise coordinate
\beq
z^*(z)=\int_0^z \frac{d \tilde{z}}{f( \tilde{z})} e^{\frac{\chi( \tilde{z})}{2}} \, ,
\label{tortoise-zstar}
\eeq
in which the radial-time part of the metric is conformally flat
\beq
ds^2= \frac{1}{z^2} \left( f e^{-\chi} (-dt^2 +(dz^*)^2 ) 
+dx^2 +dy^2 \right) \, ,
\eeq
and the coordinates $u$ and $v$
\beq
v=t-z^* \, , \qquad u= t+z^* \, .
\label{coordinate-cono-luce-u-v}
\eeq

The time dependence of the action complexity depends on the type of Penrose
diagram that we are considering, which can have a singularity with
upper bending (type $U$) or lower bending (type $D$).
It is useful to introduce a quantity to discriminate the two kinds of behaviour.
An ingoing null geodesic which leaves the boundary at $t_b=0$
reaches the singularity at a time $t_{\infty}$ given by
\beq
t_{\infty}=z^*(\infty)=\int_0^\infty \frac{e^{\frac{\chi}{2}} }{f} d \tilde{z} \, ,
\eeq
which is a convergent integral at $z \to \infty$, as can be shown using
the asymptotic Kasner behaviour in eq.~(\ref{kasner-approx}).
This shows that the integral is converging at large $z$.
If the quantity $t_{\infty}$ is positive, the solution is of type $D$.
Instead,  if $t_{\infty}$ is negative, the solution is of type $U$.
In both cases, there is a critical boundary time $t_c$ at which the structure of the WDW patch changes
in a discontinuous way. 

\subsection{RN case}

The time dependence of action complexity for the RN black hole was studied in
\cite{Carmi:2017jqz}. We briefly review their results  in appendix \ref{appe-RN-action-complexity},
where we give a closed-form expression for the time dependence of the complexity.
   In general, the details of the evolution 
 during the transient time range  area  function the counterterm scale $L_{\rm ct}$.
 The complexity rate at late time is instead independent of $L_{\rm ct}$.

\begin{figure}
\begin{center}
\includegraphics[scale=0.6]{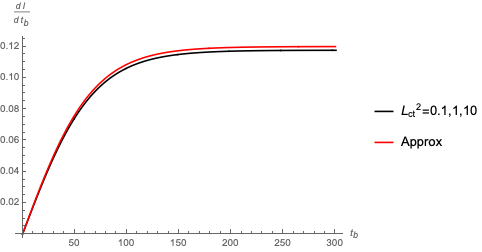} \qquad
\includegraphics[scale=0.6]{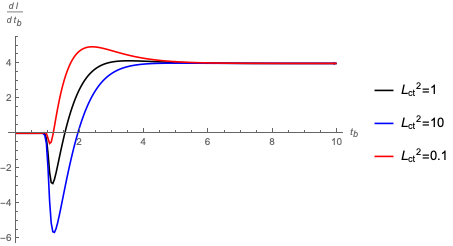} 
\caption{
Action  complexity rate as a function of time for some examples of RN case, with $V_0/(16 \pi G)=1$.
Top: almost extremal case with $T/\mu=1.4 \cdot 10^{-3}$, $\rho=3.43$ and $z_h=1$. The approximate analytical expression 
in eq. (\ref{RN-neraby-extremality-rate-CA}) in red is compared 
with numerical one, in black. In the scale of the plot, the curves with ${L}_{\rm ct}^2=0.1,1,10$
all coincide with the black curve.
Bottom: case with $T/\mu=0.28$, $\rho=0.8$ and $z_h=1$  (for various ${L}_{\rm ct}^2$). 
}
\label{action-rate-RN-1}
\end{center}
\end{figure}

Nearby the extremal limit, at small $T/\mu$, we have that
the time dependence of the rate of the  action complexity is a smooth monotonic
function of the time, and it is at first approximation
independent of the counterterm scale $L_{\rm ct}$.
By expanding the results in \cite{Carmi:2017jqz} (see appendix \ref{appe-RN-action-complexity-extremal})
we find the compact expression
 \beq
\frac{d I}{d t_b } \approx \frac{3 \, V_0}{4 \pi G}  \frac{z_c-z_h}{ z_h^4} \tanh \frac{3 t_b (z_c-z_h)}{2 z_h^2} \, .
\label{RN-neraby-extremality-rate-CA}
\eeq
For a comparison with numerical results, see the top panel of  figure \ref{action-rate-RN-1}.

As we increase $T/\mu$, the qualitative features of the plot change.
For $T/\mu \to \infty$, the Schwarzschild case must be reproduced.
In this case, the complexity rate is zero up to the critical time $\hat{t}_c$ in eq (2.9) of
\cite{Carmi:2017jqz}
\beq
\hat{t}_c=\frac{1}{2 T} \frac{1}{\sqrt{3}} =\frac{2 \pi }{3 \sqrt{3}} z_h  \approx 1.21 z_h \, .
\label{tempo-critico-RN}
\eeq
where $T$ is the temperature $T=3/(4 \pi z_h)$.
 In appendix \ref{appe-RN-action-complexity-Sch} we check
that this is indeed the case, by studying the $z_c/z_h \to \infty$ of the RN case.
Just after  $\hat{t}_c$, the rate drops to $- \infty$ in a discontinuous way, and after that raises to 
 approach a positive constant at late time.

 At large but finite $T/\mu$, we have that the action rate 
 is still to a good approximation constant up to a timescale $t_b \approx \hat{t}_c$.
 Then, there is a sudden  drop of the action rate to a large negative value
 (which in the limit $T/\mu \to \infty$ tends to $-\infty$).
After that the action rate raises  and   approaches 
 a positive constant at $t_b \to \infty$.
  Figure \ref{action-rate-RN-1} shows and example of this behavior from the numerical solution.

\begin{figure}[h]
\begin{center}
\includegraphics[scale=0.50]{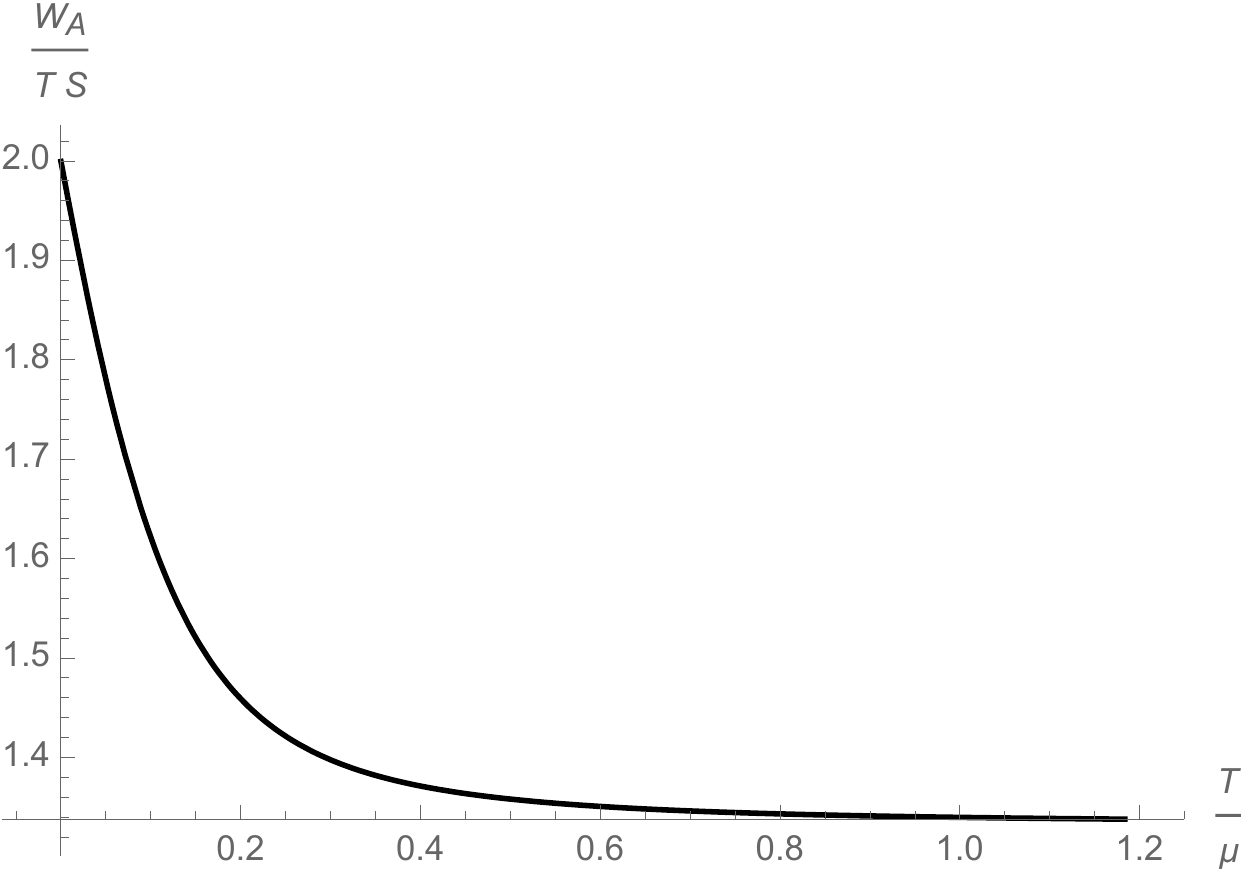}
\includegraphics[scale=0.50]{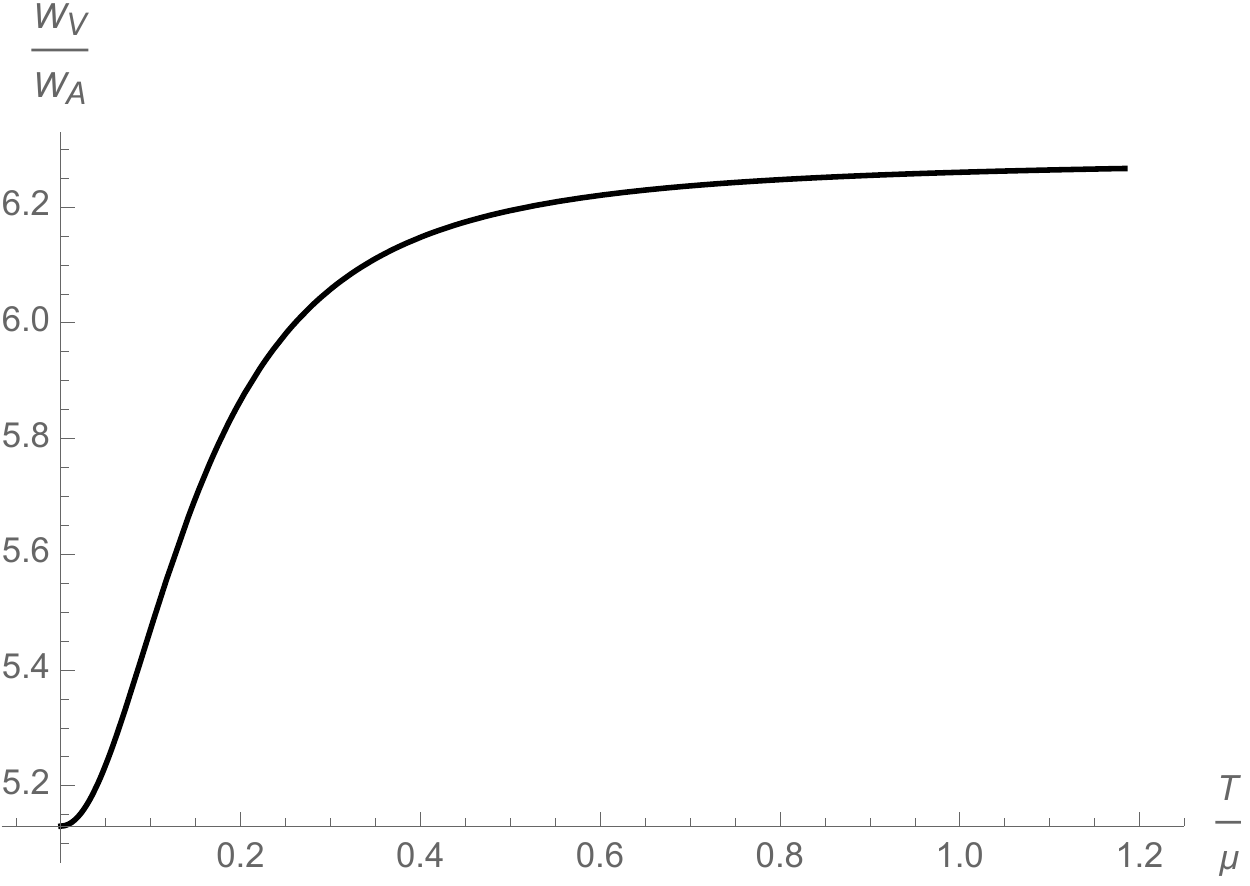}
\caption{ Left: Plot of ${W_A}/{TS}$  as a function of $T/\mu$ for the RN solution.
Right: Ratio between the asymptotic CV and the asymptotic CA rate
 as a function of $T/\mu$ for the RN solution.
Combining eqs.   (\ref{volume-rate-Schwarzchild}), (\ref{volume-rate-extremal})
and (\ref{action-rate-late-time-RN}), 
we find that for $T/\mu \to \infty$ we have $W_V/W_A= 2 \pi$,
while for $T/\mu \to 0$ we have $W_V/W_A= 2 \pi \sqrt{2/3}$.
This is in agreement with numerical calculations for generic $T/\mu$.
}
\label{CA-asintotico-RN}
\end{center}
\end{figure}

At late time, from a direct calculation it follows that the bulk contribution 
in eq. (\ref{RN-Bulk}) vanishes, i.e.  ${d I_V}/{d t_b} \approx 0$.
Also,  we can approximate eq.~(\ref{RN-Joint-and-Counterterm}) as
\beq
\frac{d I_J}{d t_b} \approx
  \frac{V_0}{8 \pi G} 
  \left(  \left. \frac{ f_{\rm RN}'  }{2 z^2}  \right|_{z=z_c}  -  
   \left. \frac{  f_{\rm RN}' }{2 z^2} \right|_{z=z_h}
  \right) \, .
\eeq
Defining a formal temperature  $T_c$ and an entropy $S_c$ computed
on the Cauchy horizon
\beq
T_c=- \frac{1}{4 \pi} f_{\rm RN}'(z_c) 
 =-\frac{(z_c-z_h)(3 z_h^2+2 z_c z_h +z_c^2)}{4 \pi z_c z_h^3}\, , 
\qquad S_c=\frac{1}{4 G} \frac{V_0}{z_c^2} \, ,
\label{TS-cauchy}
\eeq
the action rate  at late time is
\beq
W_A=\lim_{t_b \to \infty} \frac{d I_{\rm WDW}}{d t_b } 
=\frac{V_0}{4 \pi G} \frac{z_c^3-z_h^3}{ z_c^3 z_h^3}
= T \, S \, - T_c \, S_c 
 \, ,
 \label{action-rate-late-time-RN}
\eeq
where $T$ and $S$ are the RN temperature and entropy in eqs  (\ref{carica-totale}) and
(\ref {RN-temp}). 
A plot of the asymptotic action rate is shown on the left hand side of figure \ref{CA-asintotico-RN}.
On the right hand side of the same figure, we compare the action and the volume rates
for different $T/\mu$. 

\subsection{Type $D$ }

This case corresponds to a singularity with a lower bending and to  $t_{\infty}>0$. 
Here the WDW starts at $t_b=0$ in the  shape \exa, with two sides on the past and future singularity.
There is a  critical time:
\beq
t_c=2 z^*(\infty) =2 t_{\infty} \, .
\label{termpo-critico-tipo-D}
\eeq
For $t_b < t_c$, the WDW remains hexagonal,
see figure \ref{fig-caso1} on the left. 
In this case the complexity rate is zero, because the time dependence contribution from regions $1$ and
$3$ in figure turns out to cancel each other, and the contribution from region $2$ is
time independent. See appendix \ref{appe-dettagli-azione-D-presto} for the details of the calculation.
The same cancellation is present also for the Schwarzschild case \cite{Carmi:2017jqz}.

\begin{figure}[h]
\begin{center}
\includegraphics[scale=0.4]{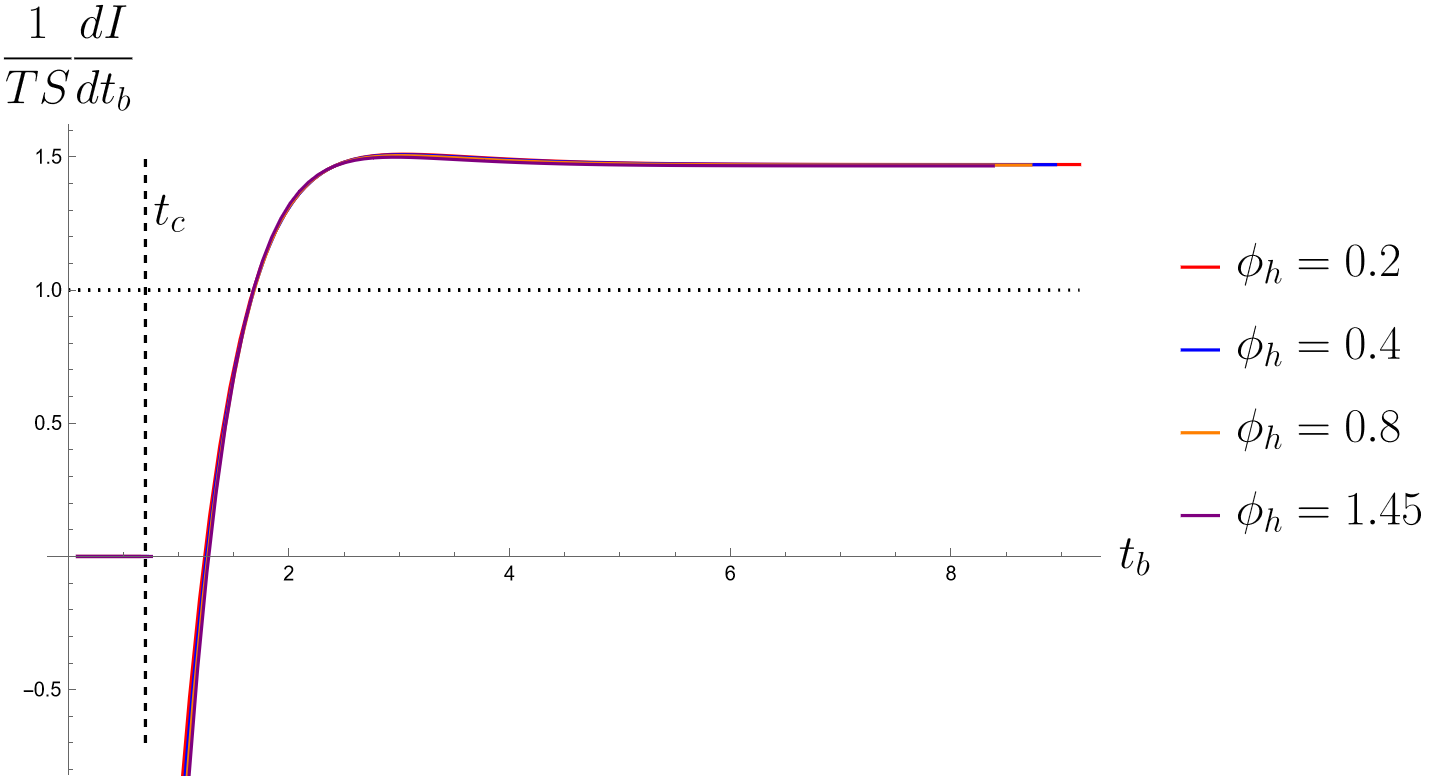}
\qquad
\includegraphics[scale=0.4]{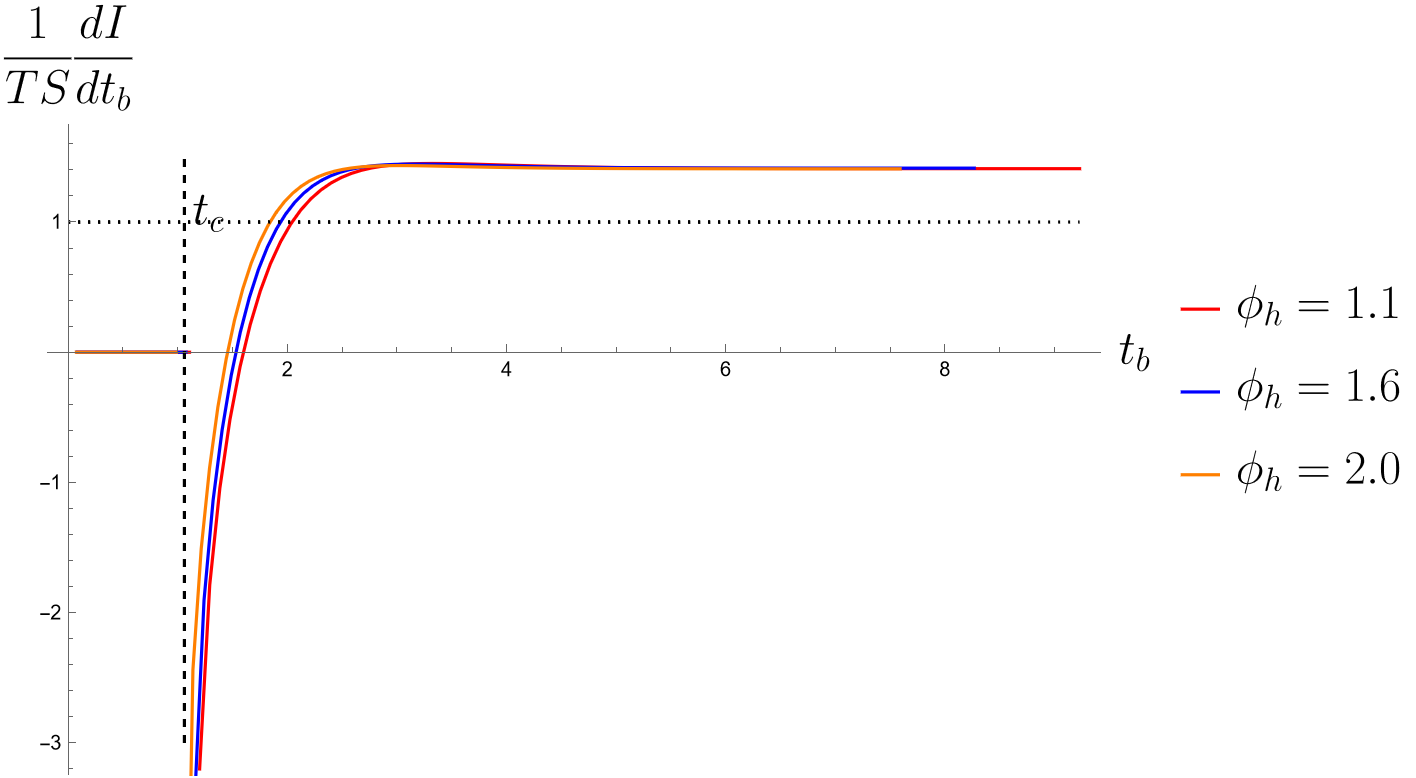}
\caption{
Time dependence of the action complexity rate in various type D examples.
Top: $q=0$ and  $\rho=2$.
Down:   $q=1$ and $\rho=0.6$.
In these examples we set $z_h=1$ and ${L}_{\rm ct}=1$.
}
\label{action-time-dep-a}
\end{center}
\end{figure}

At $t_b = t_c$ the shape becomes \pentau with the lower tip touching the past singularity.
 For $t_b> t_c$  the shape remains  \pentau with only one side at the future singularity. 
 See figure \ref{fig-caso1} on the right.
  Here the complexity rate is non zero and has a non trivial time dependence on $t_b$.
First of all, we should find the $z_{m1}$ of the lower tip of the WDW as a function of $t_b$.
For $t_b=t_c$, we have $z_{m1} \rightarrow \infty$ since it is touching the past singularity. 
For $t_b \rightarrow \infty$,  we have $z_{m1}=z_h$.
For generic time $t>t_c$, we can find
$z_{m1}(t_b)$ by inverting the equation:
\beq
z^*(z_{m1})= \frac{t_b}{2}   \, ;
\eeq
it is the same as the equation (\ref{tipsforrn}) for the RN case, but now only the lower tip is relevant. 
For the complexity rate we should evaluate three contributions
\beq
 \frac{d I }{d t_b}=
 \frac{d \,  I_V }{d t_b}+\frac{d \, I_{\rm GHY} }{d t_b}+
\frac{d \, (I_J+I_{\rm ct})}{d t_b} \, .
\label{rateIafter}
\eeq
A detailed derivation is provided in appendix \ref{appe-dettagli-azione-D-U-tardi}.
The bulk contribution is given by
\beq
 \frac{d \,  I_V }{d t_b}
 =  \frac{V_0}{16 \pi G_N}
  \int^\infty_{z_{m1}} dz  \, s(z)  \, , 
\eeq
 where $s(z)$ is given by eq.~(\ref{eq-s-generic-2}).
The GHY contribution is a constant and gives:
\beq
\frac{d I_{\rm GHY} }{d t_b}=\frac{V_0}{16 \pi G} \lim_{z \to \infty }A(z) \, , 
\label{ghyafter}
\eeq
 where $A(z)$ is given by eq.~(\ref{A-arbitrary}).
Using the asymptotic Kasner solution in eq.~(\ref{kasner-approx})
in the expression (\ref{A-arbitrary}) we find
\beq
\label{A-asintotico}
A(\infty) 
=f _0 e^{-\frac{\chi_0}{2}} (3+\a^2) \, .
\eeq
 The joints and counterterm contributions are
\beq
 \frac{d (I_J+I_{\rm ct})}{d t_b}
=\frac{V_0}{16 \pi G} 
 \left\{ f e^{-\frac{\chi}{2}} \left[ \frac{2}{z^3} 
  \log \left(   -4 f    \,  {L}_{\rm ct}^2  
  \right)
 -\frac{1}{z^2}   \frac{d}{d z} 
 \left( \log \frac{-f e^{-\chi}}{z^2 } \right)
  \right]
  \right\}_{z=z_{m1}} 
\label{rateIafterJct}
\eeq
Examples of total complexity rate as function of time are shown 
in figure \ref{action-time-dep-a}.

It is interesting to investigate more in detail the behaviour of the complexity rate
nearby the critical time $t_c$ in eq. (\ref{termpo-critico-tipo-D}).
Before $t_c$, the action rate vanishes.
Just after the critical time $t_c$, the joint inside the past horizon sits nearby the singularity,
in the Kasner region at $z_{m} \rightarrow \infty$. Using the approximations in
eq.~(\ref{kasner-approx}), we find the negative log divergence in the joint part of the action rate
\beq
\frac{d I_J}{d t_b} \approx
  -\frac{V_0}{8 \pi G} A(\infty) \, \log z_{m}  \, .
\eeq
where $A(\infty)$  is given in eq.~(\ref{A-asintotico}).
Combining eq.~(\ref{time-derivative-zm}) with the $z \to \infty $ behaviour in (\ref{kasner-approx}), we obtain
\beq
\frac{d z_{m1}}{d t_b} 
 \approx
\frac{-f_0  e^{-\frac{\chi_0}{2}}  \,  z_{m1}^{3}  }{2}  \, .
\eeq
Integrating this differential equation nearby critical time (where $ \Delta t_b =t_b-t_c>0$), we find:
\beq
z_{m1}=\frac{1}{\sqrt{f_0  e^{-\frac{\chi_0}{2}} \Delta t_b } }  \, ,
\eeq
which gives the following behaviour as a function of time
\beq
\frac{d I_J}{d t_b} \approx  \frac{V_0}{16 \pi G} A(\infty) \, \log \Delta t_b \, .
\eeq
So, for small  $\Delta t_b >0$ the rate diverges $\to - \infty$. This divergence is visible in the 
type $D$ example of figure \ref{action-time-dep-a} on as $t \to t_c^+$.
The divergence at $t_b=t_c$ is similar to the one that we have for the Schwarzschild case
for $t_b=\hat{t}_c$ in eq. (\ref{tempo-critico-RN}).

%%%%%%%%%%%%%%%%%%%%%%%%%%%%

\subsection{Type $U$ }

\begin{figure}[h]
\begin{center} 
\includegraphics[width=0.8\textwidth]{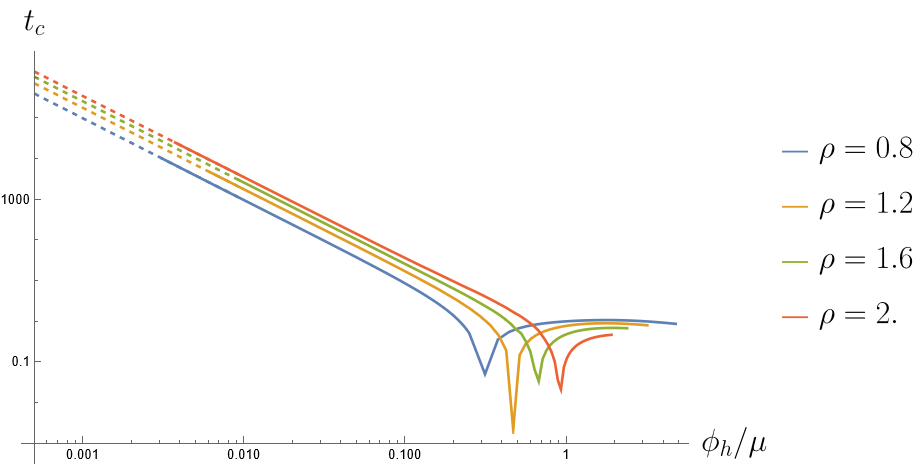}
\caption{Critical time for the $q=0$ case as function of $\phi_h$, for several $\rho$ and $z_h=1$. For each curve, the minimum in $t_c$ separates the solutons of types U and D. Dashed lines are extrapolations following the law $t_c = {\rm const.} \times \left(\phi_h/\mu\right)^{-2.003_3}$.
}
\label{critical-time}
\end{center}
\end{figure}

 This case corresponds to $t_{\infty}<0$ and to a singularity with an upper bending.
This situation is realised in the limit of $\phi_h \to 0$,
 which is closer to the unperturbed RN solution.
Here the WDW starts at $t_b=0$ in the shape \dia.
There is a  critical time, for which the WDW changes shape, given by 
\beq
t_c=- 2 z^*(\infty) = -  2 t_{\infty} \, .
\label{crittimeII}
\eeq
As we approach the limit $\phi_h \to 0$, 
the critical time $t_c$ goes to infinity.
In appendix \ref{appe-stima-tempo-critico}
we perform an analytic estimate of the divergence of the critical time, which give the result
\beq
t_c =O\left( \frac{1}{\phi_h^2} \right)
 \, .
 \label{stima-tempo-critico-small-phih}
\eeq
This is consistent with the numerical calculations, see figure \ref{critical-time}.

\begin{figure}[h]
\begin{center}
\includegraphics[scale=0.4]{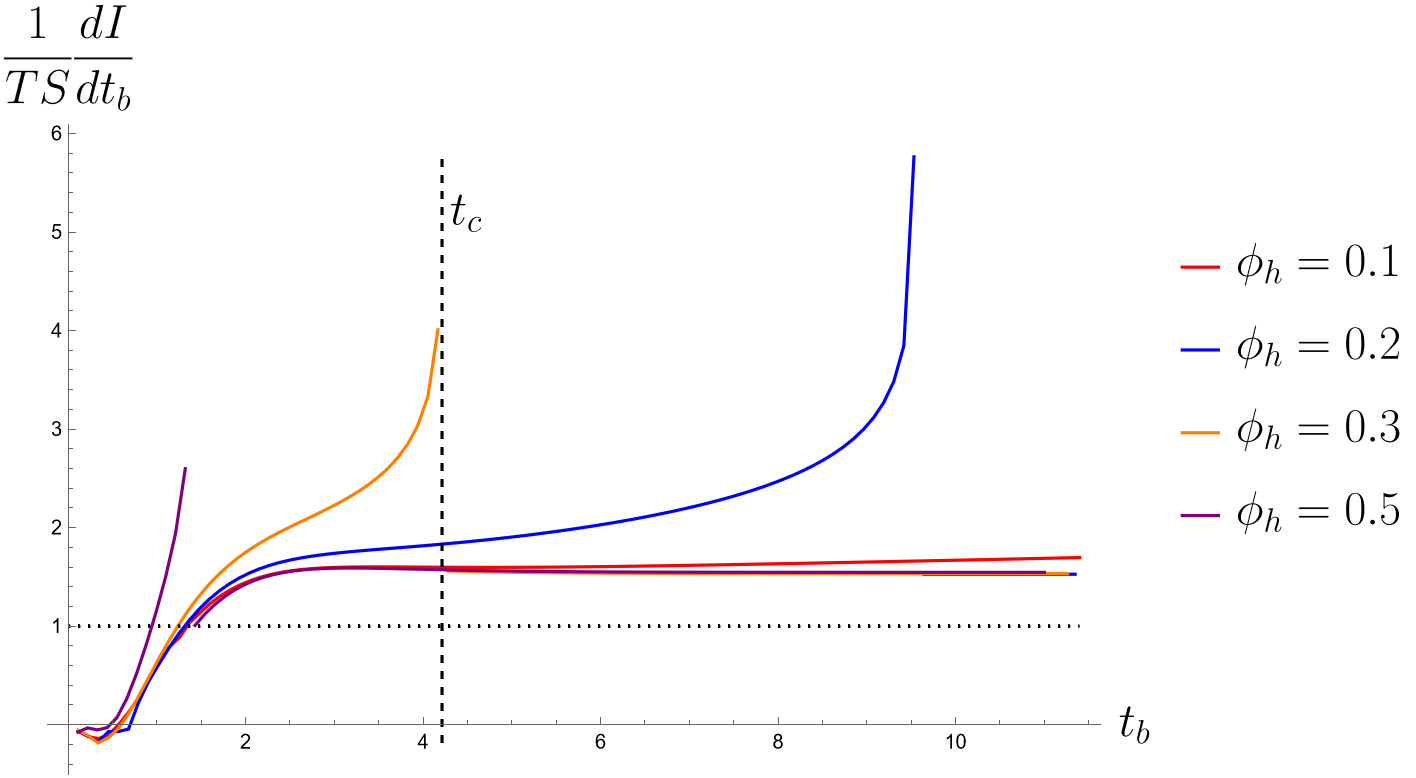}
\caption{
Time dependence of the action complexity rate in various type U examples
 for various $\phi_h$, with $q=0$, $z_h=1$ and ${L}_{\rm ct}=1$. }
\label{action-time-dep-b}
\end{center}
\end{figure}

\begin{figure}[h]
\begin{center}
\includegraphics[scale=0.33]{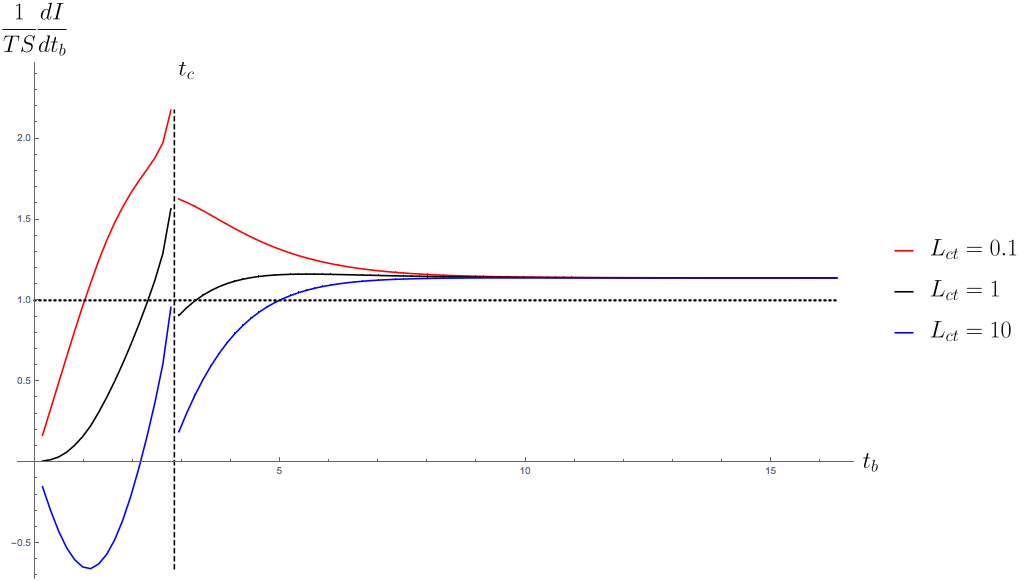}
\caption{
Time dependence of the action complexity rate  as a function  of $t_b$ for various ${L}_{\rm ct}$
 with $q=0$, $\phi_h=0.7$,
$z_h=1$ and $\rho=2.2$.}
\label{varyingLtilde}
\end{center}
\end{figure}

For $t< t_c$, , the WDW is a diamond, see figure \ref{fig-caso2} on the left  for an example.
 It is important to know first the coordinate of the two joints inside the horizons
 as a function of the boundary time $t_b$. Let us denote respectively by $z_{m1}$
 and $z_{m2}$ the coordinates of the joints inside the white and the black hole horizons.
They can be found by solving the equations
\beq
z^*(z_{m1})= \frac{t_b}{2} \, , \qquad
z^*(z_{m2})= -\frac{t_b}{2} \, .
\label{eqtipsII}
\eeq
For the complexity rate we should evaluate two contributions:
\beq
 \frac{d I }{d t_b}=
 \frac{d \,  I_V }{d t_b}+
 \frac{d( I_J+I_{\rm ct})}{d t_b} \, .
\eeq
See appendix \ref{appe-dettagli-azione-U-presto} for details.
The bulk contribution is
\beq
 \frac{ d \,  I_V}{d t_b}= 
 \frac{V_0}{16 \pi G_N} \int^{z_{m2}}_{z_{m1}} s(z) dz \,  ,
 \eeq
 $s(z)$ is given by eq.~(\ref{eq-s-generic-2}).
 The joints and the counterterm give
 \bea
\frac{d( I_J+I_{\rm ct})}{d t_b } &=&
\frac{V_0}{16 \pi G} \left(
 \left\{ f e^{-\frac{\chi}{2}} \left[ \frac{2}{z^3} 
  \log \left(   -4 f    \,  {L}_{\rm ct}^2  
  \right)
 -\frac{1}{z^2}   \frac{d}{d z} 
 \left( \log \frac{-f e^{-\chi}}{z^2 } \right)
  \right]
  \right\}_{z=z_{m1}} 
  \right. \nl 
&& - \left. 
 \left\{ f e^{-\frac{\chi}{2}} \left[ \frac{2}{z^3} 
  \log \left(   -4 f    \,  {L}_{\rm ct}^2  
  \right)
 -\frac{1}{z^2}   \frac{d}{d z} 
 \left( \log \frac{-f e^{-\chi}}{z^2 } \right)
  \right]
  \right\}_{z=z_{m2}}  \right)
\eea
At $t_b = t_c$ the shape  is still \dia with the higher tip touching the future  singularity. For $t_b> t_c$  the shape becomes 
 \pentau with only one side at the future singularity. See figure \ref{fig-caso2} on the right  for an example. 
For $t>t_c$, only $z_{m1}$ exists and can be found solving the first of (\ref{eqtipsII}). 
During this stage, the structure of WDW patch is the same of case $D$ and also the expression for the 
complexity rate eqs.~(\ref{rateIafter})-(\ref{rateIafterJct}).  
Examples of complexity rate in case $U$ for different $\phi_h$
as function of time are given in figure \ref{action-time-dep-b}. 
In figure \ref{varyingLtilde} we show a plot of total complexity rate for the uncharged case as a function of choice of ${L}_{\rm ct}$.

It is interesting to investigate the behaviour of the complexity rate
nearby the critical time in eq. \ref{crittimeII}.
Just before the critical time $t_c$ , as the  joint inside the future horizon approaches the singularity, the action rate
get a positive divergent contribution
\beq
 \frac{d I_J}{d t_b} \approx 
\frac{V_0}{8 \pi G} \, 
 A(\infty) \, \log z_{m2} \, ,
\eeq
where $A(\infty)$  is given in eq.~(\ref{A-asintotico}).
Just after $t_c$ there is no divergence in the rate.
Combining eq.~(\ref{time-derivative-zm-12})  with the $z \to \infty $ behaviour in (\ref{kasner-approx}), we obtain
\beq
\frac{d z_{m2}}{d t_b}
 \approx \frac{f_0  e^{-\frac{\chi_0}{2}}  \,  z_{m2}^{3}  }{2} \, .
\eeq
Following the same steps as for case $D$, we obtain
\beq
z_{m1}=\frac{1}{\sqrt{-f_0  e^{-\frac{\chi_0}{2}} \Delta t_b } } \, , 
\eeq
which gives, as a function of time
\beq
\frac{d I_J}{d t_b} \approx - \frac{V_0}{16 \pi G} A(\infty) \, \log (-\Delta t_b) \, .
\eeq
So, for small  $\Delta t_b < 0$ the rate diverges $\to  \infty$.
This divergence is visible in the type $U$ example of figure \ref{action-time-dep-b} on as $t \to t_c^-$.

The divergence of the action complexity rate at the critical time $t_c$ then is determined
by the quantity $A(\infty)$ in eq.  (\ref{A-asintotico}), which contains the
parameters of the metric in the $z \to \infty$ Kasner regime,
 defined in eq. (\ref{kasner-approx}).
So one may think that the coefficient of the logarithmic 
divergence of the complexity rate at $t_b=t_c$
is directly related to the Kasner exponent in the black hole interior.
This does not seem to be the case, at least in the regime of small $\phi_h$.
In particular, let us distinguish to cases:
\begin{itemize}
\item for $q=0$ and $\phi_h \to 0$, we have that $A(\infty)$ is not 
directly sensitive to the Kasner parameter $\a$ (which in this limit tends to infinity), but to a quantity
that can be defined in term of the unperturbed RN geometry.
This can be shown combining  eqs. (\ref{A-asintotico}) and  (\ref{relazione-alpha-f0-chi}), which give
\beq
A( \infty) = \frac{(z_c-z_h)(z_c^2 +2 z_c z_h +3 z_h^2)}{ z_c^3 z_h^3 }=
- \frac{16 \pi G}{V_0} T_c S_c \, ,
\label{A-small-phih}
\eeq
where $T_c$ and $S_c$ are the formal temperature and entropy computed
on the Cauchy horizon of the RN solution, see eq. (\ref{TS-cauchy}).
\item for $q \neq 0$ and $\phi_h \to 0$, we have that $A(\infty)$  vanishes.
This can be checked from eq. (\ref{f0-chi0-A}) and eq. (\ref{f0-chi0-B})
we find that for the final Kasner region $f_0 e^{-\chi_0/2} \to 0$.
Also, as the limit $\phi_h \to 0$ is approached,
 $\a$ oscillates many times, remaining  finite.
 From from eq.  (\ref{A-asintotico}) we find
 \beq
 \lim_{\phi_h \to 0} A(\infty)=
 \lim_{\phi_h \to 0}  f _0 e^{-\frac{\chi_0}{2}} (3+\a^2) =0 \, .
 \label{barbatruccone}
\eeq
So in this limit the divergence of the complexity rate at $t_b=t_c$
tends to disappear.
\end{itemize}

\begin{figure}[h]
\begin{center} 
\includegraphics[width=0.8\textwidth]{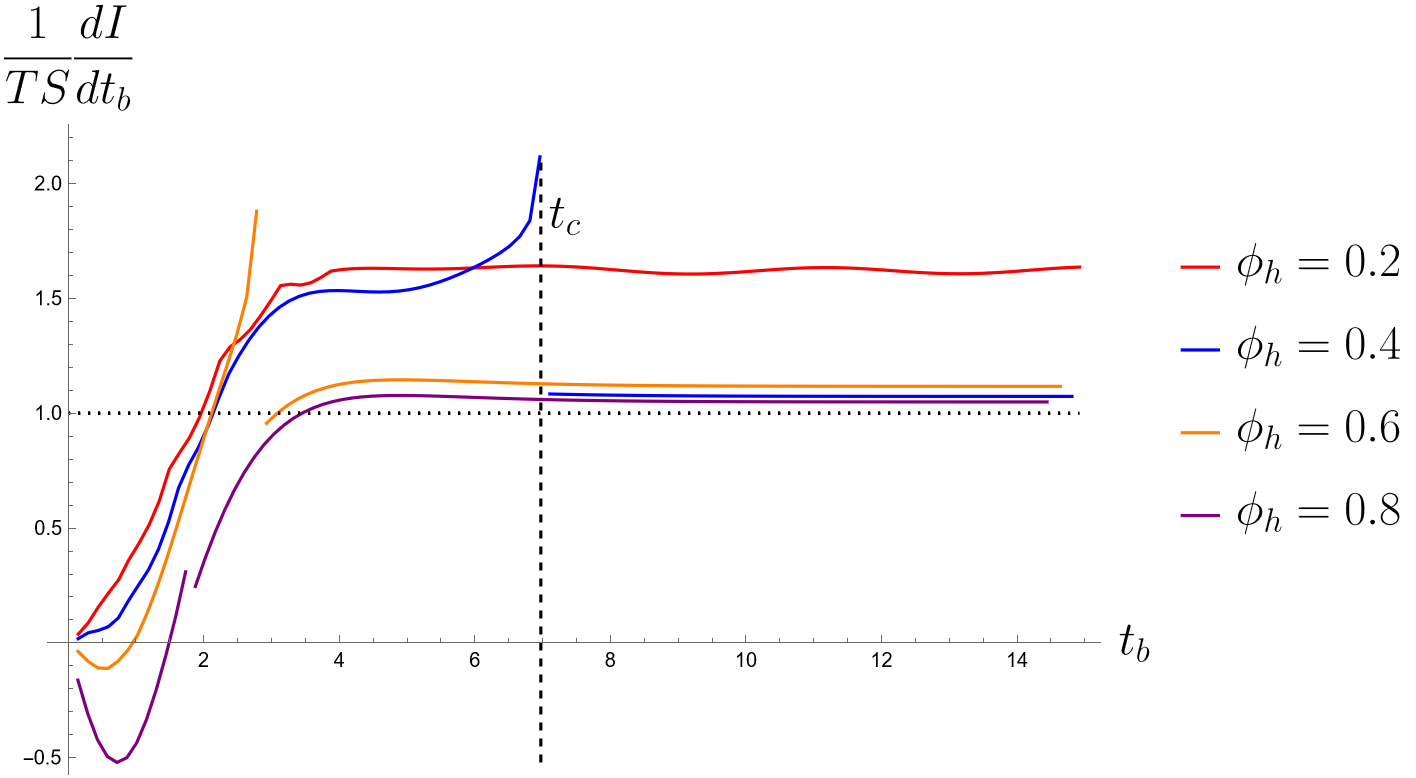}
\caption{Complexity rate for $q=1$ and $\rho=2$ as a function of $t_b$ for various values of $\phi_h$  and ${L}_{\rm ct}=1$. The divergence in the complexity rate occurs at $t_b=t_c$ and is shown here for a particular value of $\phi_h=0.4$. Type U cases.}
\label{totalactioncomplexityvsphih-carico}
\end{center}
\end{figure}

In figure \ref{totalactioncomplexityvsphih-carico}
we plot for the complexity rate in some type U examples,
with charged scalar field $q=1$.

\subsection{Asymptotic complexity rate}

In the late time limit $z_{m1} \to z_h$ and  the contribution in eq. (\ref{rateIafterJct}) is
\beq
\left. \frac{d (I_J+I_{ct})}{d t_b} \right|_{t_b \to \infty}
  = -  \frac{V_0}{8 \pi G}  \left\{ \frac{ e^{-\frac{\chi}{2}}}{2 z^2} f'
  \right\}_{z=z_h}=T S \, ,
\eeq
where $T$ ans $S$ are respectively the temperature and the entropy
\beq
T=- \frac{1}{4 \pi} f'(z_h)  e^{-\frac{\chi(z_h)}{2}} \, ,
 \qquad S=\frac{1}{4 G} \frac{V_0}{z_h^2} \ .
\eeq
We can write the asymptotic action rate including also the volume term as
\beq
W_A=\lim_{t_b \to \infty}  \frac{d I_{\rm WDW}}{d t_b } =
 T S + \frac{ V_0}{16 \pi G}  \left(
 A(\infty)
+ \int^\infty_{z_h}  \, s(z) \, dz 
 \right) \, .
 \label{rate-action-asi}
\eeq

For small $\phi_h$,  the bulk contribution  in eq. (\ref{rate-action-asi}) tends to zero.
To show this, let us separate the bulk integral in to pieces:
first, the integral of $s(z)$ up to the Cauchy horizon of the undeformed RN.
This integral is exactly zero if evaluated on the RN solution, and so it is approximately
zero for $\phi_h \to 0$:
\beq
\int_{z_h}^{z_c} s(z) \, dz = 0 \, .
\eeq
Then, there is the integral for $z>z_c$, which, using eq. (\ref{eq-s-generic-2}),
 is given by the approximation:
\beq
\int_{z_c}^{\infty} s(z) \, dz 
=\int_{z_c}^{\infty} \frac{e^{-\frac{\chi}{2}}}{z^4} \le -6 -2 \phi^2 +\frac12  \tilde{\rho}^2 z^4  \ri  dz \, .
\label{bulk-limite-piccolo-scalare}
\eeq
This integral is converging because at large $z$
\beq
s(z) \approx  \frac12 e^{-\frac{\chi_0}{2}} \frac{-12+z^4 \rho^2 - 8 \a^2 (\log z)^2 }{z^{4+\a^2}} 
\eeq
and $\a^2>1$ for the final Kasner region.
Moreover, for $\phi_h \to 0$ we have that $\chi$ tends to $\infty$ very fast for $z>z_c$,
and so the integral (\ref{bulk-limite-piccolo-scalare}) tends to zero.

\begin{figure}[h]
\begin{center}
\includegraphics[width=0.45\textwidth]{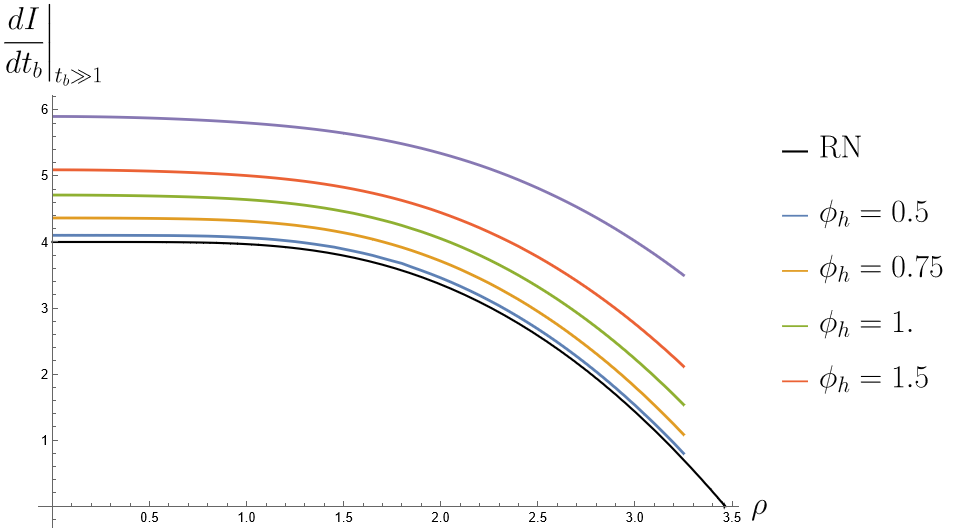}
\qquad
\includegraphics[width=0.45\textwidth]{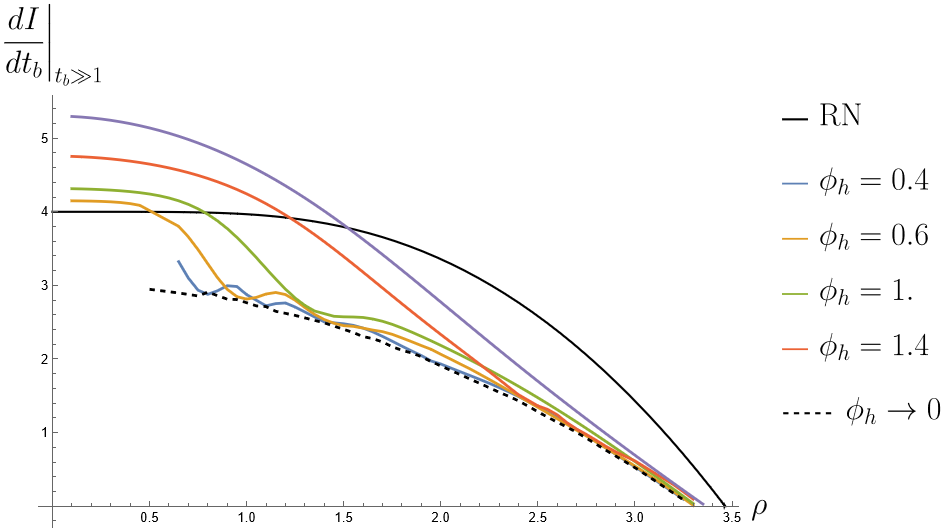}
\caption{ 
Asymptotic complexity rate  $W_A$  as function of $\rho$ for a few values
of $\phi_h$, with $16 \pi G=1$, $V_0=1$ and $z_h=1$.
 For comparison, in black we show $W_A$ for the RN black hole.
Left:  neutral scalar case. In this case $W_A$ approaches the RN limit for $\phi_h \to 0$.
Right: charged scalar case with $q=1$. In this case $W_A$ approaches $TS$, which is shown
in the dashed line, in the $\phi_h \to 0$ limit. }
\label{check-formula-analitica}
\end{center}
\end{figure}

We give now an analytic expression for the asymptotic complexity rate
in the $\phi_h \to 0$ limit. We distinguish two case:
\begin{itemize}
\item
For $q=0$ and in  the $\phi_h \to 0$ limit,
 we have that the contribution due to $A(\infty)$
reproduces the contribution of the second joint $z_{m2}$ for the RN, see eq. (\ref{A-small-phih}).
So, for $q=0$ and in  the $\phi_h \to 0$ limit, we  recover the asymptotic complexity rate
of the RN solution eq. (\ref{action-rate-late-time-RN}), i.e.
\beq
W_A = W_A^{(RN)}=T \, S \, - T_c \, S_c 
 \, ,
 \label{action-rate-neutral-scalar-case}
\eeq
\item
For $q \neq 0$ and $\phi_h \to 0$,   from eq. (\ref{barbatruccone})
that the contribution due to $A(\infty)$ vanishes.
In this case, we find
\beq
W_A= T S \, .
\label{action-rate-charged-scalar case}
\eeq
\end{itemize}
For generic  $\phi_h$, the asymptotic complexity rate can be computed numerically
using eq. (\ref{rate-action-asi}). 
We numerically checked that in the $\phi_h \to 0$ limit the analytic expressions
(\ref{action-rate-neutral-scalar-case}) and (\ref{action-rate-charged-scalar case})
are reproduced, see figure \ref{check-formula-analitica} for some illustrative plots.

\begin{figure}[h]
\begin{center}
\includegraphics[scale=0.3]{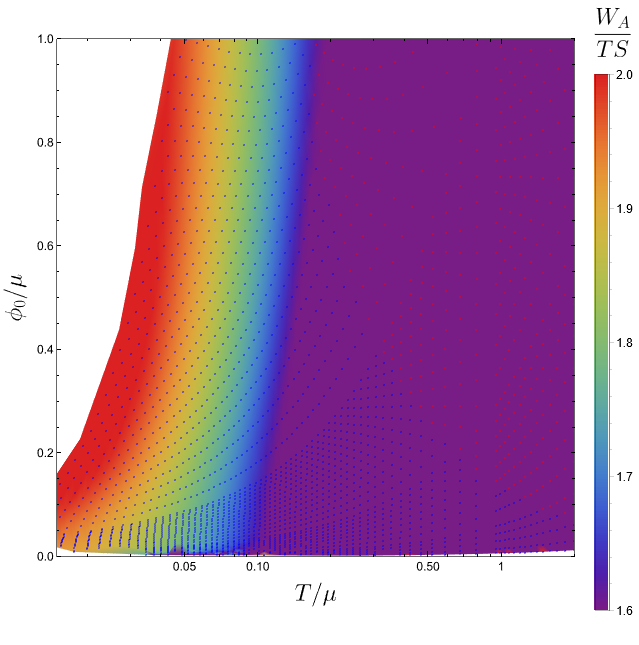}
\qquad
\includegraphics[scale=0.3]{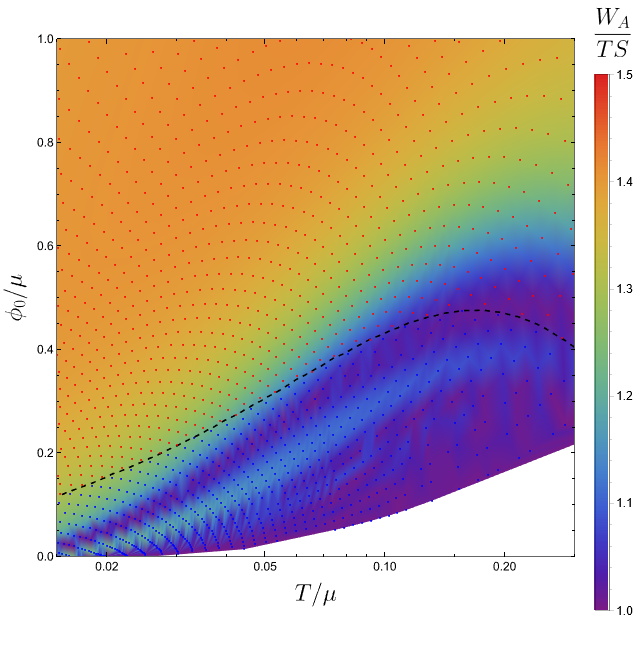}
\caption{ Left: neutral scalar case $q=0$.
Right: charged scalar case with $q=1$.
Red points denote type D solutions, black points denote type U solutions. 
The white region is not covered by our numerics. }
\label{azione-asintotica-parameter-space}
\end{center}
\end{figure}

\begin{figure}
\begin{center}
\begin{tabular}{cc}
\includegraphics[width=0.45\textwidth]{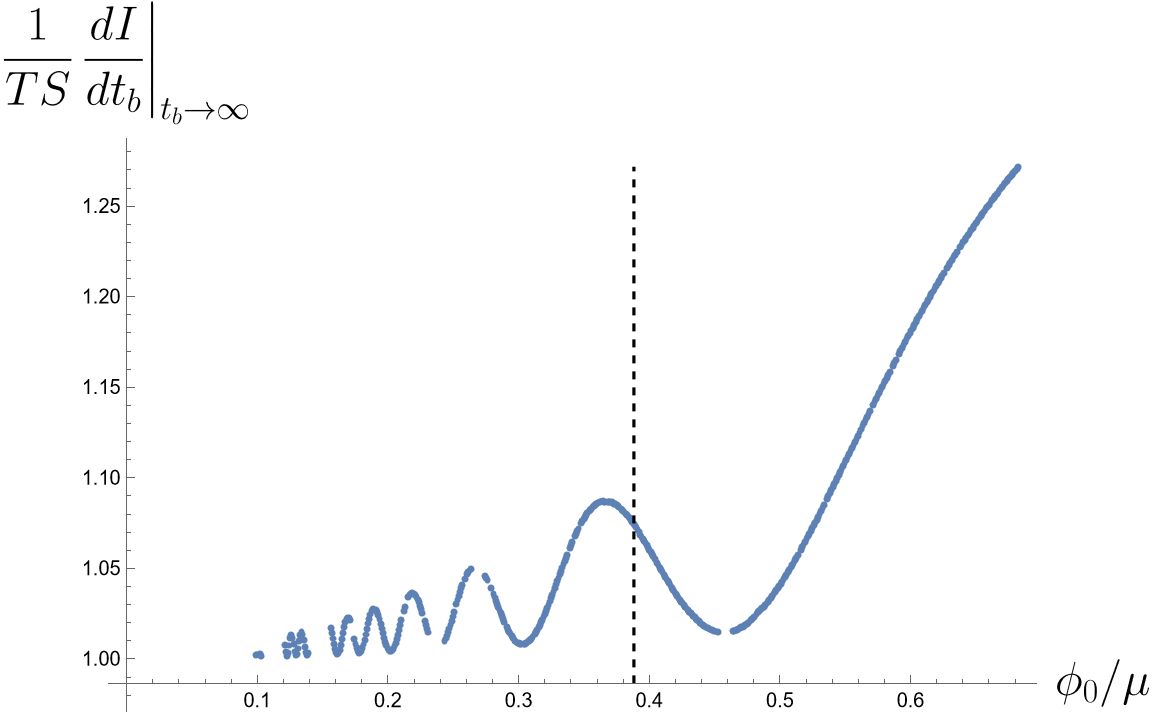}
&
\includegraphics[width=0.51\textwidth]{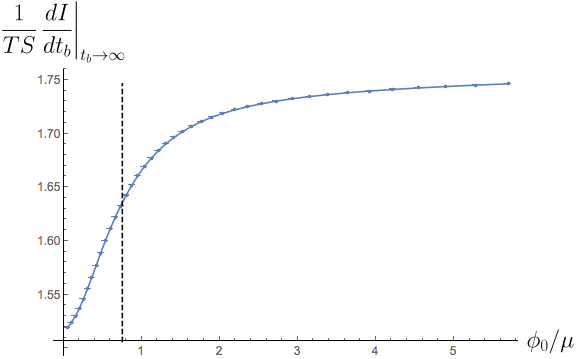}
\end{tabular}
\caption{ 
Left: Asymptotic action complexity for charged solutions with $q=1$ at constant $T/\mu=0.15$. The dashed line corresponds to the Temperature at which the solution changes type. Right: Asymptotic action complexity for uncharged solution at $T/\mu = 0.25$.
}
\label{asymptoticactionatconstT}
\end{center}
\end{figure}

We show the result 
of a scan in the parameter space of the model in
figure \ref{azione-asintotica-parameter-space}.
This confirms the expectation that the order of magnitude
of $W_A$ is $T S$. 
In figure \ref{asymptoticactionatconstT} we show 
the dependence of $W_A$ on $\phi_0/\mu$ at constant 
$T/\mu$. For $T/\mu=0.15$, we see  that $W_A$
has an interesting oscillatory behavior  as a function of $\phi_0/\mu$.
The amplitude of these oscillations vanishes for $\phi_0 \to 0$.
%We see from this figure there are interesting oscillations in $W_A$ for $\phi_0 \to 0$.
It is tempting to associate these oscillations to the Josephson oscillations of the scalar field,
 since these do not appear in the neutral scalar case, where the solution profile of the scalar field has no oscillations.

It has been conjectured   \cite{Brown:2015lvg} that the Lloyd bound is 
saturated by the uncharged planar BH in AdS$_{d+1}$, i.e.
\beq
\frac{d \, I_{WDW}}{d t_b } \leq {2 M} \, ,
\label{lloyd-bound-CA}
\eeq
where $M$ is the mass of the BH. 
This bound in eq. (\ref{lloyd-bound-CA}) is violated by the action conjecture
\cite{Carmi:2017jqz,Yang:2017czx}, because the asymptotic value (which is supposed to saturate
the bound) is approached from above. 
In the examples studied in this paper, we have that 
for type $U$ black holes the complexity rate approaches infinity at critical time,
and so this also provides another example of violation of the bound. 

\begin{figure}[h]
\begin{center}
\includegraphics[scale=0.25]{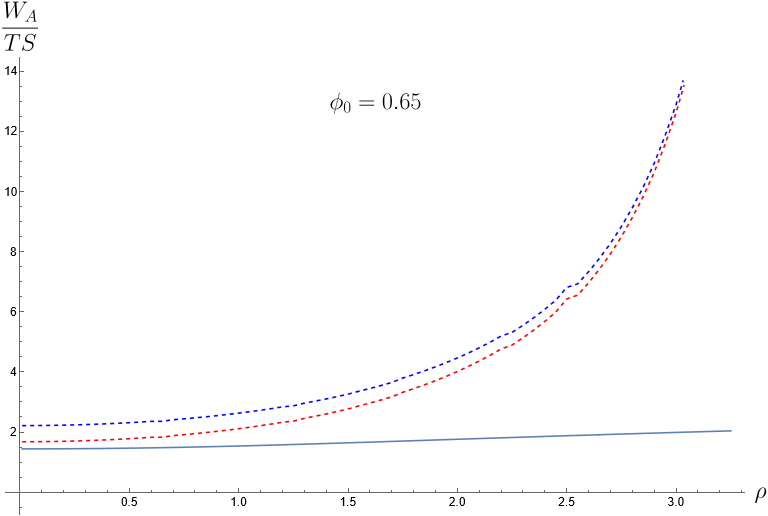}
\qquad
\includegraphics[scale=0.25]{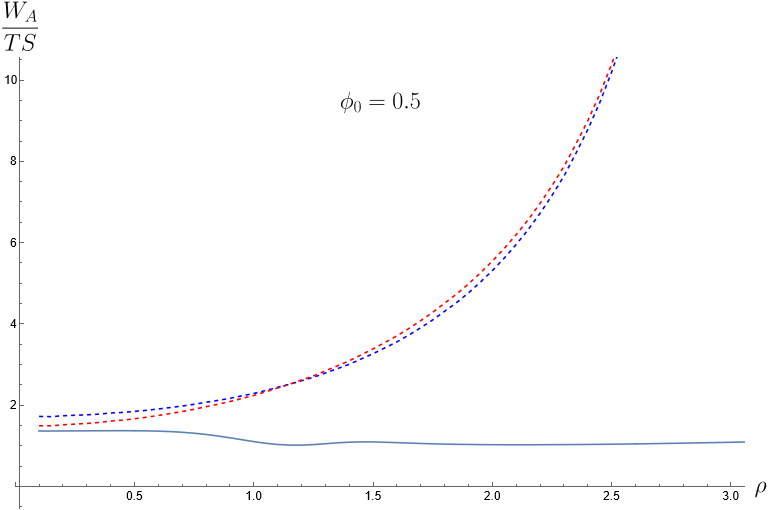}
\caption{ 
In these pictures,   we show $W_A/(T S)$ in solid line, $(2 M_{D})/(TC)$ in dashed blue line
and $(2 M_{N})/(TS)$ in dashed red line 
(see eqs. (\ref{rate-action-asi}) and (\ref{lloyd-azione-asintotico}))
as a function of $\rho$ for a fixed $\phi_0$ and $z_h=1$.
Left: neutral scalar. Right: charged $q=1$ scalar.
  }
\label{lloyd-action}
\end{center}
\end{figure}

One may wonder if a weaker version of the Lloyd bound 
holds for the asymptotic complexity rate $W_A$, i.e.
\beq
W_A \leq 2 \, M \, .
\label{lloyd-azione-asintotico}
\eeq
Violations to the bound of eq. (\ref{lloyd-azione-asintotico}) have been previously found 
in \cite{Couch:2017yil,Swingle:2017zcd,An:2018xhv,Alishahiha:2018tep,Mahapatra:2018gig,Babaei-Aghbolagh:2021ast,Avila:2021zhb}.
In our model, we checked that the bound given by eq. (\ref{lloyd-azione-asintotico})
holds in all the parameter space that we explored, with both choices 
$M=M_D,M_N$ corresponding to Dirichlet and Neumann boundary conditions.
See figure \ref{lloyd-action} for some sample plot.

%%%%%%%%%%%%%%%%
%%%%%%%%%%%%

\begin{figure}[h]
\begin{center}
\includegraphics[width=0.6\textwidth]{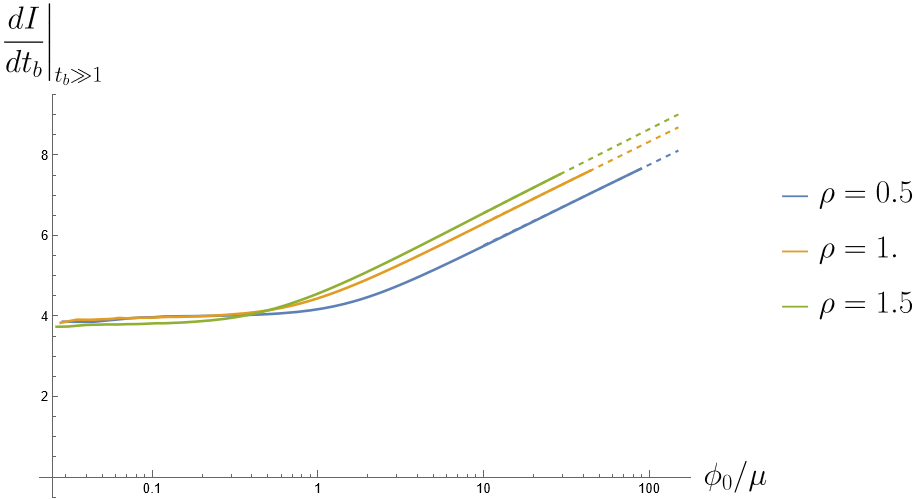}
\caption{Asymptotic value of complexity at late time as function of the scalar source, for $z_h$. 
The dashed lines are extrapolations following the law $\left.\frac{dI}{dt_b}\right|_{t_b\gg1} = {\rm const.} + 0.88_1 \log \tfrac{\phi_0}{\mu}$.}
\label{asymvssource}
\end{center}
\end{figure}

We see from figure \ref{asymvssource} that, for large $\phi_0/\mu$ the asymptotic action complexity rate has a logarithmic dependence on $\phi_0/\mu$ for fixed values of $\rho$ and $z_h$.

\acknowledgments

This work of R.A. and S.B. is supported by the INFN special research project grant  ``GAST"  (Gauge and String Theories). G.T is funded by the Fondecyt Regular grant number 1200025. The work of E.R. is partially supported by the Israeli Science Foundation Center of Excellence.

\appendix

%%%%%%%%%%%%%%%%%%%%%%%%%%%%%%%%%%%%%%%%%%%%%

\section{Conservation of electric flux in the $q=0$ case}
\label{stokes-appendix}

From Stokes theorem, we have
\beq
\int_{\Sigma} D_\b F^{\a \b} n_\a \, \sqrt{|h|} d^3 y =
\int_{\p \Sigma} F^{\a \b} \, n_{[\a} r_{\b ]} \sqrt{\sigma} d^2 \theta
\label{stokes}
\eeq
where $n^\a, r^\b$ are unit normals,
${\Sigma}$ denotes a codimension 1 submanifold,
$\p {\Sigma}$  denotes the boundary of $\Sigma$, which is codimension 2,
 and $\sqrt{|h|} d^3 y$ and $ \sqrt{\sigma} d^2 \theta$
are the induced metric measures on codimension $1$ and $2$ submanifolds.
For $q=0$ we have from Maxwell eqs. that $D_\b F^{\a \b} =0$.
Then, the right hand side of eq. (\ref{stokes}) is zero if integrated 
on a closed boundary $\p \Sigma$.
The only non-vanishing component of the gauge field strength
is 
\[
F^{t z} = a'(z) e^\chi z^4 \, ,
\]
Using the unit normals and codimension 2 measure
\[
n_\a=(1,0,0,0) \frac{\sqrt{|f|} e^{-\frac{\chi}{2}} }{z} \, , \qquad
r_\a=(0,1,0,0) \frac{1}{z \sqrt{|f|}} \, , \qquad
\sqrt{\sigma} d^2 \theta = \frac{dx \, dy}{z^2} \, ,
\]
we get
\beq
a'(z) e^{\frac{\chi}{2}} = {\rm constant} = -\rho \, .
\eeq
 
%%%%%%%%%%%%%%%%%%%%%%%%%%%%%% 
 
\section{Numerical techniques} 

\label{numerical-appendix}

For the neutral scalar case, we solve the system in eq. (\ref{eq-diff-carica-zero}).
 Introducing both an IR and a UV cutoff, $z_{\rm max}$ and $z_\epsilon$ respectively, 
 we then choose a value $z_0 > z_h$ and split the full domain in two parts, $(z_\epsilon, z_0)$ 
 and $(z_0, z_{\rm max})$. We solve the problem in the small-$z$ regime imposing at the horizon
\begin{align}
f(z_h)&=0\,, & \qquad & \chi(z_h)=\chi_h\,,\nl
%a(z_h)&=0\,, & \qquad & a'(z_h)=a_1\,,\\
\phi(z_h)&=\phi_h\,, & \qquad & \phi'(z_h)= -\frac{8 \phi _h}{z_h \left(\rho ^2 z_h^4-4 \phi _h^2-12\right)}  \,.
\end{align}
The parameter $\chi_h$ is not physical. Once a solution is found,
 it can be removed by a reparameterization of the time coordinate $t$ at infinity.
 We can define
\begin{equation}
\hat{\chi}_h=\chi_h-\chi(0) \, . % \qquad \hat{a}_1=e^{\chi(0)/2 } a_1 \,.
\end{equation}
Then, solving again with an initial condition $\chi(z_h)=\hat{\chi}_h $,
we can find the solution with the usual normalization of the time coordinate, where $\chi(0)=0$.

To determine the solution for the charged scalar case we take the full system of equations \eqref{full-system-eom}.
 We again split the full domain in two parts, $(z_\epsilon, z_0)$ 
 and $(z_0, z_{\rm max})$. We solve the system in the small-$z$ regime imposing at the horizon
\begin{align}
f(z_h)&=0\,, & \qquad & \chi(z_h)=\chi_h\,,\nl
a(z_h)&=0\,, & \qquad & a'(z_h)=a_1\,,\nl
\phi(z_h)&=\phi_h\,, & \qquad & \phi'(z_h)=\frac{8 \phi _h}{z_h \left( - a_1^2 e^{\chi _h} z_h^4+4 \phi _h^2+12\right)}\,.
\end{align}
 We can then perform a reparameterization of the time coordinate $t$ at infinity,
\begin{equation}
\hat{\chi}_h=\chi_h-\chi(0) \, , \qquad \hat{a}_1=e^{\chi(0)/2 } a_1 \,.
\end{equation}
and solve again with the initial condition
\begin{equation}
\chi(z_h)=\hat{\chi}_h \, , \qquad  a'(z_h)=\hat{a}_1 \, .
\end{equation}
This procedure fixes the normalization of the time coordinate with $\chi(0)=0$.

The value of $z_0$ is dynamically chosen such that $f(z_0) < 0$ with $|f(z_0)| \approx \mathcal{O}(1)$. 
Then, for the large-$z$ regime we rewrite our equations in terms of $\log \left|f\right|$ and compactify
 the $z$ coordinate letting $z \mapsto \tfrac{2}{\pi}\arctan \log z$. 
 This puts the boundary at the finite value of $z=1$. The boundary conditions are set at $z = z_0$
  and used to impose continuity and smoothness of the solutions.

The procedure described above was implemented in Wolfram \emph{Mathematica 13} using the {\tt NDSolve} framework\footnote{For best results, it is convenient to use the stiffness-switching method when solving for the small-$z$ region.}. In practice, we set $z_h = \chi_h = 1$, $z_\epsilon = 10^{-10}$ and $\tfrac{2}{\pi}\arctan z_{\rm max} \in [0.95, 0.99]$ depending on the physical parameters, leading to $z_{\rm max}/z_h \approx 10^5$ to $10^{27}$. Once numerical solutions have been found all relevant physical quantities can be computed straightforwardly. Those defined at $z = 0$ are in practice evaluated at $z_\delta = 10^{-5}$ to avoid boundary effects near $z = z_\epsilon$. All numerical integrations are perfomed using a standard locally adaptive method in the same variables as the solutions are found, so that $z$ is compactified (not compactified) for $z > z_0$ ($z<z_0$). Furthermore, for the computation of action complexities, which involves integrals that are divergent near $z = z_h$, we introduce another cutoff at $|z-z_h| = 10^{-6}$.

The value of $\alpha$ coming into the definition of the Kasner exponent \eqref{kasner-approx} is found by fitting the functional forms \eqref{kasner-approx} to the solutions at the regions of interest, \emph{i.e.} $z \approx z_{\rm max}$ and possibly an intermediate regime $z_h \ll z_{\rm int} \ll z_{\rm max}$ before a Kasner inversion, when present. Error bars are computed from these fits and also the variation of results obtained using the functions $f$, $\phi$ and $\chi$.

%%%%%%%%%%%%%%%%%%%%%%%%%%%%%%%%%%%

\section{Penrose diagrams}

\label{penrose-appendix}

Let us start from the metric (\ref{metric1}) and let us pass to lightlike coordinates
$u$, $v$, as defined in eq. (\ref{coordinate-cono-luce-u-v}).
We  now introduce the   lightcone coordinates $(U,V)$.
Let us define
\beq
\kappa =-f'(z_h) e^{-\frac{\chi(z_h)}{2}} >0 \, ,
\eeq
(remember that $f'(z_h)<0$).
The quantity $\kappa$ is proportional to the temperature of the BH.
We pass to the coordinates
\beq
U=\exp \le -\frac{u}{2}  \kappa \ri \, , \qquad
V=\exp \le  \frac{v}{2}  \kappa \ri \, .
\eeq
These coordinates are related to the original ones $(z,t)$ as follows
\beq
U V = e^{- \kappa \, z^* (z)} \, , \qquad \frac{V}{U} = e^{\kappa \, t } \, .
\eeq
The metric in the $U$, $V$ coordinates is
\beq
ds^2=  \frac{1}{z^2} \le f e^{-\chi} \frac{4}{\kappa^2 }   \exp ( \kappa \, z^*)  d U \, d V +dx^2 +dy^2 \ri \, .
\eeq
Nearby $z \to z_h $, we have $f \approx f'(z_h) (z-z_h) $ and $ z^*(z)$ is divergent.
The divergence can be estimated as follows
\beq
 z^*(z)  \approx  \frac{e^{\frac{\chi(z_h)}{2}}}{f'(z_h )}  \int_0^z \frac{d \tilde{z}}{\tilde{z}-z_h}  \approx
  \frac{1}{\kappa} \log \frac{1}{|z-z_h|}  \, ,
\eeq
in such a way that the metric nearby $z_h$ is
\beq
ds^2 \approx  \frac{1}{z_h^2} \le  \frac{4}{f'(z_h) }  {\rm sign} \, {  ( z-z_h) }  d U \, d V +dx^2 +dy^2 \ri \, 
\eeq
So the coordinate $(U,V)$ are also non-continuous at the horizon.
We can now introduce the smooth Kruskal coordinates $(T,R)$ in this way:
\bea
{\rm for} \, \,\,\, && z< z_h \, , \qquad T= \frac{U-V}{2}  \, , \qquad R= \frac{U+V}{2} \, ,
\nl
{\rm for} \, \,\,\,  && z> z_h \, , \qquad T= \frac{U+V}{2}  \, , \qquad R= \frac{U-V}{2} \, .
\eea
Indeed, nearby the horizon, both from outside and from the inside, we find
\beq
ds^2 \approx    \frac{1}{z_h^2} \le  - \frac{2}{f'(z_h) } \big(-dT^2+dR^2 \big)+dx^2 +dy^2 \ri \ 
\eeq

At this point we introduce
\beq
\hat{U}={R+T} \, , \qquad \hat{V}={R-T}
\eeq
in such a way that the metric nearby the horizon is
\beq
ds^2 \approx    \frac{1}{z_h^2} \le  - \frac{8}{f'(z_h) } \, d \hat{U} d \hat{V} +dx^2 +dy^2 \ri \, .
\eeq
Then we pass to compact variables:
\beq
\tilde{U}= \tan^{-1} \hat{U} \, , \qquad  \tilde{V}= \tan^{-1} \hat{V}
\label{convenzioni-penrose}
\eeq
with
\beq
- \frac{\pi}{2} <\tilde{U}, \tilde{V} < \frac{\pi}{2}
\eeq
and then we define the coordinates of the Penrose diagram
\beq
\tilde{T}=\frac{\tilde{U}-\tilde{V}}{2} \, ,
\qquad
\tilde{R}=\frac{\tilde{U}+\tilde{V}}{2} \, .
\eeq
The choice of the function $\tan^{-1}$ in eq. (\ref{convenzioni-penrose})
is a convention, and can be replaced by any other $C^\infty$
functions which maps the real line to the segment $(-\pi/2,\pi/2)$.

We finally get the expression which relate the coordinates $\tilde{T}$ and $\tilde{R}$ of the Penrose diagram
to the original coordinates $t$, $z$
\begin{itemize}
\item for $z<z_h$, then we get
\beq
\tan \le \frac{\tilde{R} +\tilde{T}}{2} \ri \tan \le \frac{\tilde{R} -\tilde{T}}{2} \ri = e^{-\kappa \, z^*} \, , \qquad
\frac{\tan \le \frac{\tilde{R} -\tilde{T}}{2} \ri}{\tan \le \frac{\tilde{R} +\tilde{T}}{2} \ri }=e^{\kappa \, t} \, ,
\label{penroA}
\eeq
which, for  $\tilde{R},\tilde{T} >0  $ should be in the quadrant $\tilde{R}>\tilde{T}$
\item for $z>z_h$, then we get
\beq
\tan \le \frac{\tilde{R} +\tilde{T}}{2} \ri \tan \le \frac{\tilde{T} -\tilde{R}}{2} \ri = e^{-\kappa \, z^*} \, , \qquad
\frac{\tan \le \frac{\tilde{T} -\tilde{R}}{2} \ri}{\tan \le \frac{\tilde{R} +\tilde{T}}{2} \ri }=e^{\kappa \, t} \, .
\label{penroB}
\eeq
which, for  $\tilde{R},\tilde{T} >0  $ should be in the quadrant $\tilde{T}>\tilde{R}$
\end{itemize}

The boundary of  AdS is realised for $z^*=0$, which is at
\beq
\tilde{R}= \pm \frac{\pi}{2} \, .
\eeq
The horizons are at $z^* \to \infty$,
\beq
\tilde{T}=\pm \tilde{R} \, . 
\eeq
The value $z^*(\infty)$ determines the "concavity" of the singularity.
If $z^*(\infty)=0$, the singularity is at $\tilde{T}=\frac{\pi}{2}$.
If $z^*(\infty)>0$, the concavity is below the line $\tilde{T}=\frac{\pi}{2}$
(type $D$).
If $z^*(\infty)<0$, the concavity is above the line $\tilde{T}=\frac{\pi}{2}$
(type $U$).

%%%%%%%%%%%%%%%%%%%%%%%%%%%%%%%%%%%%%%%%%%

\section{Holographic renormalisation}
\label{app-masses}

We take the $\phi$ expansion around the boundary $z=0$ as
in eq. (\ref{expansion-phi-boundary}).
Let us expand  the metric profile functions as follows
\beq
f=1+f_2 z^2 + f_3 z^3 + f_4 z^4 + \dots
\qquad
\chi=\chi_2 z^2+\chi_3 z^3 + \dots 
\eeq
The equations of motion fix:
\beq
\chi_2=\frac{\phi_0^2}{2} \, , \qquad \chi_3=\frac{4 \phi_0 \phi_1}{3} \, , \qquad
f_2=\frac{\phi_0^2}{2} \, .
\eeq
In Fefferman-Graham (FG) coordinates, the metric has the following form
\beq
ds^2= \frac{d \hat{z}^2}{\hat{z}^2}+ \frac{1}{\hat{z}^2} g_{ab}(\hat{z},x^a) d x^a d x^b
  \, ,
\eeq
where the boundary coordinates are
$ x^a=(t,x,y)$.
A direct calculation gives
\beq
z=\hat{z}+\frac{f_2 }{4} \hat{z}^3+\frac{f_3 }{6} \hat{z}^4
+\frac{1}{16} \left(f_2^2+2 f_4\right) \hat{z} ^5 + \dots \, ,
\eeq
The energy-momentum tensor (with Dirichlet boundary conditions)
 can be obtained from the results in \cite{Balasubramanian:1999re,deHaro:2000vlm,Caldarelli:2016nni}
\beq
T_{mn}^{(D)}=\frac{1}{8 \pi G } \,  \lim_{\hhz \rightarrow 0}
\frac{1}{\hhz} \le K_{mn} -g_{mn} K - 2 \, g_{mn} 
-\frac{1}{2} \, g_{mn}  \,  \phi^2    \ri \, .
\eeq
This gives energy density and pressure
\beq
\mathcal{E}=T_{00}^{(D)}=\frac{\phi_0 \phi_1- f_3}{8 \pi G} \, \qquad
p=T_{ii}^{(D)}=\frac{\phi_0 \phi_1-f_3/2}{8 \pi G} \, .
\eeq
which gives the Dirichlet mass
\beq
M_D=\frac{V_0}{8 \pi G}  \le - f_3 +\phi_0 \phi_1 \ri \, .
\label{mass-D}
\eeq
The energy momentum tensor with Neumann boundary condition instead is
\beq
T_{mn}^{(N)}=T_{mn}^{(D)}-\eta_{ij} \frac{\phi_0 \phi_1}{8 \pi G} \, ,
\eeq
which gives the Neumann mass
\beq
M_N=\frac{V_0}{8 \pi G}  \le - f_3 +2 \phi_0 \phi_1 \ri \, .
\label{mass-N}
\eeq
If we consider the case of a system with spontaneous symmetry breaking
(zero sources) with Dirichlet or Neumann boundary conditions,
as in \cite{Yang:2019gce},
indeed both the definitions of mass give the same result.

%%%%%%%%%%%%%%%%%%%%%%%%%%%%%%%%%%%%%%%%%%

\section{Terms in the WDW action}
\label{appterms}

We discuss here in detail the evaluation of the terms in the 
action of the WDW patch (\ref{actionfour}).

\subsection{Bulk action density $I_{V}$}

The bulk term is given by the Lagrangian of the model in eq. (\ref{lagrangiana-modello}).
Using the Einstein equations, we can express the scalar curvature as follows
\beq
R =-12-4 \phi^2 + D_\mu \phi (D^\mu \phi)^* \, ,
\eeq
Using the ansatz in eqs. (\ref{metric1}) and (\ref{ansatz1}), we find
\beq
-\frac{1}{4} F_{\mu \nu} F^{\mu \nu} = \frac{z^4 e^\chi (a')^2}{2} \, , \qquad
 D_\mu \phi (D^\mu \phi)^*=f z^2 (\phi')^2 -  q^2 \frac{z^2 e^\chi}{f} a^2 \phi^2 \, .
\eeq
The on-shell bulk action density then is
\beq
\mathcal{L}=\frac{1}{16 \pi G} \le -6+\frac{ (a')^2 z^4 e^\chi}{2} -2 \phi^2 \ri \, .
\eeq
and the bulk action is
\beq
I_V =  \int d^4x \sqrt{-g} \,  \mathcal{L} = \frac{1}{16 \pi G}  \int d^4x \, s(z) \, , \qquad
s(z)= - \frac{e^{-\frac{\chi}{2}}}{z^4} \le 6 + 2 \phi^2 \ri  +   \frac{ e^{\frac{\chi}{2}}  (a')^2  }{2}       \, ,
\label{eq-s-generic}
\eeq
Using  eq. (\ref{rho-tilda}), we find
\beq
s(z) = 
\frac{e^{-\frac{\chi}{2}}}{z^4} \le -6 -2 \phi^2 +\frac12  \tilde{\rho}^2 z^4  \ri \, .
\label{eq-s-generic-2}
\eeq

\subsection{Gibbons-Hawking-York term  $I_{\rm GHY}$}

The Gibbons-Hawking-York term is 
\beq
\label{ghyformula}
I_{\rm GHY}=\mp \frac{1}{8 \pi G} \int_\p d^3 x \sqrt{|h|} K \, ,
\eeq
where $-$ sign is chosen for space-like and $+$ for time-like boundaries.  $h_{\alpha\beta}$ is the induced metric on the manifold $\p$ as embedded in the space-time and $K$ is the extrinsic curvature.
For diagonal metric, one can use a useful formula for the extrinsic curvature 
\beq
\label{Kformula}
K= \frac{n^\mu \p_\mu \sqrt{|h|} }{\sqrt{|h|}} \, , 
\eeq
where $n^\a$ is the unit outward-pointing normal. 
The contributions that we need for the WDW patch are on portions space-like surfaces at the singularity eventually present
at constant $z \to \infty$. Thus we always use  the minus sign in (\ref{ghyformula}) and $h$ is positive.
We find:
\bea
&&h_{\alpha\beta }= {\rm diag}\left(\frac{-f e^{-\chi}}{z^2},\frac{1}{z^2} ,\frac{1}{z^2} \right) \ ,\qquad \sqrt{h} =\frac{e^{-\frac{\chi}{2}} \sqrt{-f} }{z^3} \, ,\nonumber \\
&& 
 \ n^\mu=\left(0,- z \sqrt{-f}  ,0,0\right) \, .
\eea
Using (\ref{Kformula}) we obtain 
\beq
 \sqrt{h} K=- z \sqrt{-f}
  \p_z \le \frac{e^{-\frac{\chi}{2}} \sqrt{-f} }{z^3} \ri =
\frac{e^{-\frac{\chi}{2}}}{2 z^3}\left(z f' -( 6 + z \chi') f\right) \, .
\label{termine-K-GHY-appe-C}
\eeq
It is useful to introduce
\beq
A(z)=2  \sqrt{h} K =\frac{e^{-\frac{\chi}{2}}}{ z^2} \le  f'  -  \chi' f - \frac{6 f}{z} \ri \, .
\label{A-arbitrary}
\eeq
To compute the contribution of GHY
 we need to determine
$A(\infty)=\lim_{z \to \infty } A(z) $.

\subsection{Joint terms  $I_{\rm J}$}

For joints between space-like and time-like boundary surfaces, the action contribution
was studied in \cite{Hayward:1993my}; the case of joint involving null surfaces was discussed in
\cite{Lehner:2016vdi}. The joint action is
\beq
I_{\rm J}= \frac{1}{8 \pi G} \int_{\Sigma} d \theta \sqrt{\sigma} \, \mathfrak{a} \, ,
\label{jjoints}
\eeq
where $ \sigma $ is the  determinant of induced metric on the joint
and $ \mathfrak{a} $ depends on normals in the following way.
Let us denote $k^\mu$ the future directed null normal to a null surface,
 $n^\mu$ the normal to a space-like surface 
and $s^\mu$ the normal to a time-like surface, both directed outwards the volume of interest. 
In the case of intersection of two null surfaces with normals  $k^\mu_1$ and $k^\mu_2$:
\beq
\mathfrak{a} = \eta  \log \left| \frac{k_1 \cdot k_2}{2} \right| \, ,
\label{jnn}
\eeq
while in the case of intersection of a null surface  
and a space-like surface  or a time-like surface with normal:
\beq
\mathfrak{a} = \eta  \log \left|  k \cdot n \right| \, , \qquad
\mathfrak{a} = \eta \log \left|  k \cdot s \right| \, .
\label{jns}
\eeq
The $\eta$ is a sign defined as follows.
In eqs. (\ref{jnn})-(\ref{jns}), if  the outward direction
to the region of interest is pointing along the future,
we should set $ \eta= 1 $ 
if the joint lies in the future of the spacetime volume of interest, 
and $ \eta=-1 $ if the joint lies in the past.
If instead the outward direction is pointing along the past,
we should set $ \eta= 1 $ 
if the joint lies in past of the spacetime volume of interest, 
and $ \eta=-1 $ if the joint lies in the future.
There is a subtlety for null surfaces:
the normalisation of  $k^\mu$ can not be fixed by its length and we thus need a regulator. For the cases we need in this paper
\beq
(k_1)_\mu=\Big(-1, \frac{e^{\frac{\chi}{2}} }{f},0,0 \Big) a_{\rm reg} \, ,  \qquad
(k_2)_\mu=  \Big(  1, \frac{e^{\frac{\chi}{2}} }{f} ,0,0 \Big)  a_{\rm reg} \, .
\label{null-normals}
\eeq
For this reason, the terms in  eqs.  (\ref{jnn}) and (\ref{jns}) 
contain an ambiguous coefficient $a_{\rm reg }$ inside the logarithm.
This dependence on $a_{\rm reg}$ is cancelled by the null boundary counterterm
 \cite{Carmi:2017jqz} discussed next.

\subsection{Null boundary term $I_{\rm b}$ and  counterterm  $I_{\rm ct}$}

The null boundary terms is
\beq
I_{\rm b}=\int dS \, d \l \, \sqrt{\s} \, \kappa \, .
\label{boundary-action}
\eeq
In our case, the null normals $k^\mu$ satisfy the geodesic equation
\beq
k^\mu D_\mu k^\nu = \kappa \, k^\nu \, ,
\eeq
with $\kappa=0$.
For this reason, the term in eq. (\ref{boundary-action}) is zero
 in our  parametrization.

We need also to add a counterterm \cite{Lehner:2016vdi},
 which is needed to restore parameterisation invariance
\beq
I_{\rm ct}=\frac{1}{8 \pi G} \int d \l \, dS \, \Theta \log | {L}_{\rm ct} \Theta | \, , 
\label{counter-term-action}
\eeq
where
\beq
 \Theta=\p_\l \log \sqrt{\gamma}=D_\mu k^\mu
\eeq
for each of the null boundaries with normals $k=k_1,k_2$.
For both the normals
\beq
\Theta= - 2 a_{\rm reg}  \, z \, e^{\frac{\chi}{2}} \, .
\eeq
Using geodesics equations, the integration measure in the null affine parameter $\l$ can be written as follows
\beq
d \l = \pm \frac{1}{a_{\rm reg}} \frac{dz}{z^2 e^{\frac{\chi}{2}}} \, ,
\eeq
so we get
\beq
I_{\rm ct}=\mp \frac{V_0}{4 \pi G}   \int     dz \, 
 \frac{1}{z^3}  \,  \log \left|  2 a_{\rm reg }  \,  {L}_{\rm ct} \, z \, e^{\frac{\chi}{2}}  \right|  , \, 
 \label{azione-controtermine}
\eeq
where $\pm$ depend on which of the four null boundaries we are considering.
For consistency, the dependence on $a_{\rm reg}$ should cancel with the 
joint term.
The counterterm  ${L}_{\rm ct}$
does not affect  the late-time limit of the complexity, but just
the finite-time behaviour.

%%%%%%%%%%%%%%%%%%%%%%%%%%%%%%%%%%%%%%%%%

\section{Details of the RN action complexity}

\label{appe-RN-action-complexity}

For the RN black hole the tortoise coordinate $z^*$ is given by eq.~(\ref{tortoise-zstar}), with $\chi=0$:
\beq
z_{\rm RN}^*(z)=\int_0^z \frac{d \tilde{z}}{f_{\rm RN}(\tilde{z})}  \, .
\label{z-stella-RN}
\eeq
where $f_{\rm RN}$ is in eq.~(\ref{fRN1}) or alternatively  (\ref{fRN2}).
The integral in eq. (\ref{z-stella-RN}) is divergent at the horizon $z=z_h$ 
and must be regularised using the Cauchy principal value method.

We are interested in the function $z^*_{\rm RN}(z)$  is defined in the interval $z_h \leq z \leq z_c$. 
By direct evaluation of eq. (\ref{z-stella-RN}), we can write the function $z^*(z)$
 as follows
\beq
z^*_{\rm RN}(z)=\frac{z_h z_c}{z_c-z_h} \left[ 
\frac{z_h^2}{(z_c+z_h)^2 +2 z_h^2 } \log \le 1- \frac{z}{z_c} \ri -
\frac{z_c^2}{(z_c+z_h)^2 +2 z_c^2 } \log \le  \frac{z}{z_h} -1 \ri \right] + \tilde{z}^*_{\rm RN}(z) \, ,
\label{zstar-RN-espessione-esplicita}
\eeq
where $ \tilde{z}_{\rm RN}^*(z) $ is finite in the interval $z_h \leq z \leq z_c$ and is 
{\small
\bea
 \tilde{z}^*_{\rm RN}(z) &=&\frac{z_c z_h}{2 \left(2 z_c z_h+3 z_c^2+z_h^2\right) \left(2 z_c z_h+z_c^2+3
   z_h^2\right)}
   \left\{
   \left(z_c+z_h\right)^3 \log \left(z
   \left(\frac{z+z_h}{z_h z_c}+\frac{z}{z_c^2}+\frac{z_h+z}{z_h^2}\right)+1\right)  
   \right. 
\nl
& &
\left.
-2 \, \frac{6 z_c^3 z_h+10 z_c^2 z_h^2+6 z_c z_h^3+3 z_c^4+3 z_h^4}{\sqrt{2 z_c z_h+3 z_c^2+3
   z_h^2}}
    \left[ \tan ^{-1}\left(\frac{z_c+z_h}{\sqrt{2 z_c z_h+3 z_c^2+3 z_h^2}}\right) 
   \right. \right. 
\nl
& & \left. \left.   
    -\tan^{-1}\left(\frac{z_c z_h \left(z_c+z_h\right)+2 z \left(z_c
   z_h+z_c^2+z_h^2\right)}{z_c z_h \sqrt{2 z_c z_h+3 z_c^2+3 z_h^2}}\right) \right]
\right\}
\eea
}
For $z \to z_h$, the function $z^*_{\rm RN}$ diverges to $+ \infty$, while for
 $z \to z_c$ it diverges to $- \infty$.

Here it is important to know first the coordinate of the two joints inside the horizons
 as a function of the boundary time $t_b$. Let us denote respectively by $z_{m1}$
 and $z_{m2}$ the coordinates of the joints inside the white and the black hole horizons.
They can be found by solving the equations
\beq
z^*_{\rm RN}(z_{m1})= \frac{t_b}{2} \, , \qquad
z^*_{\rm RN}(z_{m2})= -\frac{t_b}{2} \, .
\label{tipsforrn}
\eeq
At late time, $z_{m1}=z_h$ and $z_{m2}=z_c$.
The WDW patch is always of a diamond form \dia, and never reaches the singularity.
 Its boundaries are only null-like so there is no GHY term in the action. 
The rate of increase of the WDW action is then given by
\beq
\frac{d I_{\rm WDW}}{d t_b } =\frac{d I_V}{d t_b} +\frac{d (I_J+ I_{\rm ct})}{d t_b} \, ,
\label{action-rate-RN}
\eeq
where
%from (\ref{eq-s-carica-zero}),
\beq
 \frac{ d \, I_V}{d t_b}= \frac{V_0}{16 \pi G_N} \int^{z_{m2}}_{z_{m1}} s_{\rm RN}(z) dz \, ,
 \qquad
 s_{\rm RN}=\frac{1}{z^4} \le -6+\frac12  \rho^2 z^4   \ri  \, ,
   \label{RN-Bulk}
\eeq
and
\bea
\frac{d( I_J+I_{\rm ct})}{d t_b } &=&
\frac{V_0}{16 \pi G} \left(
 \left\{ f_{\rm RN}  \left[ \frac{2}{z^3} 
  \log \left(   -4 f_{\rm RN}    \,  {L}_{\rm ct}^2  
  \right)
 -\frac{1}{z^2}   \frac{d}{d z} 
 \left( \log \frac{-f_{\rm RN} }{z^2 } \right)
  \right]
  \right\}_{z=z_{m1}} 
  \right. \nl
%%%%
&& - \left. 
 \left\{ f_{\rm RN}  \left[ \frac{2}{z^3} 
  \log \left(   -4 f_{\rm RN}    \,  {L}_{\rm ct}^2  
  \right)
 -\frac{1}{z^2}   \frac{d}{d z} 
 \left( \log \frac{-f_{\rm RN}}{z^2 } \right)
  \right]
  \right\}_{z=z_{m2}}  \right) \, .
  \label{RN-Joint-and-Counterterm}
\eea

\subsection{Extremal limit}

\label{appe-RN-action-complexity-extremal}

In the extremal limit $z_c \to z_h$, we can find some compact
 approximate expressions. In this limit, the contribution of the function $ \tilde{z}^* (z) $ is negligible.
 We can approximate $z^*$ as follows:
 \beq
 z^*_{\rm RN}=\frac{z_h^2}{6 (z_c-z_h)} \log \frac{z_c-z}{z-z_h} \, .
 \eeq
 So we can find the analytical expressions
 \beq
 z_{m1}=
 \frac{z_h e^{\frac{3 t_b \left(z_c-z_h\right)}{z_h^2}}+z_c}{e^{\frac{3 t_b \left(z_c-z_h\right)}{z_h^2}}+1} \, , \qquad
 z_{m2}=\frac{z_c e^{\frac{3 t_b \left(z_c-z_h\right)}{z_h^2}}+z_h}{e^{\frac{3 t_b \left(z_c-z_h\right)}{z_h^2}}+1} \, .
 \eeq
 Plugging in the expressions for the action rate, we find eq. (\ref{RN-neraby-extremality-rate-CA}). 
Note that in this limit the dependence on ${L}_{\rm ct}$ tends to disappear.  See left of figure \ref{action-rate-RN-1}.

\subsection{Schwarzschild  limit}
\label{appe-RN-action-complexity-Sch}

In the large $z_c$ limit with fixed $z_h$, eq. (\ref{action-rate-RN})
must reproduce the case of a black brane with zero charge.
Here we will  reproduce the result  that the rate of CA is almost zero for $t_b < \hat{t}_c$ in eq. (\ref{tempo-critico-RN})
from the $z_c \to \infty$ limit of the RN expression.
In this approximation, we find:
\bea
z^*_{RN}(z) &=&z_h \left[ 
\frac{z_h^2}{z_c^2 } \log \le 1- \frac{z}{z_c} \ri -
\frac{1}{3}  \log \le  \frac{z}{z_h} -1 \ri \right] + \tilde{z}^*(z) \, ,
\nl
 \tilde{z}^*(z) & = &\frac{ z_h}{6 }
   \left\{
    \log \left(1+   \frac{z}{z_h^2} (z_h+z ) \right) 
-2 \, \sqrt{3} 
    \left[ \frac{\pi}{6}
    -\tan^{-1}\left(\frac{ z_h +2 z   }
   { z_h \sqrt{3}}\right) \right]
\right\}
\eea

In the limit of small time, one can write explicit expressions for $z_{m1}$ and $z_{m2}$.
In this limit, both $z_{m1}$ and $z_{m2}$ are nearby $z_c$.
Nearby $z_c$,  we can approximate $z^*$ as follows:
\bea
z^*_{RN}(z) &=&z_h \left\{
\frac{z_h^2}{z_c^2 } \log \le 1- \frac{z}{z_c} \ri -
\frac{1}{3}  \log \le  \frac{z_c}{z_h} -1 \ri
+ \frac{1}{6}    \log \left(1+   \frac{z_c}{z_h^2} (z_h+z_c ) \right) 
\right.   \, ,
\nl
&& \left. -\frac{1}{\sqrt{3}} 
    \left[ \frac{\pi}{6}
    -\tan^{-1}\left(\frac{ z_h +2 z_c   }
   { z_h \sqrt{3}}\right) \right]  \right\}  \approx
  \nl
 &\approx& 
\frac{z_h^3}{z_c^2 } \log \le 1- \frac{z}{z_c} \ri  + \frac{\pi}{3 \sqrt{3}} z_h +
\frac{z_h^3}{2 z_c^2} \, .
\label{zstar-small-time}
\eea
This solution then is valid up to the time given by the plateau of the function in eq. (\ref{zstar-small-time}), i.e.
\beq
\hat{t}_c=\frac{2 \pi}{3 \sqrt{3}} z_h +
\frac{z_h^3}{ z_c^2} \, ,
\eeq
which indeed reproduces the result in eq. (\ref{tempo-critico-RN}).

We can then solve for $z_{m1}$ and $z_{m2}$:
\beq
z_{m1}=
z_c \left(1 - e^{ \frac{t_b z_c^2}{2 z_h^3}-\frac{\pi  z_c^2}{3 \sqrt{3} z_h^2}-\frac{1}{2}  } \right) \, , \qquad
z_{m2}=\left(1 - e^{-\frac{t_b z_c^2}{2 z_h^3}-\frac{\pi  z_c^2}{3 \sqrt{3} z_h^2}-\frac{1}{2}} \right) \approx z_c \, .
\eeq
At very early time, both $z_{m1} \approx z_{m2} \approx z_c$. 
At time of order $\hat{t}_c$, $z_{m1}$ drops rather suddenly, while $z_{m2}$ remains approximately constant.
The complexity rate  then is
\beq
 \frac{ d \, I }{d t_b} \approx  \frac{V_0}{16 \pi G_N} \frac{1}{z_h^3} \, w \, e^{-w/2} \, , \qquad
  w=\frac{2 \sqrt{3} \pi}{9} \frac{z_c^2}{z_h^2} +1 - \frac{z_c ^2}{ z_h^3} t_b \, .
\eeq
The complexity rate is then exponentially suppressed for $t < \hat{t}_c$.

%%%%%%%%%%%%%%%%%%%%%%%%%%%%%%%%%%%%%%%%%%

\section{Details of action calculation}

\label{appe-dettagli-azione}

\subsection{Type $D$ before $t_c$}

\label{appe-dettagli-azione-D-presto}

{\bf Bulk term:}
Let us split the integral in the three regions $1,2,3$,
see fig. \ref{fig-caso1} on the left:
\bea
I_V^1 &=& V_0 \int^\infty_{z_h} dz \int_0^{\frac{t_b}{2}+z^*} dt \, s(z)=
V_0 \int^\infty_{z_h} dz \left[ \frac{t_b}{2}+z^*(z) \right] s(z) \, ,
\nl
I_V^2 &=& V_0 \int^{z_h}_{\epsilon} dz \int_{\frac{t_b}{2}-z^*}^{\frac{t_b}{2}+z^*} dt \, s(z)=
V_0 \int^{z_h}_{\epsilon} dz [ 2 z^*(z) ] s(z) \, ,
\nl
I_V^3 &=& V_0 \int^\infty_{z_h} dz \int_{\frac{t_b}{2}-z^*}^0 dt \, s(z)=
V_0 \int^\infty_{z_h} dz \left[ -\frac{t_b}{2}+z^*(z) \right] s(z) \, .
\label{bulk-D-case-before-tc}
\eea
The sum of the three terms is time-independent.

{\bf GHY term:} From eq. (\ref{termine-K-GHY-appe-C}),
 there are two contributions, respectively nearby the future and past singularity,
that we cutoff at $z=z_{\max}$
\bea
I_{\rm GHY} ^1 &=& 2 V_0  \sqrt{|h|}  K(z_{\max}) \le z^*(z_{\max}) +\frac{t_b}{2} \ri
\, , \nl
I_{\rm GHY} ^2 &=& 2 V_0   \sqrt{|h|} K(z_{\max}) \le z^*(z_{\max}) -\frac{t_b}{2} \ri
  \, .
\eea
The total is time-independent.

{\bf  Joint and counterterm contributions:} These contributions are time-independent.

%%%%%%

\subsection{Type $U$ before $t_c$}

\label{appe-dettagli-azione-U-presto}

Here it is important to know first the coordinate of the two joints inside the horizons
 as a function of the boundary time $t_b$. Let us denote respectively by $z_{m1}$
 and $z_{m2}$ the coordinates of the joints inside the white and the black hole horizons.
They can be found by solving the equations
\beq
z^*(z_{m1})= \frac{t_b}{2} \, , \qquad
z^*(z_{m2})= -\frac{t_b}{2} \, .
\label{zstella-U-before-tc}
\eeq
The time derivative gives
\beq
\frac{d z_{m1}}{d t_b}=\frac12 \frac{1}{\frac{d z^*}{d z_{m1}}} = \frac{f(z_{m1}) e^{-\chi(z_{m1})/2}}{2} \, ,
\qquad
\frac{d z_{m2}}{d t_b}=- \frac{f(z_{m2}) e^{-\frac{\chi(z_{m2})}{2}}}{2} \, .
\label{time-derivative-zm-12}
\eeq

{\bf Bulk term:}
Let us split the integral in the three regions $1,2,3$,
see figure \ref{fig-caso2} on the left:
\bea
I_V^1 &=& V_0 \int^{z_{m2}}_{z_h} dz \int_0^{\frac{t_b}{2}+z^*} dt \, s(z)=
V_0 \int^{z_{m2}}_{z_h} dz \left[ \frac{t_b}{2}+z^*(z) \right] s(z) \, ,
\nl
I_V^2 &=& V_0 \int^{z_h}_{\epsilon} dz \int_{\frac{t_b}{2}-z^*}^{\frac{t_b}{2}+z^*} dt \, s(z)=
V_0 \int^{z_h}_{\epsilon} dz [ 2 z^*(z) ] s(z) \, ,
\nl
I_V^3 &=& V_0 \int^{z_{m1}}_{z_h} dz \int_{\frac{t_b}{2}-z^*}^0 dt \, s(z)=
V_0 \int^{z_{m1}}_{z_h} dz \left[ -\frac{t_b}{2}+z^*(z) \right] s(z) \, .
\eea
The time derivative of the sum of these three terms 
(including a factor of $2$) gives
\beq
\frac{1}{V_0} \frac{ d \,  I_V}{d t_b}
= \int^{z_{m2}}_{z_{m1}} s(z) dz 
 +\frac{d z_{m1}}{d t_b}
 [-t_b  +2 z^* (z_{m1})]  \, s(z_{m1})
+\frac{d z_{m2}}{d t_b}
[ t_b +2 z^* (z_{m2})] \, s(z_{m2})  \, .
\eeq
From eq. (\ref{zstella-U-before-tc}), we finally find
\beq
\frac{1}{V_0} \frac{ d \, \Delta I_V}{d t_b}= \int^{z_{m2}}_{z_{m1}} s(z) dz \, .
\eeq

{\bf Joint contributions:}
The joint contributions come from eq.  (\ref{jjoints}).
 Both the future and the past joint have $\eta=1$ and they give
the following contribution 
\beq
I_J = - \frac{V_0}{8 \pi G} \frac{1}{z_{m1}^2} \log \left| \frac{f e^{-\chi}}{z^2 a_{\rm reg}^2} \right|_{z=z_{m1}}
- \frac{V_0}{8 \pi G} \frac{1}{z_{m2}^2} \log \left| \frac{f e^{-\chi}}{z^2 a_{\rm reg}^2} \right|_{z=z_{m2}} \, ,
\eeq
The time derivative gives
\[
\frac{d I_J}{d t_b}
=\frac{V_0}{8 \pi G}
 \left\{
 \left[ \frac{f e^{-\frac{\chi}{2}}}{2}
 \frac{d}{d z} \left( \frac{1}{z^2} \log \frac{-f e^{-\chi}}{z^2 a_{\rm reg}^2} \right)\right]_{z=z_{m2}}
-   \left[ \frac{f e^{-\frac{\chi}{2}}}{2}
 \frac{d}{d z} \left( \frac{1}{z^2} \log \frac{-f e^{-\chi}}{z^2 a_{\rm reg}^2} \right) \right]_{z=z_{m1}}
  \right\} \, .
\]

{\bf Null boundaries counterterms:} From eq. (\ref{azione-controtermine}), including a factor of $2$
\[
\frac{d I_{ct}}{d t_b }  =
  \frac{V_0}{4 \pi G} \left( \left\{  f e^{-\frac{\chi}{2}}
 \frac{1}{z^3}  \,  \log \left(  2 a_{\rm reg }  \,  {L}_{\rm ct} \, z \, e^{\frac{\chi}{2}}  \right)
 \right\}_{z=z_{m1}} 
-   \left\{  f e^{-\frac{\chi}{2}}
 \frac{1}{z^3}  \,  \log \left(  2 a_{\rm reg }  \,  {L}_{\rm ct} \, z \, e^{\frac{\chi}{2}}  \right)
 \right\}_{z=z_{m2}} 
\right) \, .
\]
Combining the  counterterm with the joint term, we get
\bea
\frac{d( I_J+I_{ct})}{d t_b } &=&
\frac{V_0}{16 \pi G} \left(
 \left\{ f e^{-\frac{\chi}{2}} \left[ \frac{2}{z^3} 
  \log \left(   -4 f    \,  {L}_{\rm ct}^2  
  \right)
 -\frac{1}{z^2}   \frac{d}{d z} 
 \left( \log \frac{-f e^{-\chi}}{z^2 } \right)
  \right]
  \right\}_{z=z_{m1}} 
  \right. \nl
%%%%
&& - \left. 
 \left\{ f e^{-\frac{\chi}{2}} \left[ \frac{2}{z^3} 
  \log \left(   -4 f    \,  {L}_{\rm ct}^2  
  \right)
 -\frac{1}{z^2}   \frac{d}{d z} 
 \left( \log \frac{-f e^{-\chi}}{z^2 } \right)
  \right]
  \right\}_{z=z_{m2}}  \right) \, ,
\eea
where the dependence on $a_{\rm reg}$ 
simplifies, due to reparametrization invariance.

%%%%%%%

\subsection{After $t_c$}

\label{appe-dettagli-azione-D-U-tardi}

After $t_c$ the expressions for case $U$
are the same as for case $D$, with
$z_m=z_{m1}$, see figures  \ref{fig-caso1} 
and  \ref{fig-caso2} on the right.
To find the time dependence of action, it is important to evaluate
the value of $z_m$ as a function of the boundary time;
this can be found by solving the equation
\beq
z^*(z_m)= \frac{t_b}{2} \, .
\label{zstar-after-tc-appe}
\eeq
Taking time derivative 
\beq
\frac{d z_m}{d t_b}=\frac12 \frac{1}{\frac{d z^*}{d z_m}} = \frac{f(z_m) e^{-\chi(z_m)/2}}{2} \, .
\label{time-derivative-zm}
\eeq

{\bf Bulk term:} The first two terms $I_V^{1,2}$ remain the same as in eq. (\ref{bulk-D-case-before-tc}).
The third term is:
\beq
I_V^3=V_0 \int^{z_m}_{z_h} dz \int_{\frac{t_b}{2}-z^*}^0 dt \, s(z)=
V_0 \int^{z_m}_{z_h} dz \left[ -\frac{t_b}{2}+z^*(z) \right] s(z) \, .
\eeq
where $z_m$ is the coordinate of the lower joint of the WDW patch
(which is a function of $t_b$).
The time derivative of the total bulk action is
\beq
\frac{1}{V_0} \frac{d \, \Delta I_V }{d t_b}=
  \int^\infty_{z_m} dz  \, s(z) +
 \frac{d \, z_m}{ d t_b} \left(2 z^*(z_m)- t_b  \right) s(z_m)
 =  \int^\infty_{z_m} dz  \, s(z)  \, ,
\eeq
where we used eq. (\ref{zstar-after-tc-appe}).

{\bf GHY term:} There is just one contribution,  nearby the future singularity ,
that we cutoff at $z=z_{\max}$
\beq
I_{\rm GHY}^1 = 2 V_0   \sqrt{|h|} K(z_{\max}) (z^*(z_{\max}) +\frac{t_b}{2}) 
=V_0  A(z_{\max})
\left( z^*(z_{\max}) +\frac{t_b}{2} \right)
\, ,
\eeq
where $A(z)$ is given in eq. (\ref{A-arbitrary})
The action rate (including the factor of $2$) is
\beq
\frac{1}{V_0} \frac{d I_{\rm GHY} }{d t_b}=A(z_{\rm max}) \, , 
\eeq
which should be computed in the limit  $z_{\rm max} \rightarrow \infty$.

{\bf  Joint contributions:} 
There is only a contribution which is time dependent, 
which is the one located at $z=z_m$.
The light-like normals are:
\beq
k_1=( - \, dt + \frac{1}{f} e^{\frac{\chi}{2}} \, dz ) a_{\rm reg} \, ,  \qquad
k_2=  ( \, dt + \frac{1}{f} e^{\frac{\chi}{2}} \, dz )  a_{\rm reg} \, , 
\eeq
The join term  in eq. (\ref{jnn}) then gives
\beq
I_J = - \frac{V_0}{8 \pi G} \frac{1}{z_m^2} \log \left| \frac{f e^{-\chi}}{z^2 a_{\rm reg}^2} \right|_{z=z_m} \, ,
\eeq
where we took $\eta=1$.

At late time $z_m \rightarrow z_h$,  but  the term does not
approach to constant because it is log divergent.
The time derivative is:
\beq
\frac{d I_J}{d t_b}
 =  - \frac{V_0}{16 \pi G} 
\left\{ f e^{-\frac{\chi}{2}}
  \frac{d}{d z} \left( \frac{1}{z^2} \log \frac{-f e^{-\chi}}{z^2 a_{\rm reg}^2}\right)
  \right\}_{z=z_m} \, .
  \label{time-der-joint-late-time}
\eeq

{\bf Null boundaries counterterms:} From eq. (\ref{azione-controtermine}), including a factor of $2$
\beq
\frac{I_{ct}}{d t_b }
 =  \frac{V_0}{4 \pi G} \left\{  f e^{-\frac{\chi}{2}}
 \frac{1}{z^3}  \,  \log \left(  2 a_{\rm reg }  \,  {L}_{\rm ct} \, z \, e^{\frac{\chi}{2}}  \right)
 \right\}_{z=z_m} 
\eeq
This contribution vanishes at late time, as $z_m \to z_h$.
Combining the joint  term and the counterterm, we find
\beq
\frac{d( I_J+I_{ct})}{d t_b }
=\frac{V_0}{16 \pi G} 
 \left\{ f e^{-\frac{\chi}{2}} \left[ \frac{2}{z^3} 
  \log \left(   -4 f    \,  {L}_{\rm ct}^2  
  \right)
 -\frac{1}{z^2}   \frac{d}{d z} 
 \left( \log \frac{-f e^{-\chi}}{z^2 } \right)
  \right]
  \right\}_{z=z_m} 
\eeq
Note that the dependence on $a_{\rm reg}$ 
simplifies, due to reparametrization invariance.

\section{An estimate of $t_c$ for small $\phi_h$}

\label{appe-stima-tempo-critico}

This estimate is independent from the mass of the scalar,
because is derived using the behaviour nearby the collapse of the ER bridge,
which is universal in the scalar mass \cite{Hartnoll:2020rwq,Hartnoll:2020fhc}.
It also applies to the $q \neq 0$ case.

In the regime of small $\phi_h$, we have that the back reaction is small just
up to $z < z_c - \epsilon $, where $\epsilon  $ should go to zero for $\phi_h \to 0$.
In this limit we expect that the collapse of Einstein-Rosen bridge is very fast in the coordinate $z$, and so 
we expect that for $z > z_c - \epsilon $ we are already in the Kasner regime.
So we may say that:
\bea
z^*(z) &=& z^*_{\rm RN}(z) \qquad {\rm for} \qquad z < z_i=z_c - \epsilon   \, , \nl
z^*(z) &=& z^*_{\rm RN}(z_c-\epsilon)+ \frac{1}{2} \frac{e^{\frac{\chi_0}{2}}}{f_0} \le \frac{1}{z^2} - \frac{1}{z_c^2} \ri
\qquad {\rm for} \qquad z > z_i   \, , 
\label{zstar-approx-small-phih}
\eea
where we have used the Kasner approximation eq.~(\ref{kasner-approx}) for 
$z > z_c - \epsilon $. The critical time is given by (\ref{crittimeII}).

We can estimate $\epsilon$ using equation (\ref{soluzione-stoppacciosa}), 
in which $\tilde{A}=O(\phi_h^2)$ and $\tilde{B},\tilde{C}=O(\phi_h^0)$.
The location of $\epsilon$ is determined by the coordinate in which the terms
proportional to $\log H$ and the one proportional to $H$  are of the same order.
This happens for 
\beq
O(\phi_h) <\epsilon < O(\phi_h^2) \, .
\eeq
Using  eq. (\ref{zstar-RN-espessione-esplicita}),
we can approximate $z^*_{\rm RN}$    nearby $z_c$ as follows:
\beq
 z^*_{\rm RN}(z_c-\epsilon) \approx
\frac{z_h  z_c}{z_c-z_h}
\frac{1}{\left(\frac{z_c}{z_h} +1\right)^2 +2  } \log \frac {\epsilon}{z_c}  
\eeq
From eq. (\ref{zstar-approx-small-phih}), the critical time is given by
\beq
t_c = -2 z^*(\infty) =
2 \frac{z_h  z_c}{z_c-z_h}
\frac{1}{\left(\frac{z_c}{z_h} +1\right)^2 +2  } \log \le \frac {\epsilon}{z_c}  \ri
+ \frac{1}{f_0 e^{\frac{-\chi_0}{2}}}  \frac{1}{z_c^2} \, .
\label{tempo-critico-approssimazione}
\eeq
From eq. (\ref{f0-chi0-A}), we find that the 
first term in eq. (\ref{tempo-critico-approssimazione}) is subleading compared to the second,
and that the critical time scales as in eq. (\ref{stima-tempo-critico-small-phih}).
This behaviour is consistent with a numerical analysis, see figure \ref{critical-time}.


\begin{thebibliography}{99}

%\cite{Hartnoll:2020rwq}
\bibitem{Hartnoll:2020rwq}
S.~A.~Hartnoll, G.~T.~Horowitz, J.~Kruthoff and J.~E.~Santos,
``Gravitational duals to the grand canonical ensemble abhor Cauchy horizons,''
JHEP \textbf{10} (2020), 102
%doi:10.1007/JHEP10(2020)102
[arXiv:2006.10056 [hep-th]].
%11 citations counted in INSPIRE as of 10 Feb 2021

%\cite{Hartnoll:2020fhc}
\bibitem{Hartnoll:2020fhc}
S.~A.~Hartnoll, G.~T.~Horowitz, J.~Kruthoff and J.~E.~Santos,
``Diving into a holographic superconductor,''
SciPost Phys. \textbf{10} (2021), 009
%doi:10.21468/SciPostPhys.10.1.009
[arXiv:2008.12786 [hep-th]].
%4 citations counted in INSPIRE as of 10 Feb 2021


%\cite{Ryu:2006bv}
\bibitem{Ryu:2006bv}
S.~Ryu and T.~Takayanagi,
``Holographic derivation of entanglement entropy from AdS/CFT,''
Phys. Rev. Lett. \textbf{96} (2006), 181602
doi:10.1103/PhysRevLett.96.181602
[arXiv:hep-th/0603001 [hep-th]].
%2914 citations counted in INSPIRE as of 24 Jan 2022

%\cite{Hubeny:2007xt}
\bibitem{Hubeny:2007xt}
V.~E.~Hubeny, M.~Rangamani and T.~Takayanagi,
``A Covariant holographic entanglement entropy proposal,''
JHEP \textbf{07} (2007), 062
doi:10.1088/1126-6708/2007/07/062
[arXiv:0705.0016 [hep-th]].
%1318 citations counted in INSPIRE as of 24 Jan 2022

%\cite{Susskind:2014moa}
\bibitem{Susskind:2014moa}
L.~Susskind,
``Entanglement is not enough,''
Fortsch. Phys. \textbf{64} (2016), 49-71
doi:10.1002/prop.201500095
[arXiv:1411.0690 [hep-th]].
%266 citations counted in INSPIRE as of 29 Jan 2022

\bibitem{Susskind:2014rva}
  L.~Susskind,
   ``Computational Complexity and Black Hole Horizons,''
  [Fortsch.\ Phys.\  {\bf 64} (2016) 24]
   Addendum: Fortsch.\ Phys.\  {\bf 64} (2016) 44
 % doi:10.1002/prop.201500093, 10.1002/prop.201500092
  [arXiv:1403.5695 [hep-th], arXiv:1402.5674 [hep-th]].
  %%CITATION = doi:10.1002/prop.201500093, 10.1002/prop.201500092;%%
  %150 citations counted in INSPIRE as of 14 Apr 2018

\bibitem{Nielsen1}
Michael~A.~Nielsen,
  ``A geometric approach to quantum circuit lower bounds,''
  Quantum Information \& Computation,
Volume 6 Issue 3, May 2006,
Pages 213-262, [arXiv:quant-ph/0502070]

%\cite{Brown:2016wib}
\bibitem{Brown:2016wib}
A.~R.~Brown, L.~Susskind and Y.~Zhao,
``Quantum Complexity and Negative Curvature,''
Phys. Rev. D \textbf{95} (2017) no.4, 045010
doi:10.1103/PhysRevD.95.045010
[arXiv:1608.02612 [hep-th]].
%113 citations counted in INSPIRE as of 07 Feb 2022

%\cite{Brown:2017jil}
\bibitem{Brown:2017jil}
A.~R.~Brown and L.~Susskind,
``Second law of quantum complexity,''
Phys. Rev. D \textbf{97} (2018) no.8, 086015
doi:10.1103/PhysRevD.97.086015
[arXiv:1701.01107 [hep-th]].
%194 citations counted in INSPIRE as of 07 Feb 2022

%\cite{Brown:2019whu}
\bibitem{Brown:2019whu}
A.~R.~Brown and L.~Susskind,
``Complexity geometry of a single qubit,''
Phys. Rev. D \textbf{100} (2019) no.4, 046020
doi:10.1103/PhysRevD.100.046020
[arXiv:1903.12621 [hep-th]].
%37 citations counted in INSPIRE as of 07 Feb 2022

%\cite{Auzzi:2020idm}
\bibitem{Auzzi:2020idm}
R.~Auzzi, S.~Baiguera, G.~B.~De Luca, A.~Legramandi, G.~Nardelli and N.~Zenoni,
``Geometry of quantum complexity,''
Phys. Rev. D \textbf{103} (2021) no.10, 106021
doi:10.1103/PhysRevD.103.106021
[arXiv:2011.07601 [hep-th]].
%16 citations counted in INSPIRE as of 07 Feb 2022

%\cite{Brown:2021euk}
\bibitem{Brown:2021euk}
A.~R.~Brown,
``A Quantum Complexity Lowerbound from Differential Geometry,''
[arXiv:2112.05724 [hep-th]].
%1 citations counted in INSPIRE as of 07 Feb 2022

%\cite{Basteiro:2021ene}
\bibitem{Basteiro:2021ene}
P.~Basteiro, J.~Erdmenger, P.~Fries, F.~Goth, I.~Matthaiakakis and R.~Meyer,
``Quantum Complexity as Hydrodynamics,''
[arXiv:2109.01152 [hep-th]].
%3 citations counted in INSPIRE as of 16 May 2022

%\cite{Jefferson:2017sdb}
\bibitem{Jefferson:2017sdb}
R.~Jefferson and R.~C.~Myers,
``Circuit complexity in quantum field theory,''
JHEP \textbf{10} (2017), 107
doi:10.1007/JHEP10(2017)107
[arXiv:1707.08570 [hep-th]].
%288 citations counted in INSPIRE as of 07 Feb 2022

%\cite{Chapman:2017rqy}
\bibitem{Chapman:2017rqy}
S.~Chapman, M.~P.~Heller, H.~Marrochio and F.~Pastawski,
``Toward a Definition of Complexity for Quantum Field Theory States,''
Phys. Rev. Lett. \textbf{120} (2018) no.12, 121602
doi:10.1103/PhysRevLett.120.121602
[arXiv:1707.08582 [hep-th]].
%255 citations counted in INSPIRE as of 07 Feb 2022


  
  %\cite{Khan:2018rzm}
\bibitem{Khan:2018rzm}
R.~Khan, C.~Krishnan and S.~Sharma,
``Circuit Complexity in Fermionic Field Theory,''
Phys. Rev. D \textbf{98} (2018) no.12, 126001
doi:10.1103/PhysRevD.98.126001
[arXiv:1801.07620 [hep-th]].
%166 citations counted in INSPIRE as of 07 Feb 2022

%\cite{Hackl:2018ptj}
\bibitem{Hackl:2018ptj}
L.~Hackl and R.~C.~Myers,
``Circuit complexity for free fermions,''
JHEP \textbf{07} (2018), 139
doi:10.1007/JHEP07(2018)139
[arXiv:1803.10638 [hep-th]].
%150 citations counted in INSPIRE as of 07 Feb 2022

%\cite{Caputa:2017urj}
\bibitem{Caputa:2017urj}
P.~Caputa, N.~Kundu, M.~Miyaji, T.~Takayanagi and K.~Watanabe,
``Anti-de Sitter Space from Optimization of Path Integrals in Conformal Field Theories,''
Phys. Rev. Lett. \textbf{119} (2017) no.7, 071602
doi:10.1103/PhysRevLett.119.071602
[arXiv:1703.00456 [hep-th]].
%187 citations counted in INSPIRE as of 07 Feb 2022

%\cite{Caputa:2018kdj}
\bibitem{Caputa:2018kdj}
P.~Caputa and J.~M.~Magan,
``Quantum Computation as Gravity,''
Phys. Rev. Lett. \textbf{122} (2019) no.23, 231302
doi:10.1103/PhysRevLett.122.231302
[arXiv:1807.04422 [hep-th]].
%119 citations counted in INSPIRE as of 07 Feb 2022

%\cite{Erdmenger:2020sup}
\bibitem{Erdmenger:2020sup}
J.~Erdmenger, M.~Gerbershagen and A.~L.~Weigel,
``Complexity measures from geometric actions on Virasoro and Kac-Moody orbits,''
JHEP \textbf{11} (2020), 003
doi:10.1007/JHEP11(2020)003
[arXiv:2004.03619 [hep-th]].
%39 citations counted in INSPIRE as of 07 Feb 2022

%\cite{Flory:2020eot}
\bibitem{Flory:2020eot}
M.~Flory and M.~P.~Heller,
``Geometry of Complexity in Conformal Field Theory,''
Phys. Rev. Res. \textbf{2} (2020) no.4, 043438
doi:10.1103/PhysRevResearch.2.043438
[arXiv:2005.02415 [hep-th]].
%35 citations counted in INSPIRE as of 07 Feb 2022

%\cite{Chagnet:2021uvi}
\bibitem{Chagnet:2021uvi}
N.~Chagnet, S.~Chapman, J.~de Boer and C.~Zukowski,
``Complexity for Conformal Field Theories in General Dimensions,''
Phys. Rev. Lett. \textbf{128} (2022) no.5, 051601
doi:10.1103/PhysRevLett.128.051601
[arXiv:2103.06920 [hep-th]].
%25 citations counted in INSPIRE as of 07 Feb 2022

%\cite{Koch:2021tvp}
\bibitem{Koch:2021tvp}
R.~d.~Koch, M.~Kim and H.~J.~R.~Van Zyl,
``Complexity from spinning primaries,''
JHEP \textbf{12} (2021), 030
doi:10.1007/JHEP12(2021)030
[arXiv:2108.10669 [hep-th]].
%7 citations counted in INSPIRE as of 16 May 2022

%\cite{Stanford:2014jda}
\bibitem{Stanford:2014jda}
  D.~Stanford and L.~Susskind,
  ``Complexity and Shock Wave Geometries,''
  Phys.\ Rev.\ D {\bf 90} (2014) no.12,  126007
 % doi:10.1103/PhysRevD.90.126007
  [arXiv:1406.2678 [hep-th]].
  %%CITATION = doi:10.1103/PhysRevD.90.126007;%%
  %119 citations counted in INSPIRE as of 14 Apr 2018

%\cite{Brown:2015bva}
\bibitem{Brown:2015bva}
A.~R.~Brown, D.~A.~Roberts, L.~Susskind, B.~Swingle and Y.~Zhao,
``Holographic Complexity Equals Bulk Action?,''
Phys. Rev. Lett. \textbf{116} (2016) no.19, 191301
doi:10.1103/PhysRevLett.116.191301
[arXiv:1509.07876 [hep-th]].
%549 citations counted in INSPIRE as of 28 Jan 2022

%\cite{Brown:2015lvg}
\bibitem{Brown:2015lvg}
A.~R.~Brown, D.~A.~Roberts, L.~Susskind, B.~Swingle and Y.~Zhao,
``Complexity, action, and black holes,''
Phys. Rev. D \textbf{93} (2016) no.8, 086006
doi:10.1103/PhysRevD.93.086006
[arXiv:1512.04993 [hep-th]].
%376 citations counted in INSPIRE as of 30 Mar 2021


%\cite{Couch:2016exn}
\bibitem{Couch:2016exn}
J.~Couch, W.~Fischler and P.~H.~Nguyen,
``Noether charge, black hole volume, and complexity,''
JHEP \textbf{03} (2017), 119
doi:10.1007/JHEP03(2017)119
[arXiv:1610.02038 [hep-th]].
%164 citations counted in INSPIRE as of 29 Jan 2022

%\cite{Belin:2021bga}
\bibitem{Belin:2021bga}
A.~Belin, R.~C.~Myers, S.~M.~Ruan, G.~S\'arosi and A.~J.~Speranza,
``Complexity Equals Anything?,''
[arXiv:2111.02429 [hep-th]].
%4 citations counted in INSPIRE as of 01 Jan 2022

%\cite{Susskind:2014jwa}
\bibitem{Susskind:2014jwa}
L.~Susskind and Y.~Zhao,
``Switchbacks and the Bridge to Nowhere,''
[arXiv:1408.2823 [hep-th]].
%159 citations counted in INSPIRE as of 06 Feb 2022

%\cite{Carmi:2016wjl}
\bibitem{Carmi:2016wjl}
D.~Carmi, R.~C.~Myers and P.~Rath,
``Comments on Holographic Complexity,''
JHEP \textbf{03} (2017), 118
doi:10.1007/JHEP03(2017)118
[arXiv:1612.00433 [hep-th]].
%239 citations counted in INSPIRE as of 07 Feb 2022

%\cite{Reynolds:2016rvl}
\bibitem{Reynolds:2016rvl}
A.~Reynolds and S.~F.~Ross,
``Divergences in Holographic Complexity,''
Class. Quant. Grav. \textbf{34} (2017) no.10, 105004
doi:10.1088/1361-6382/aa6925
[arXiv:1612.05439 [hep-th]].
%103 citations counted in INSPIRE as of 07 Feb 2022

%\cite{Akhavan:2019zax}
\bibitem{Akhavan:2019zax}
A.~Akhavan and F.~Omidi,
``On the Role of Counterterms in Holographic Complexity,''
JHEP \textbf{11} (2019), 054
doi:10.1007/JHEP11(2019)054
[arXiv:1906.09561 [hep-th]].
%15 citations counted in INSPIRE as of 29 Jun 2022


%\cite{Omidi:2020oit}
\bibitem{Omidi:2020oit}
F.~Omidi,
``Regularizations of Action-Complexity for a Pure BTZ Black Hole Microstate,''
JHEP \textbf{07} (2020), 020
doi:10.1007/JHEP07(2020)020
[arXiv:2004.11628 [hep-th]].
%9 citations counted in INSPIRE as of 29 Jun 2022

%\cite{Susskind:2015toa}
\bibitem{Susskind:2015toa}
L.~Susskind,
``The Typical-State Paradox: Diagnosing Horizons with Complexity,''
Fortsch. Phys. \textbf{64} (2016), 84-91
doi:10.1002/prop.201500091
[arXiv:1507.02287 [hep-th]].
%62 citations counted in INSPIRE as of 01 May 2022

%\cite{Cai:2016xho}
\bibitem{Cai:2016xho}
R.~G.~Cai, S.~M.~Ruan, S.~J.~Wang, R.~Q.~Yang and R.~H.~Peng,
``Action growth for AdS black holes,''
JHEP \textbf{09} (2016), 161
doi:10.1007/JHEP09(2016)161
[arXiv:1606.08307 [gr-qc]].
%153 citations counted in INSPIRE as of 18 Dec 2021

%\cite{Carmi:2017jqz}
\bibitem{Carmi:2017jqz}
D.~Carmi, S.~Chapman, H.~Marrochio, R.~C.~Myers and S.~Sugishita,
 ``On the Time Dependence of Holographic Complexity,''
JHEP \textbf{11} (2017), 188
%doi:10.1007/JHEP11(2017)188
[arXiv:1709.10184 [hep-th]].
%175 citations counted in INSPIRE as of 10 Feb 2021

%\cite{Yang:2017czx}
\bibitem{Yang:2017czx}
R.~Q.~Yang, C.~Niu, C.~Y.~Zhang and K.~Y.~Kim,
``Comparison of holographic and field theoretic complexities for time dependent thermofield double states,''
JHEP \textbf{02} (2018), 082
doi:10.1007/JHEP02(2018)082
[arXiv:1710.00600 [hep-th]].
%106 citations counted in INSPIRE as of 30 Jan 2022



%\cite{Auzzi:2018zdu}
\bibitem{Auzzi:2018zdu}
R.~Auzzi, S.~Baiguera and G.~Nardelli,
``Volume and complexity for warped AdS black holes,''
JHEP \textbf{06} (2018), 063
doi:10.1007/JHEP06(2018)063
[arXiv:1804.07521 [hep-th]].
%21 citations counted in INSPIRE as of 06 May 2022

%\cite{Auzzi:2018pbc}
\bibitem{Auzzi:2018pbc}
R.~Auzzi, S.~Baiguera, M.~Grassi, G.~Nardelli and N.~Zenoni,
``Complexity and action for warped AdS black holes,''
JHEP \textbf{09} (2018), 013
doi:10.1007/JHEP09(2018)013
[arXiv:1806.06216 [hep-th]].
%29 citations counted in INSPIRE as of 27 Jan 2022

%\cite{Bernamonti:2021jyu}
\bibitem{Bernamonti:2021jyu}
A.~Bernamonti, F.~Bigazzi, D.~Billo, L.~Faggi and F.~Galli,
``Holographic and QFT complexity with angular momentum,''
JHEP \textbf{11} (2021), 037
doi:10.1007/JHEP11(2021)037
[arXiv:2108.09281 [hep-th]].
%5 citations counted in INSPIRE as of 18 Dec 2021




%\cite{Iliesiu:2021ari}
\bibitem{Iliesiu:2021ari}
L.~V.~Iliesiu, M.~Mezei and G.~S\'arosi,
``The volume of the black hole interior at late times,''
[arXiv:2107.06286 [hep-th]].
%20 citations counted in INSPIRE as of 01 May 2022

%\cite{Maldacena:2001kr}
\bibitem{Maldacena:2001kr}
J.~M.~Maldacena,
``Eternal black holes in anti-de Sitter,''
JHEP \textbf{04} (2003), 021
doi:10.1088/1126-6708/2003/04/021
[arXiv:hep-th/0106112 [hep-th]].
%1125 citations counted in INSPIRE as of 05 Feb 2022

%\cite{Simpson:1973ua}
\bibitem{Simpson:1973ua}
M.~Simpson and R.~Penrose,
``Internal instability in a Reissner-Nordstrom black hole,''
Int. J. Theor. Phys. \textbf{7} (1973), 183-197
doi:10.1007/BF00792069
%155 citations counted in INSPIRE as of 03 Feb 2022


\bibitem{Chandrasekhar-Hartle}
S.~Chandrasekhar
and J.~B.~Hartle,
``On crossing the Cauchy horizon of a Reissner-Nordstrom black hole,''
Proc. R. Soc. London A 384 (1982), 301-315
doi:10.1098/rspa.1982.0160







%\cite{Hartnoll:2008vx}
\bibitem{Hartnoll:2008vx}
S.~A.~Hartnoll, C.~P.~Herzog and G.~T.~Horowitz,
``Building a Holographic Superconductor,''
Phys. Rev. Lett. \textbf{101} (2008), 031601
%doi:10.1103/PhysRevLett.101.031601
[arXiv:0803.3295 [hep-th]].
%1385 citations counted in INSPIRE as of 17 Aug 2021

%\cite{Hartnoll:2008kx}
\bibitem{Hartnoll:2008kx}
S.~A.~Hartnoll, C.~P.~Herzog and G.~T.~Horowitz,
``Holographic Superconductors,''
JHEP \textbf{12} (2008), 015
%doi:10.1088/1126-6708/2008/12/015
[arXiv:0810.1563 [hep-th]].
%1052 citations counted in INSPIRE as of 17 Aug 2021



%\cite{Yang:2019gce}
\bibitem{Yang:2019gce}
R.~Q.~Yang, H.~S.~Jeong, C.~Niu and K.~Y.~Kim,
 ``Complexity of Holographic Superconductors,''
JHEP \textbf{04} (2019), 146
%doi:10.1007/JHEP04(2019)146
[arXiv:1902.07586 [hep-th]].
%23 citations counted in INSPIRE as of 10 Feb 2021

\bibitem{Lloyd}
S. Lloyd, ``Ultimate physical limits to computation," 
Nature 406 (2000), no. 6799 1047-1054.

%\cite{Fidkowski:2003nf}
\bibitem{Fidkowski:2003nf}
L.~Fidkowski, V.~Hubeny, M.~Kleban and S.~Shenker,
 ``The Black hole singularity in AdS / CFT,''
JHEP \textbf{02} (2004), 014
%doi:10.1088/1126-6708/2004/02/014
[arXiv:hep-th/0306170 [hep-th]].
%269 citations counted in INSPIRE as of 25 Feb 2021


%\cite{Barbon:2015ria}
\bibitem{Barbon:2015ria}
J.~L.~F.~Barbon and E.~Rabinovici,
``Holographic complexity and spacetime singularities,''
JHEP \textbf{01} (2016), 084
doi:10.1007/JHEP01(2016)084
[arXiv:1509.09291 [hep-th]].
%73 citations counted in INSPIRE as of 06 May 2022

%\cite{Bolognesi:2018ion}
\bibitem{Bolognesi:2018ion}
S.~Bolognesi, E.~Rabinovici and S.~R.~Roy,
``On Some Universal Features of the Holographic Quantum Complexity of Bulk Singularities,''
JHEP \textbf{06} (2018), 016
doi:10.1007/JHEP06(2018)016
[arXiv:1802.02045 [hep-th]].
%25 citations counted in INSPIRE as of 04 May 2022

%\cite{An:2022lvo}
\bibitem{An:2022lvo}
Y.~S.~An, L.~Li, F.~G.~Yang and R.~Q.~Yang,
``Interior Structure and Complexity Growth Rate of Holographic Superconductor from M-Theory,''
[arXiv:2205.02442 [hep-th]].
%0 citations counted in INSPIRE as of 06 May 2022

 
%\cite{Breitenlohner:1982jf}
\bibitem{Breitenlohner:1982jf}
P.~Breitenlohner and D.~Z.~Freedman,
``Stability in Gauged Extended Supergravity,''
Annals Phys. \textbf{144} (1982), 249
doi:10.1016/0003-4916(82)90116-6
%1551 citations counted in INSPIRE as of 29 Jan 2022


%\cite{Cai:2020wrp}
\bibitem{Cai:2020wrp}
R.~G.~Cai, L.~Li and R.~Q.~Yang,
``No Inner-Horizon Theorem for Black Holes with Charged Scalar Hairs,''
JHEP \textbf{03} (2021), 263
doi:10.1007/JHEP03(2021)263
[arXiv:2009.05520 [gr-qc]].
%20 citations counted in INSPIRE as of 06 May 2022

%\cite{An:2021plu}
\bibitem{An:2021plu}
Y.~S.~An, L.~Li and F.~G.~Yang,
``No Cauchy horizon theorem for nonlinear electrodynamics black holes with charged scalar hairs,''
Phys. Rev. D \textbf{104} (2021) no.2, 024040
doi:10.1103/PhysRevD.104.024040
[arXiv:2106.01069 [gr-qc]].
%9 citations counted in INSPIRE as of 16 May 2022

%\cite{Frenkel:2020ysx}
\bibitem{Frenkel:2020ysx}
A.~Frenkel, S.~A.~Hartnoll, J.~Kruthoff and Z.~D.~Shi,
``Holographic flows from CFT to the Kasner universe,''
JHEP \textbf{08} (2020), 003
%doi:10.1007/JHEP08(2020)003
[arXiv:2004.01192 [hep-th]].
%5 citations counted in INSPIRE as of 27 Apr 2021

%\cite{Henneaux:2022ijt}
\bibitem{Henneaux:2022ijt}
M.~Henneaux,
``The final Kasner regime inside black holes with scalar or vector hair,''
JHEP \textbf{03} (2022), 062
doi:10.1007/JHEP03(2022)062
[arXiv:2202.04155 [hep-th]].
%1 citations counted in INSPIRE as of 06 May 2022

%\cite{Mansoori:2021wxf}
\bibitem{Mansoori:2021wxf}
S.~A.~H.~Mansoori, L.~Li, M.~Rafiee and M.~Baggioli,
``What's inside a hairy black hole in massive gravity?,''
JHEP \textbf{10} (2021), 098
doi:10.1007/JHEP10(2021)098
[arXiv:2108.01471 [hep-th]].
%6 citations counted in INSPIRE as of 16 May 2022

%\cite{Sword:2021pfm}
\bibitem{Sword:2021pfm}
L.~Sword and D.~Vegh,
``Kasner geometries inside holographic superconductors,''
JHEP \textbf{04} (2022), 135
doi:10.1007/JHEP04(2022)135
[arXiv:2112.14177 [hep-th]].
%2 citations counted in INSPIRE as of 06 May 2022

%\cite{Caceres:2022smh}
\bibitem{Caceres:2022smh}
E.~Caceres, A.~Kundu, A.~K.~Patra and S.~Shashi,
``Trans-IR Flows to Black Hole Singularities,''
[arXiv:2201.06579 [hep-th]].
%2 citations counted in INSPIRE as of 16 May 2022


%\cite{Banados:1992wn}
\bibitem{Banados:1992wn}
M.~Banados, C.~Teitelboim and J.~Zanelli,
``The Black hole in three-dimensional space-time,''
Phys. Rev. Lett. \textbf{69} (1992), 1849-1851
doi:10.1103/PhysRevLett.69.1849
[arXiv:hep-th/9204099 [hep-th]].
%2945 citations counted in INSPIRE as of 26 Jan 2022

%\cite{Balasubramanian:1999zv}
\bibitem{Balasubramanian:1999zv}
V.~Balasubramanian and S.~F.~Ross,
``Holographic particle detection,''
Phys. Rev. D \textbf{61} (2000), 044007
doi:10.1103/PhysRevD.61.044007
[arXiv:hep-th/9906226 [hep-th]].
%213 citations counted in INSPIRE as of 24 Jan 2022


%\cite{Festuccia:2005pi}
\bibitem{Festuccia:2005pi}
G.~Festuccia and H.~Liu,
``Excursions beyond the horizon: Black hole singularities in Yang-Mills theories. I.,''
JHEP \textbf{04} (2006), 044
%doi:10.1088/1126-6708/2006/04/044
[arXiv:hep-th/0506202 [hep-th]].
%122 citations counted in INSPIRE as of 26 Feb 2021

%\cite{Festuccia:2008zx}
\bibitem{Festuccia:2008zx}
G.~Festuccia and H.~Liu,
 ``A Bohr-Sommerfeld quantization formula for quasinormal frequencies of AdS black holes,''
Adv. Sci. Lett. \textbf{2} (2009), 221-235
%doi:10.1166/asl.2009.1029
[arXiv:0811.1033 [gr-qc]].
%69 citations counted in INSPIRE as of 26 Feb 2021


%\cite{Kraus:2002iv}
\bibitem{Kraus:2002iv}
P.~Kraus, H.~Ooguri and S.~Shenker,
 ``Inside the horizon with AdS / CFT,''
Phys. Rev. D \textbf{67} (2003), 124022
%doi:10.1103/PhysRevD.67.124022
[arXiv:hep-th/0212277 [hep-th]].
%203 citations counted in INSPIRE as of 26 Feb 2021


%\cite{Hartman:2013qma}
\bibitem{Hartman:2013qma}
T.~Hartman and J.~Maldacena,
 ``Time Evolution of Entanglement Entropy from Black Hole Interiors,''
JHEP \textbf{05} (2013), 014
%doi:10.1007/JHEP05(2013)014
[arXiv:1303.1080 [hep-th]].
%430 citations counted in INSPIRE as of 26 Feb 2021



%\cite{Gubser:2009cg}
\bibitem{Gubser:2009cg}
S.~S.~Gubser and A.~Nellore,
``Ground states of holographic superconductors,''
Phys. Rev. D \textbf{80} (2009), 105007
doi:10.1103/PhysRevD.80.105007
[arXiv:0908.1972 [hep-th]].
%178 citations counted in INSPIRE as of 14 Jan 2022


%\cite{Chapman:2018dem}
\bibitem{Chapman:2018dem}
S.~Chapman, H.~Marrochio and R.~C.~Myers,
``Holographic complexity in Vaidya spacetimes. Part I,''
JHEP \textbf{06} (2018), 046
doi:10.1007/JHEP06(2018)046
[arXiv:1804.07410 [hep-th]].
%126 citations counted in INSPIRE as of 27 Jan 2022

%\cite{Chapman:2018lsv}
\bibitem{Chapman:2018lsv}
S.~Chapman, H.~Marrochio and R.~C.~Myers,
``Holographic complexity in Vaidya spacetimes. Part II,''
JHEP \textbf{06} (2018), 114
doi:10.1007/JHEP06(2018)114
[arXiv:1805.07262 [hep-th]].
%113 citations counted in INSPIRE as of 27 Jan 2022





%\cite{Couch:2017yil}
\bibitem{Couch:2017yil}
J.~Couch, S.~Eccles, W.~Fischler and M.~L.~Xiao,
``Holographic complexity and noncommutative gauge theory,''
JHEP \textbf{03} (2018), 108
doi:10.1007/JHEP03(2018)108
[arXiv:1710.07833 [hep-th]].
%62 citations counted in INSPIRE as of 30 Jan 2022

%\cite{Swingle:2017zcd}
\bibitem{Swingle:2017zcd}
B.~Swingle and Y.~Wang,
``Holographic Complexity of Einstein-Maxwell-Dilaton Gravity,''
JHEP \textbf{09} (2018), 106
doi:10.1007/JHEP09(2018)106
[arXiv:1712.09826 [hep-th]].
%84 citations counted in INSPIRE as of 30 Jan 2022



%\cite{An:2018xhv}
\bibitem{An:2018xhv}
Y.~S.~An and R.~H.~Peng,
``Effect of the dilaton on holographic complexity growth,''
Phys. Rev. D \textbf{97} (2018) no.6, 066022
doi:10.1103/PhysRevD.97.066022
[arXiv:1801.03638 [hep-th]].
%72 citations counted in INSPIRE as of 30 Jan 2022

%\cite{Alishahiha:2018tep}
\bibitem{Alishahiha:2018tep}
M.~Alishahiha, A.~Faraji Astaneh, M.~R.~Mohammadi Mozaffar and A.~Mollabashi,
``Complexity Growth with Lifshitz Scaling and Hyperscaling Violation,''
JHEP \textbf{07} (2018), 042
doi:10.1007/JHEP07(2018)042
[arXiv:1802.06740 [hep-th]].
%80 citations counted in INSPIRE as of 30 Jan 2022

%\cite{Mahapatra:2018gig}
\bibitem{Mahapatra:2018gig}
S.~Mahapatra and P.~Roy,
``On the time dependence of holographic complexity in a dynamical Einstein-dilaton model,''
JHEP \textbf{11} (2018), 138
doi:10.1007/JHEP11(2018)138
[arXiv:1808.09917 [hep-th]].
%42 citations counted in INSPIRE as of 16 May 2022

%\cite{Babaei-Aghbolagh:2021ast}
\bibitem{Babaei-Aghbolagh:2021ast}
H.~Babaei-Aghbolagh, D.~M.~Yekta, K.~Velni Babaei and H.~Mohammadzadeh,
``Complexity growth in Gubser\textendash{}Rocha models with momentum relaxation,''
Eur. Phys. J. C \textbf{82} (2022) no.4, 383
doi:10.1140/epjc/s10052-022-10253-9
[arXiv:2112.10725 [hep-th]].
%2 citations counted in INSPIRE as of 16 May 2022

%\cite{Avila:2021zhb}
\bibitem{Avila:2021zhb}
D.~\'Avila, C.~D\'\i{}az, Y.~D.~Olivas and L.~Pati\~no,
``Insensitivity of the complexity rate of change to the conformal anomaly and Lloyd\textquoteright{}s bound as a possible renormalization condition,''
Phys. Rev. D \textbf{104} (2021) no.6, 066011
doi:10.1103/PhysRevD.104.066011
[arXiv:2104.12796 [hep-th]].
%0 citations counted in INSPIRE as of 16 May 2022


%\cite{Balasubramanian:1999re}
\bibitem{Balasubramanian:1999re}
V.~Balasubramanian and P.~Kraus,
``A Stress tensor for Anti-de Sitter gravity,''
Commun. Math. Phys. \textbf{208} (1999), 413-428
doi:10.1007/s002200050764
[arXiv:hep-th/9902121 [hep-th]].
%1695 citations counted in INSPIRE as of 01 May 2022

%\cite{deHaro:2000vlm}
\bibitem{deHaro:2000vlm}
S.~de Haro, S.~N.~Solodukhin and K.~Skenderis,
``Holographic reconstruction of space-time and renormalization in the AdS / CFT correspondence,''
Commun. Math. Phys. \textbf{217} (2001), 595-622
doi:10.1007/s002200100381
[arXiv:hep-th/0002230 [hep-th]].
%1540 citations counted in INSPIRE as of 01 May 2022

%\cite{Caldarelli:2016nni}
\bibitem{Caldarelli:2016nni}
M.~M.~Caldarelli, A.~Christodoulou, I.~Papadimitriou and K.~Skenderis,
``Phases of planar AdS black holes with axionic charge,''
JHEP \textbf{04} (2017), 001
doi:10.1007/JHEP04(2017)001
[arXiv:1612.07214 [hep-th]].
%42 citations counted in INSPIRE as of 01 May 2022


\bibitem{Hayward:1993my}
  G.~Hayward,
   ``Gravitational action for space-times with nonsmooth boundaries,''
  Phys.\ Rev.\ D {\bf 47} (1993) 3275.
 % doi:10.1103/PhysRevD.47.3275
  %%CITATION = doi:10.1103/PhysRevD.47.3275;%%
  %106 citations counted in INSPIRE as of 29 May 2018


\bibitem{Lehner:2016vdi}
  L.~Lehner, R.~C.~Myers, E.~Poisson and R.~D.~Sorkin,
  ``Gravitational action with null boundaries,''
  Phys.\ Rev.\ D {\bf 94} (2016) no.8,  084046
  doi:10.1103/PhysRevD.94.084046
  [arXiv:1609.00207 [hep-th]].
  %%CITATION = doi:10.1103/PhysRevD.94.084046;%%
  %78 citations counted in INSPIRE as of 29 May 2018  



\end{thebibliography}
\end{document}